\documentclass[11pt]{article}

\usepackage[utf8]{inputenc}
\usepackage[a4paper, total={5in, 8in}]{geometry}

\usepackage[noblocks]{authblk}

\usepackage{amsmath}
\usepackage{graphicx}
\usepackage{float}
\usepackage{xcolor,colortbl}
\usepackage{placeins}

\definecolor{Gray}{gray}{0.85}
\definecolor{LightCyan}{rgb}{0.88,1,1}

\usepackage{psfrag,epsf}
\usepackage{enumerate}
\usepackage{natbib}
\usepackage{url} 

\usepackage{wrapfig}

\usepackage{graphicx}
\usepackage{tikz}
\usetikzlibrary{fit,positioning,arrows,automata}
\usetikzlibrary{shapes,shadows,arrows,positioning,graphs}

\usepackage{amssymb,verbatim,color}
\usepackage{bm,lscape}
\usepackage{wasysym}
\usepackage{bbm}
\RequirePackage[colorlinks,citecolor=blue,urlcolor=blue]{hyperref}
\usepackage{theorem}
\usepackage{subcaption}
\usepackage{algorithm}
\usepackage[noend]{algpseudocode}

\def\bSig\mathbf{\Sigma}

\definecolor{darkblue}{rgb}{0.0, 0.0, 0.55}

\newcommand{\beq}{\begin{equation}}
	\newcommand{\eeq}{\end{equation}}

\theoremstyle{plain}

\allowdisplaybreaks

\title{Change point detection in  
dynamic Gaussian graphical models:   the impact of COVID-19 pandemic on the US stock market}

\author[1]{Beatrice Franzolini}
\author[2]{Alexandros Beskos}
\author[3]{Maria De Iorio}
\author[4]{Warrick Poklewski Koziell} 
\author[5]{Karolina Grzeszkiewicz}
\affil[1,3]{Singapore Institute for Clinical Sciences, 
	 Agency for Science, Technology and Research, Singapore, Republic of Singapore}
\affil[2,3,4]{Department of Statistical Science, University College London, UK}
\affil[3]{Yong Loo Lin School of Medicine, National University of Singapore, Singapore, Republic of Singapore}
\affil[5]{Yale-NUS College, Singapore, Republic of Singapore}
\affil[2]{The Alan Turing Institute, London, UK}
\date{}

\begin{document}

	\maketitle

\begin{abstract}
Reliable estimates of volatility and correlation are fundamental in economics and
 finance for understanding the impact of macroeconomics events on the market and guiding future investments and policies. Dependence across financial returns is likely to be subject to sudden structural changes, especially in correspondence with major global events, such as the COVID-19 pandemic. In this work, we are interested in capturing abrupt changes over time in the {conditional} dependence across US industry stock portfolios, over a time horizon that covers the COVID-19 pandemic. The selected stocks give a comprehensive picture of the US stock market. 
To this end, we develop a Bayesian multivariate stochastic volatility model based on a time-varying sequence of graphs capturing the evolution of the dependence structure. The model builds on the Gaussian graphical models and the random change points literature. In
 particular, we treat the number, the position of change points, and the graphs as  object of posterior inference, allowing for sparsity in graph recovery and change point detection. The high dimension of the parameter space poses complex computational challenges. However, the model admits a hidden Markov model formulation. This leads to the development of an efficient computational strategy, based on a combination of sequential Monte-Carlo and Markov chain Monte-Carlo techniques. Model and computational development are widely applicable, beyond the scope of the application of interest in this work.
\end{abstract}
\section{Introduction}
\label{sec:intro}
Understanding the temporal evolution of the dependence structure among time series  is a fundamental topic in many fields, such as psychology \citep{williams2021bayesian}, speech recognition \citep{bilmes2004graphical}, genomics \citep{yin2011sparse}, and, in particular, finance. In this latter context, estimates of volatility and correlation of different financial instruments are largely used for portfolio allocation, option-pricing, and to draw conclusions about the impact of macroeconomic events on the markets with the goal of guiding future investments and policies.
In particular, estimates of correlation are key to minimise the risk of investment portfolios and define hedging strategies \citep[see, among others,][]{lien2002evaluating, lee2010regime, thampanya2020asymmetric, dutta2021climate}. Changes in correlation modify the return/risk profile of the investments and are of interest to both investors and policy makers. To understand how to better prepare for and deal with future major global events, it is important to estimate the impact of the COVID-19 pandemic on the volatility and the dependence structure of financial instruments \citep{just2020stock,sakurai2020has,alqaralleh2021evidence,guidolin2021boosting,yousfi2021effects, derbali2022covid, dey2022impacts}.
{Global ``catastrophic" events, such as financial crises, often lead to sudden changes in the dependence structure across. Financial markets' reaction to the pandemic appears to be no exception: around the end of February 2020 the Dow Jones and S\&P 500 fell by 11\% and
12\%, respectively, marking the biggest weekly decline  since the financial crisis of
2008, to the point that  the Financial Times described such decline as the ``quickest correction since the Great Depression". Standard statistical approaches assume time-varying dependence to change smoothly over time, which appears to be an unrealistic assumption when investigating  financial shocks.}
In this manuscript, we develop statistical machinery to detect abrupt changes in the correlation structure among time series. Such machinery is employed to detect the impact of the COVID-19 pandemic on the US stock market and, in particular, on cross-industry relationships. 

 There exists a vast literature on  models for time-varying second moments. More specifically, there are two main approaches: \emph{conditional volatility models}, as the well-known ARCH and GARCH  \citep{engle1982autoregressive, bollerslev1994arch, bollerslev1986generalized, engle1986modelling,bauwens2006multivariate, silvennoinen2009multivariate, boudt2019multivariate}, and \emph{stochastic volatility models} \citep[e.g.,][]{taylor1982financial,wiggins1987option, hull1987pricing, asai2006multivariate}. The former class specifies second moments at a certain time $t$ as a deterministic function of past values of observations, volatility, and possibly covariance, given model parameters. The latter assumes second moments to follow a latent stochastic process, typically of Markovian structure, so that, even conditionally on all past information, volatility and correlations are unobservable random variables evolving over time. 
While stochastic volatility models are often more flexible and may achieve better inferential performances when compared to conditional volatility approaches \citep{chan2013moving, clark2015macroeconomic}, they are more difficult to estimate since the likelihood is typically intractable, see, e.g.,~\cite{nilsson2016stochastic}. 

\begin{figure}[!t]
\centering
\resizebox{\columnwidth}{!}{%
\begin{tikzpicture}
\node[draw = white] (graph) at (1,1) {\textbf{Market n.1}};

\node[circle,minimum size=5mm,draw=black,fill=green, fill opacity=0.2,label=center:X] (X) at (0,0) {};
\node[circle,minimum size=5mm,draw=black,fill=green, fill opacity=0.2,label=center:Y] (Y) at (2,0) {};
\node[circle,minimum size=5mm,draw=black,fill=green, fill opacity=0.2,label=center:Z] (Z) at (1,-1.2) {};

\path [-] (Z) edge node { } (X);
\path [-] (Z) edge node { } (Y);

\node[draw = white] (graph) at (1,-2.2) {$X = Z +\epsilon_x$};
\node[draw = white] (grapha) at (1,-2.7) {$Y = Z +\epsilon_y$};
\node[draw = white] (graphb) at (0.63,-3.2) {$Z = \epsilon_z$};

\node[draw = white] (graph) at (6,1) {\textbf{Market n.2}};

\node[circle,minimum size=5mm,draw=black,fill=green, fill opacity=0.2,label=center:X] (X1) at (5,0) {};
\node[circle,minimum size=5mm,draw=black,fill=green, fill opacity=0.2,label=center:Y] (Y1) at (7,0) {};
\node[circle,minimum size=5mm,draw=black,fill=green, fill opacity=0.2,label=center:Z] (Z1) at (6,-1.2) {};

\path [-] (Z1) edge node { } (X1);
\path [-] (Z1) edge node { } (Y1);
\path [-] (X1) edge node { } (Y1);

\node[draw = white] (graph1) at (6,-2.2) {$X = W +\epsilon_x$};
\node[draw = white] (graph1a) at (6,-2.7) {$Y = W +\epsilon_y$};
\node[draw = white] (graph1b) at (6,-3.2) {$Z = W + \epsilon_z$};

\node[draw = white] (graph) at (11,1) {\textbf{Market n.3}};
\node[circle,minimum size=5mm,draw=black,fill=green, fill opacity=0.2,label=center:X] (X2) at (10,0) {};
\node[circle,minimum size=5mm,draw=black,fill=green, fill opacity=0.2,label=center:Y] (Y2) at (12,0) {};
\node[circle,minimum size=5mm,draw=black,fill=green, fill opacity=0.2,label=center:Z] (Z2) at (11,-1.2) {};

\path [-] (X2) edge node { } (Y2);

\node[draw = white] (graph2) at (11,-2.2) {$X = W +\epsilon_x$};
\node[draw = white] (graph2a) at (11,-2.7) {$Y = W +\epsilon_y$};
\node[draw = white] (graph2b) at (10.6,-3.2) {$Z = \epsilon_z$};

\node[draw = white] (graph) at (16,1) {\textbf{Market n.4}};
\node[circle,minimum size=5mm,draw=black,fill=green, fill opacity=0.2,label=center:X] (X3) at (15,0) {};
\node[circle,minimum size=5mm,draw=black,fill=green, fill opacity=0.2,label=center:Y] (Y3) at (17,0) {};
\node[circle,minimum size=5mm,draw=black,fill=green, fill opacity=0.2,label=center:Z] (Z3) at (16,-1.2) {};

\node[draw = white] (graph3) at (16,-2.2) {$X = \epsilon_x$};
\node[draw = white] (graph3a) at (16,-2.7) {$Y = \epsilon_y$};
\node[draw = white] (graph3b) at (16,-3.2) {$Z = \epsilon_z$};
\end{tikzpicture} }
\caption{\label{fig:toy}Toy example of graphs substructures in a financial market with three assets. Market n.1: the asset $Z$ acts as common risk factor in the market driving the dependence and is a hub.  Market n.2: the return of all three assets $X$, $Y$, and $Z$ are driven by an unobservable risk factor $W$ and the tree assets form a clique. Market n.3: $X$ and $Y$ are driven by a common risk factor, not affecting the asset $Z$. Market n.4: the three assets are independent and the graph does not include any edge. Here $\epsilon_z,\epsilon_x,\epsilon_y$ are white noises.}
\end{figure}
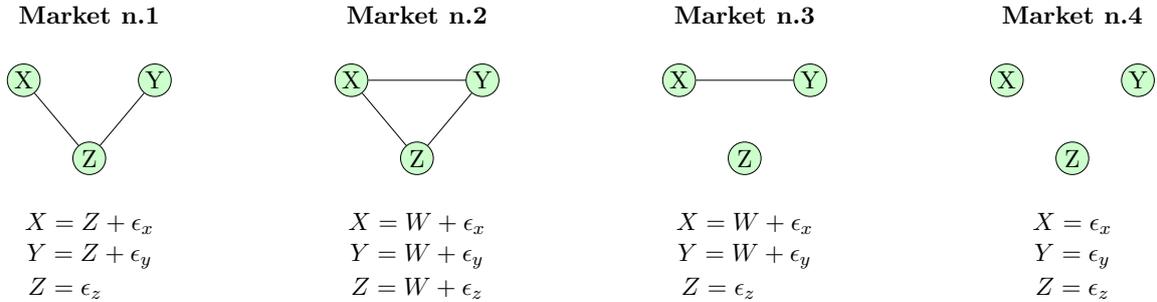

Within both classes, 
a  further distinction may be made  between models that explicitly target the covariance matrix $\Sigma_t$ and those focusing on the precision matrix $\Omega_t = \Sigma_t^{-1}$, specifically allowing for zero entries in $\Omega_t$ to favour parsimony. In this work, we develop a Bayesian stochastic volatility model for the precision matrix. {Specifically}, the precision matrix at time $t$ is modelled conditionally on a graph at time $t$, which describes the dependence structure among time series. As such, our work lies within the literature on Gaussian graphical models (GGMs) \citep[see, e.g.,][]{carvalho2007dynamic,Wang2009,prado2010time,wang2010sparse, chandra2021bayesian}. This approach presents an important advantage: GGMs target conditional independence instead of marginal, leading to possible identification of  macro-components (represented, for instance, by hubs and cliques in the graph)  and safeguarding against
spurious relationships{, in the sense that {GGMs}  aid understanding if pairwise correlations between variables can be fully or partially explained by their relationship with one or more {additional variables.}} The identification of graph substructures  is of particular interest in finance, where hubs may be interpreted as risk factors driving the market, while cliques represent financial instruments exposed to the same unobserved risk factor. See Figure~\ref{fig:toy} for a toy example clarifying the role of graph substructures in financial markets {and, in particular, {the interpretation of  hubs as risk factors}}. 
Moreover, marginal dependence and Pearson correlation simply measure the pairwise co-movement between two investments, but do not provide any indication on whether a risk factor generating the co-movement is specific to the two investments or it is common also to other financial instruments of interest. On the contrary, entries of the precision matrix represent co-movements conditionally on the effect of all the other instruments considered in the model \citep{michis2022multiscale}, so that an entry is non-zero if and only if the two returns are dependent conditionally on all other investments. 

Changes over time of second moments can be smooth or abrupt. The focus of this work is on changes of the second type. Standard versions of the models cited so far assume variances and covariances changing smoothly over time and, in particular, between any two consecutive time points. For instance, \cite{carvalho2007dynamic} propose a Bayesian dynamic stochastic volatility model based on GGMs and conditional independence. In their construction, the graph structure is kept constant over time, while the covariance matrix changes smoothly between any two consecutive time points.  However, this feature is often in contrast with what is observed in financial markets, where volatility clusters (i.e., periods with a persistent value of volatility, that are interrupted by sudden changes) and correlation breakdowns (i.e., substantial changes in correlations during stressed times and financial crises) are well documented \citep[see, for example,][]{von1989international, contessi2014did}.
\begin{table}[!tb]
\centering
\begin{tabular}{c|l|l}
Portfolio&&\\
 Name&Industry&Description\\
\hline
NoDur & Consumer Nondurables & \footnotesize{Food, Tobacco, Textiles, Apparel, Leather, Toys}\\
Durbl & Consumer Durables & \footnotesize{Cars, TVs, Furniture, Household Appliances}\\
Manuf & Manufacturing & \footnotesize{Machinery, Trucks, Planes, Chemicals, Off Furn, Paper}\\
Enrgy & Energy & \footnotesize{Oil, Gas, and Coal Extraction and Products}\\
HiTec & Business Equipment & \footnotesize{Computers, Software, and Electronic Equipment}\\
Telcm & Telecommunications & \footnotesize{Telephone and Television Transmission}\\
Shops & Shops & \footnotesize{Wholesale, Retail, and Some Services }\\
Hlth  & Health & \footnotesize{Healthcare, Medical Equipment, and Drugs}\\
Utils & Utilities & \footnotesize{Utilities}\\
\end{tabular}
\caption{\label{tab:ptf_descr} Industry portfolios descriptions. SIC codes for each portfolio are available at \url{https://mba.tuck.dartmouth.edu/pages/faculty/ken.french/Data_Library/det_10_ind_port.html}}
\end{table}

\begin{figure}[!t]
\centering
\includegraphics[width=\textwidth]{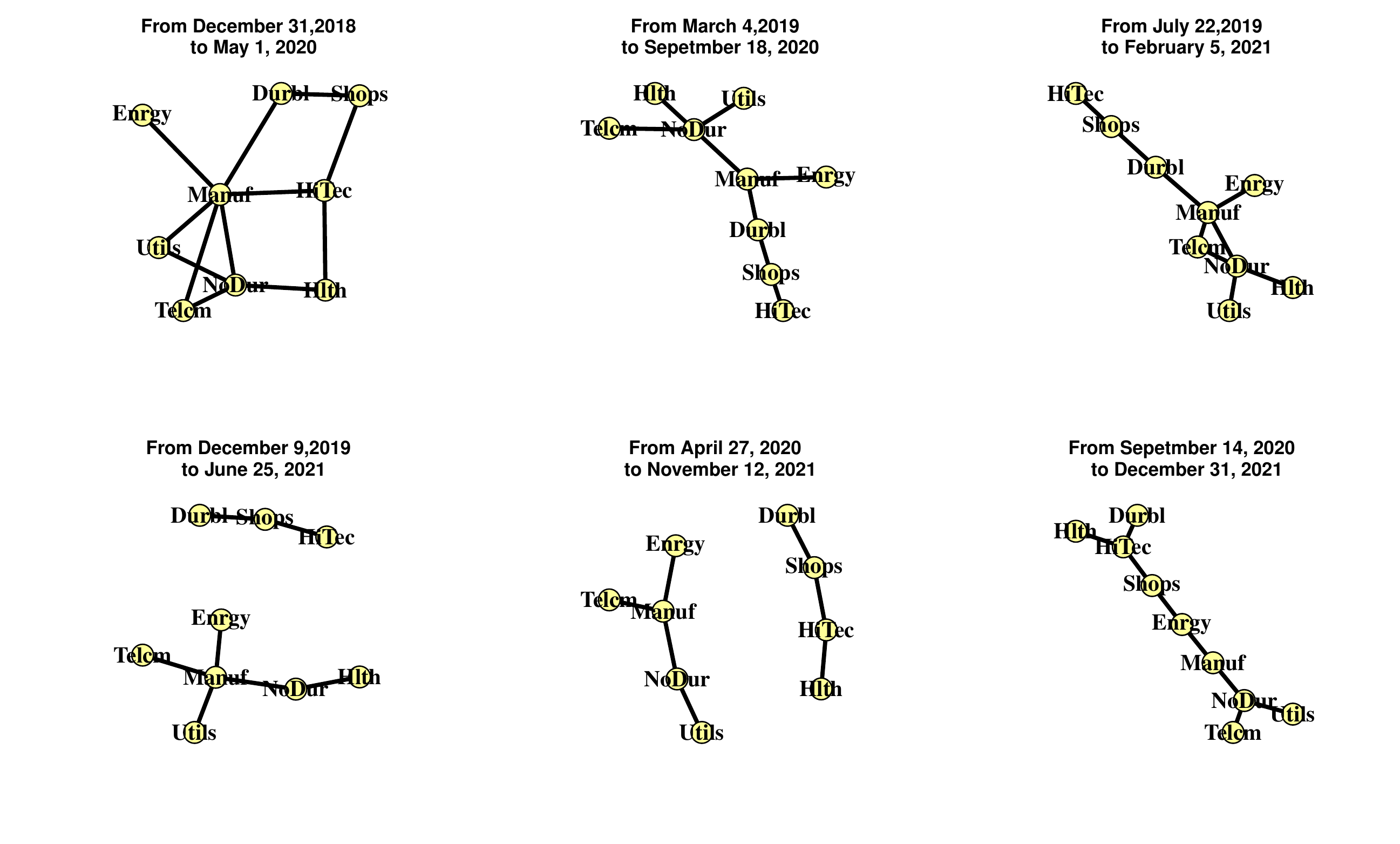}
\caption{\label{fig:moving_window} Moving window: graph estimated with the \textsf{R} package \textsf{GGMselect} \citep{GGMselect}. The four central graphs are computed on 80 time points, while the first and the last graphs refer respectively to the the first 70 and last 68 weeks. The window is moved by 20 time points at the time from one graph to the next.}
\end{figure}
To detect the possible effect of COVID-19 pandemic on the US stock market, we analyse the correlation structure between nine industry portfolios, considering logarithmic weekly returns in the years 2019, 2020, and 2021.
Weekly returns are computed starting from the daily returns available at Kenneth R. French's Data Library. 
{The Kenneth R. and French's Data Library provides also portfolio returns corresponding to more fragmented definitions of industries. 
However, our goal is to detect possible macro effects of the pandemic in the market, and for this reason, we focus on the industry portfolios described in Table~\ref{tab:ptf_descr}. Thanks to diversification within the same portfolio, the corresponding returns are less volatile and appropriately represent the tendency of a whole industry. More details about the construction of the portfolios can be found in Section~\ref{sec:real}.}

Figure~\ref{fig:moving_window} shows empirical estimates of a graph describing the conditional dependence  structure over time. The estimates are obtained using a moving window of 80 weeks, shifting in steps of 20 weeks from one graph to the next. Graph estimates are obtained using an adaptive lasso approach as implemented in the \textsf{R} package \textsf{GGMselect} \citep{GGMselect}. From {this} preliminary analysis, changes in dependence are already evident, as well as the role of hub of the manufacturing and consumer non-durables\ industries. However, it is difficult to determine whether the pandemic had an effect on the overall  structure and when. Moreover, it is well known that edge estimation in GGMs is sensitive to sample size and this empirical approach is highly dependent on the arbitrary choice of the window size (and corresponding sample size). To achieve our inferential goals, a sound modelling strategy is needed to be able to  effectively infer  the existence and location of change points due to sudden changes, still borrowing information across the entire time horizon.

Sudden changes in volatility and dependence have been modelled generalizing either conditional or stochastic volatility models with the introduction of Markov switching regimes \citep[see, among others,][]{so1998stochastic, haas2004new, bianchi2019modeling, caporale2019modelling}.  However, in a frequentist framework, Markov switching approaches require an arbitrary choice of the number of different regimes {and, consequently, ad hoc criteria for model choice \citep[see also][]{cribben2017estimating}. A full Bayesian model for evolving graphs has been introduced by \cite{warnick2018bayesian}; similarly to Markov switching models, here the graph evolves assuming one possible state out of a finite number of exchangeable (not consecutive) states. Still in a Bayesian framework, \cite{schwaller2017exact} develop a strategy for change point detection in graphical models, which, in order to preserve computational tractability, assumes independent consecutive graphs. Their method has the often unrealistic implication that the graphs are estimated independently without borrowing information across the entire time horizon. Recently, \cite{keshavarz2020sequential} have proposed an accurate algorithmic procedure, which employs multiple frequentist tests to detect abrupt changes in the precision matrix of a GGM, however, the procedure does not provide estimates of the graph's structure.} 
Alternatively, penalised likelihood techniques have been successfully employed for estimating dynamic GGMs \citep{zhou2010time,kolar2012estimating,danaher2014joint,yang2015fused,gibberd2017regularized,hallac2017network,roy2017change,bybee2018change,cribben2019change,yang2020estimating, liu2021simultaneous}, however, such approaches do not allow for uncertainty quantification on the number and temporal location of the abrupt changes and the graph topology. {Similar limitations are shared also by algorithmic approaches, as the one proposed by \cite{anastasiou2022cross}.} {A detailed comparison between our contribution and penalised likelihood approaches is provided  in {Sections~\ref{sec:simmodel} and~\ref{sec:real}.}}  

In this work, we introduce a Bayesian dynamic GGM  to detect abrupt changes in the conditional dependence structure between time series. Our proposal is a piece-wise constant stochastic volatility model. It favours sparsity at three levels by explicitly penalizing: (i) the number of change points; (ii) the number of edges within each graph; (iii) the number of edges which are activated (appear) and deactivated (disappear) at each change point.
In a Bayesian framework, it is straightforward, at least in principle, to perform posterior  inference also on the number of change points and on their location. {Finally, we note that our model does not assume global Gaussianity, which would imply the existence of a single Gaussian distribution for the entire temporal span. The assumption of global Gaussianity poses challenges in the analysis of financial returns, which are typically characterised by heavy tails and changes of behaviour. In our setup, we assume local Gaussianity, between two consecutive change points. Our assumption on the return distribution possibly accommodates the  excess of kurtosis typically observed in financial returns' empirical distributions. More precisely, the introduction of change points allows the Multivariate Gaussian distribution to change along the overall time horizon so that the observed empirical distribution can be thought of as been drawn from a mixture of Gaussians, which can accommodate heavy tails \citep{cui2012semiparametric}. }

The paper is structured as follows. In Section~\ref{sec:DGGM},  the dynamic GGM is presented. Section~\ref{sec:alg} contains a discussion of the computational challenges, the proposed algorithm, and a simulation study to assess the performance of the sequential Monte-Carlo procedure. Results on simulated data and on the US stock market data can be found in Sections~\ref{sec:simmodel} and~\ref{sec:real}, respectively. Section~\ref{sec:conclusion} concludes the paper with a discussion about future directions and extensions. In Supplementary Material we provide the dataset, \textsf{R} codes to reproduce all the results in this work, and additional results on the algorithm, simulation studies, and the application.
\section{The dynamic Gaussian graphical model}
\label{sec:DGGM}
We first introduce some definitions and notation.
Let $G = (V,E)$ represent an undirected, {simple, and unweighted} graph, where  $
V=\{1, \dots, p\}$, $p\ge 1$, corresponds to the set of labelled nodes 
and  $E \subseteq \{\,(h,k) \in V \times V:h<k\,\}$ the set of edges linking pairs of nodes. There is a one-to-one correspondence between 
$G$ and its $p\times p$ adjacency binary matrix 
$A$, which is defined as follows. The element $A[h,k]$ on the $h-$th row and $k-$th column is equal to 1 when an edge exists between nodes $h$ and $k$, and to 0 otherwise. Note that $A$ is symmetric  with zeros on the main diagonal, since $G$ is simple. 
When each node corresponds to a random variable, the  graph structure can be used to encode conditional independence so that an edge is present between vertices $h$ and $k$ if and only if the $h-$th and $k-$th random variables are dependent conditionally on all other variables in the graph  \citep{laur:96}. 

A powerful modelling tool is offered by GGMs, which assume that the distribution of the random variables represented by the nodes in $V$ is Multivariate Gaussian. Then, the precision matrix $\Omega$ can be modelled  conditionally on the graph, so that the presence of  an edge between two nodes in $G$ implies  a non-zero entry in the precision matrix between the corresponding random variables, while the absence of an edge implies a zero entry.
Let  the cone $M^{+}$ be the space of symmetric positive-definite matrices on 
$\mathbb{R}^{p\times p}$. For graph $G$ and adjacency matrix $A$,
$M^{+}(G)\subset M^{+}$ denotes the set of the matrices, $M$, with $M[h,k]=0$ if and only if $A[h,k]=0$, for any $h\neq k$, so that $\Omega \in M^{+}(G)$.

In a time series setting, let  $G_t = (V,\,E_t)$ describe the (conditional) dependence structure at time $t$ between $p$ time series, each corresponding to one node in $V$. We propose a prior distribution for the process $\{G_t , t\geq 1\}$, obtained by letting $t$ vary, which lies in the class of stochastic volatility models.  

Data are collected at common discrete time points  $t=1,2,\ldots,T$. We denote with $Y_{t}$ the vector of observations at time $t$ on the $p$ variables (i.e., returns at week $t$ for the considered industry portfolios) and with $Y_{1:T}=[Y_t]_{t=1}^T $ the $T\times p$ data matrix. We assume that, conditionally on a time-indexed collection of precision matrices $\{\Omega_t, t=1,\ldots,T\}$, the vectors of observations are normally distributed and independent over time, i.e.,
\begin{equation}\label{eq:obseq}
	Y_t\mid\Omega_t \overset{ind}{\sim}\text{N}_p(0,\Omega_t^{-1})\qquad \text{for } t=1,\ldots,T
\end{equation}
where $\text{N}_p(\mu,\Sigma)$ denotes a $p$-variate Gaussian distribution with mean $\mu$ and covariance matrix~$\Sigma$.

We model $\Omega_t$ conditionally on a graph at time $t$, $G_t$. 
Then,  to allow for 
time-varying dependence structure among the $p$ variables, we introduce a sequence of random change points.  A time point $t$ is said to be a \emph{change point} if the dependence structure among the $p$ observable variables changes between $t-1$ and $t$, i.e., if $G_t \neq G_{t-1}$ and, consequently, $\Omega_t \neq \Omega_{t-1}$. Let $c_{1:\kappa} = (c_1,c_2,\ldots,c_\kappa)$ be the {(possibly empty)} vector of ordered change points,  which, similarly to the precision matrices and the graphs, are unobserved. Here, ${\kappa\ge 0}$ denotes the (random) number of change points. {In what follows,  we use the conventions $c_{1:0} = \emptyset$, $c_0=1$, and  $c_{\kappa+1}=T+1$}. Note that between consecutive change points the graph and the corresponding precision matrix are kept constant. 
Given the sequence of graphs,  $G_{1:T}=\{G_{t}, t=1,\ldots,T\}$, and change points, we assume that
\begin{equation}\label{eq:stateeq1}
    \Omega_1 \mid G_{1} \sim W_{G_1}(d,D)
\end{equation}
and, for $t\geq 2$,
\begin{equation}\label{eq:stateeq2}
\Omega_t \mid \Omega_{t-1}, G_t, c_{1:\kappa} \sim \begin{cases}
	W_{G_t}(d,D),& \text{if } t\in\{c_1,\ldots,c_{\kappa}\}\\
	\delta_{\Omega_{t-1}},& \text{otherwise}
\end{cases}
\end{equation}
where, $\delta_x$ denotes the Dirac delta distribution at $x$ and $W_{G}(d,D)$ the $G$-Wishart distribution \citep[see][]{rove:02,dobra2011bayesian}, with shape parameter $d>2$ and inverse scale matrix parameter $D\in M^{+}$. 
Its density w.r.t.~the Lebesgue measure of dimension equal to the free elements of a matrix in $M^{+}(G)$ is 
\begin{align*}
	P(\Omega|G) = \frac{1}{I_{G}(d,D)}\,|\Omega|^{(d-2)/2}\exp
	\big\{-\tfrac{1}{2}\mathrm{tr}(D\,\Omega)\big\},\quad \Omega\in M^{+}(G)
\end{align*}
The {stated} constraints for hyper-parameters $d$ and $D$ 
suffice to ensure the integrability of the above density \citep{diac:79}.
The normalizing constant is equal to 
\begin{align*}
	I_{G}(d,D) = \int_{M^{+}(G)} |\Omega|^{(d-2)/2}\exp
	\big\{-\tfrac{1}{2}\mathrm{tr}(D\,\Omega)\big\}\,\mathrm{d}\Omega
\end{align*}
and will be used later to compute the marginal likelihood of the data conditionally only on the graph structure.

To complete the model, we next describe the graph dynamics. Denote with $A_t$ the adjacency matrix corresponding to $G_t$, then, for all $ h,k\in\{1,\ldots, p\}$ with $h<k$, we specify the prior  distributions
\begin{equation}\label{eq:G}
		A_1[h,k]\mid \omega \overset{iid}{\sim} \text{Bernoulli}\left(\frac{2\omega}{p-1}\right)
\end{equation}
and, for $t\geq 2$,
		\begin{equation}\label{eq:GG}
			\begin{aligned}
		A_t[h,k]\mid A_{t-1}[h,k], c_{1:\kappa},z &\overset{ind}{\sim}
		\begin{cases}
	\mid A_{t-1}[h,k] - \text{Bernoulli}\left( \frac{2z}{p-1}\right) \mid, & \text{if } t\in\{c_1,\ldots,c_k\} \\
		\delta_{A_{t-1}[h,k]},&\text{otherwise}
		\end{cases}\\
	\end{aligned}
\end{equation}
Notice that the hyper-parameter $\omega\in[0,(p-1)/2]$ controls the graph sparsity, so that the expected number of edges for the initial graph a priori equals $p\,\omega$, while the hyper-parameter $z\in[0,(p-1)/2]$ controls the impact of an event on graph structure when a change point is reached, in particular, the (a priori) expected number of edges that will change is equal to $p\,z$. 
Our prior choice is reminiscent of the one proposed in \cite{jones2005experiments}, who recommend setting a prior edge inclusion probability equal to $2/(p-1)$ so that the expected number of edges
is $p$. The extra parameters $\omega $ and $z$ allow for more control on graph sparsity and temporal dependence.  {We note that alternative priors can be employed to model the precision matrix and the graph as, for instance, the graphical horseshoe \citep{li2019graphical} and the prior proposed by \cite{banerjee2015bayesian}. The former is a prior used directly on the precision matrix, which requires (arbitrary) thresholding of its entries  in order to recover a sparse graph representation. The latter is more similar to our modelling strategy and consists of three elements: (i) Bernoulli priors for the entries of the adjacency matrix, conditionally on  a maximum number of edges; (ii) a Laplace prior on the non-zero
off-diagonal elements of the precision matrix, and (iii) an exponential prior for the {diagonal} elements, still  imposing the positive definiteness of the matrix. This construction is still computationally intensive.}

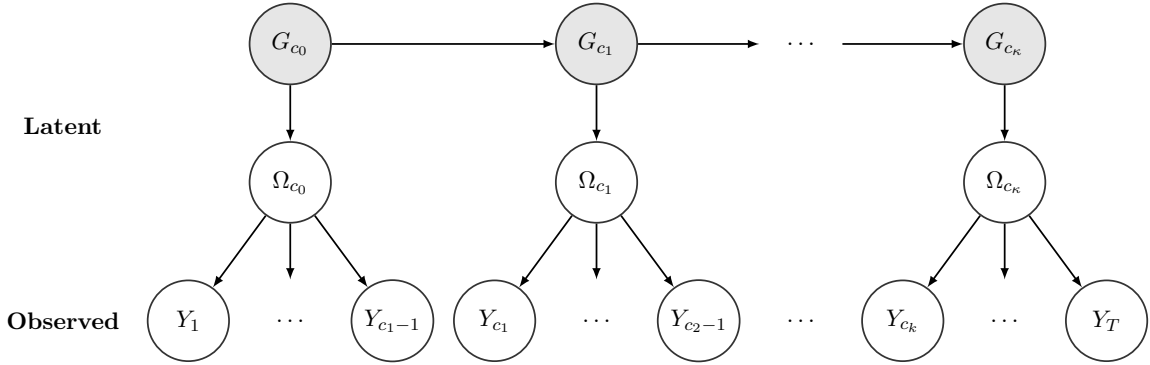
\begin{figure}
\resizebox{\textwidth}{!}{
\begin{tikzpicture}
\tikzstyle{main}=[circle, minimum size = 13mm, thick, draw =black!80, node distance = 3mm]
\tikzstyle{connect}=[-latex, thick]
\tikzstyle{box}=[rectangle, draw=black!100]
  \node[box,draw=white!100] (Latent) {\textbf{Latent}};
  \node[box,draw=white!100] (Observed) [below=2.6cm of Latent] {\textbf{Observed}};
  \node[main] (O1) [right=of Observed] {$Y_1$};
  \node[main] (O3) [right=of O1, draw=white!100] {$\ldots$};
  \node[main] (O5) [right=of O3] {$Y_{c_1-1}$};
  \node[main] (O6) [right=of O5] {$Y_{c_1}$};
  \node[main] (O7) [right=of O6, draw=white!100] {$\ldots$};
  \node[main] (O8) [right=of O7] {$Y_{c_2-1}$};
  \node[main] (O9) [right=of O8, draw=white!100] {$\ldots$};
  \node[main] (O10) [right=of O9] {$Y_{c_k}$};
  \node[main] (O11) [right=of O10, draw=white!100] {$\ldots$};
  \node[main] (O12) [right=of O11] {$Y_{T}$}; 
  \node[main] (L1) [above=0.9of O3] {$\Omega_{c_0}$};
  \node[main] (L2) [above=0.9of O7] {$\Omega_{c_1}$};
  \node[main] (Lempty) [above=0.9of O9, draw=white!100] { };
  \node[main] (L3) [above=0.9of O11] {$\Omega_{c_{\kappa}}$};
  \node[main] (G1) [above=0.9of L1,fill=black!10] {$G_{c_0}$};
  \node[main] (G2) [above=0.9of L2,fill=black!10] {$G_{c_1}$};
  \node[main] (Gempty) [above=0.9of Lempty, draw=white!100] {$\ldots$};
  \node[main] (G3) [above=0.9of L3,fill=black!10] {$G_{c_{\kappa}}$};
  \path (G1) edge [connect] (L1);
  \path (G2) edge [connect] (L2);
  \path (G3) edge [connect] (L3);
  \path (G1) edge [connect] (G2);
  \path (G2) edge [connect] (Gempty);
  \path (Gempty) edge [connect] (G3);
  \path (L1) edge [connect] (O1);
  \path (L1) edge [connect] (O3);
  \path (L1) edge [connect] (O5);
  \path (L2) edge [connect] (O6);
  \path (L2) edge [connect] (O7);
  \path (L2) edge [connect] (O8);
  \path (L3) edge [connect] (O10);
  \path (L3) edge [connect] (O11);
  \path (L3) edge [connect] (O12);
\end{tikzpicture}}
\caption{\label{fig:model} Graphical representation of the model conditionally on change points. Both graph and precision matrix are constant between change points. Observations $(Y_t)$ are independent conditionally on the model parameters over time, while are iid between change points. Moreover, the precision matrices $\{ \Omega_{c_i}, i=1,\dots,\kappa\}$ are conditionally independent given the sequence $\{G_{c_i}, i=1,\dots,\kappa\}$.}
\end{figure}

Equations~(\ref{eq:obseq}) and (\ref{eq:stateeq1})-(\ref{eq:GG}) can be viewed as observation and state dynamics, respectively, of a hidden Markov model with the unobserved signal corresponding to the pair 
$\{(G_{c_i}, \Omega_{c_i}), i=0,1,\ldots, \kappa\}$ (see Figure~\ref{fig:model} for a graphical representation).  For more details see, for example, \cite{west2006bayesian}.

Finally, the prior distribution for $c_{1:\kappa}$ is chosen as
\begin{equation}\label{eq:geometric}
	\begin{aligned}
		c_{1:\kappa}|\kappa&\sim \mathrm{Uniform}(\mathcal{T}_{\kappa\ell})\\
		\kappa \mid p_0 &\sim \mathrm{Truncated-Geometric}(p_0) \qquad \text{for } \kappa = 0,1,\ldots,K_{T\ell} 
	\end{aligned}
\end{equation}
for hyper-parameter $p_0\in(0,1)$, so that the (a priori) number of expected change points is 
\begin{equation}\label{eq:kappaEV}
\mathbb{E}[\kappa] = \frac{1-p_0}{p_0}\, \frac{1 - (1-p_0)^{K_{T\ell}}\,(p_0 \, K_{T\ell} + 1)}{1 - (1-p_0)^{K_{T\ell} + 1}}
\end{equation}
Through appropriate choice of $p_0$ in \eqref{eq:geometric}, we are able to enforce the desired level of sparsity on the number of change points.
Here, $\mathcal{T}_{\kappa\ell}$ is the space of ordered $\kappa$-tuples $c_1<\cdots<c_{\kappa}$, with $c_j$ in $\{2,\ldots,T\}$, under the minimum-span constraint that $c_{j+1}-c_{j}\ge \ell$, for any $j=0,1,\ldots,\kappa$, with the convention $c_0=1$
and $c_{\kappa+1} =T+1$. Moreover, 
 $K_{T\ell}$ is the maximum number of change points compatible with the minimum-span constraint. { Notice that for $K_{T\ell}$ of moderate size, the second term in the product in \eqref{eq:kappaEV} is negligible and the Truncated Geometric in \eqref{eq:geometric} approximates a standard Geometric distribution on $\mathbb{N}_0=\{0,1,\ldots,\}$. The imposition of the minimum-span constraint defined by $\ell$ ensures likelihood identifiability between change points, leading to more stable  computations and robust inference. Notice that, between any two change points, the sample covariance/precision  matrix has $p(p+1)/2$ entries. To guarantee likelihood identifiability of the model, we need the number of data points between any two change points, i.e., $(c_{j+1}-c_{j})$, to be greater than $p$. This poses a trade-off: the lower $\ell$, the more flexible the change-point detection procedure, and the higher $\ell$, the more stable the estimates of the precision matrices. 
In the simulation study and in the application, we set $\ell$ equal to $p + 2$. When $p = 9$, this means that each precision matrix containing $45$ entries is estimated with at least $99$ single data points.}
\section{Bayesian inference via sequential Monte-Carlo}
\label{sec:alg}
The dynamic GGM proposed herein is a hierarchical model, with the first layer represented by the change points, $c_{1:\kappa}$, the second by graphs and precision matrices $(G_{c_{j}},\Omega_{c_{j}})$, and the third by the observations.
Markov chain Monte-Carlo (MCMC) methods developed directly on such space of unobserved variables would face major challenges. We ignore the precision matrices in this discussion as they are later integrated out.
Gibbs-type approaches would involve reversible-jump MCMC \citep{green:95}, thus requiring the design of a joint update on the ``model'' 
space (as determined by the change-points) and model ``parameters'' of varying dimension (corresponding to the graphs).  This joint space is entangled, with very limited space for maneuvering, as updates on the graph space would be heavily constrained by the strong prior Markovian dependencies amongst graphs.

Instead, we perform computationally effective posterior inference for the dynamic GGM  through a tailored  sequential Monte-Carlo (SMC) algorithm. 
The proposed Particle MCMC (PMCMC) method is quite appropriate for exploiting the hidden Markov model structure conditionally on the change points, and naturally disentangles the updates on the change points and the  latent Markovian signal. See, e.g., \cite{kara:13} and \cite{pers:15} for related ideas.
The proposed PMCMC algorithm is better understood as comprised of an `outer' cycle and an `inner' cycle. In the former, the change points are updated via a reversible jump Metropolis-Hasting (M-H) algorithm. In the latter, a particle filter, of enhanced performance due to a combination of adaptive tempering, dynamic resampling, and mutation steps, is employed to sample the sequence of graphs and compute the acceptance probability of the outer algorithm. In particular, for each M-H step, the inner component provides an unbiased estimate of the conditional likelihood given the proposed change point sequence {together with a corresponding proposed graph sequence}. Adaptive tempering and resampling steps are used to improve the robustness of such estimate, while the mutation step is used both to bring particles closer to the modal region
of the likelihood and to avoid depletion of the number of unique particles which can otherwise be a consequence of successive resampling and tempering.

The overall PMCMC algorithm is well-understood as an ``exact-approximate" one, in the sense that it targets the correct posterior on the space of graphs and change points,  thanks to the unbiasedness (and positivity) of the estimator provided by the inner particle filter.

\subsection{Outer component}
The key component in the development of the M-H step is the choice of proposal distribution, $q(c'_{1:\kappa'}\mid c_{1:\kappa})$, where $c_{1:\kappa}$ and $c'_{1:\kappa'}$ are the current and proposed collections of ordered change points, respectively. Starting from $c_{1:\kappa}$, one of four alternative events (namely a birth, a death, a global move, or a local move) generates the proposed new value.
With probabilities equal to $P(B\mid c_{1:\kappa})$, $P(D\mid c_{1:\kappa})$, $P(M_{glob}\mid c_{1:\kappa})$ and $P(M_{loc}\mid c_{1:\kappa})$, one of the following four events takes place, respectively: a new change point is added to the current set (birth); a change point is removed from the current set (death);  one of the existing change points is moved to another position (global move); one of the existing change points is moved to another position in-between its neighbours (local move).    
When a new change point, $c^*$, is created, $c^*$ is chosen uniformly over
 the set $\mathsf{B}(c_{1:\kappa},\ell)\subset \{2,\ldots, T\}$ of allowed positions (i.e., satisfying the minimum-span constraint), of size
 $|\mathsf{B}(c_{1:\kappa},\ell)|=:n(c_{1:\kappa}, \ell)\ge 0$. Thus, in the birth scenario
\begin{equation}
\label{eq:Birth}
	q(c'_{1:\kappa'}\mid c_{1:\kappa}) = \frac{P(B\mid c_{1:\kappa})}{\textcolor{black}{n(c_{1:\kappa},\ell)}}, \qquad \kappa'=\kappa+1,\quad c'_{1:\kappa'}=c_{1:\kappa}\cup c^*,\quad c^*\in \mathsf{B}(c_{1:\kappa},\ell)
\end{equation}
 When a change point, $c'$, is removed, the change point is chosen uniformly among the current change points, i.e., in the death scenario 
\begin{equation}
\label{eq:Death}
	q(c'_{1:\kappa'}\mid c_{1:\kappa}) 
 = \frac{P(D\mid c_{1:\kappa})}{\kappa}, \qquad
 \kappa'=\kappa-1,\quad c'_{1:\kappa'}=c_{1:\kappa}\setminus c', \quad c'\in c_{1:\kappa}
\end{equation}
 To improve mixing and posterior  exploration, we introduce also two move steps. 
 When a change point is moved, firstly, a change point $c'\in c_{1:\kappa}$ is selected uniformly among the current change points and removed. Then, if the step is a global move a new change point is selected 
 uniformly in $\mathsf{B}(c_{1:\kappa}\!\setminus\!c',\ell)$.
 If instead the step is a local move a new change point  $c^*$ is selected with probability proportional to 
\begin{align}
\label{eq:loc}
\exp\{-\lambda\,|c^*-c'|\}\mathbbm{1}_{\{c^*\in[\bar{c}_l, \bar{c}_r]\}}
\end{align}
for algorithmic parameter $\lambda>0$,
with
 \begin{equation*}
 \bar{c}_l = c_l  +\ell,\quad  \bar{c}_r = c_r - \ell
 \end{equation*}
 where $c_l$ and $c_r$ denote the left-side and right-side neighbours of $c'$ in $c_{0:\kappa+1}$.
 Thus, the proposal kernel for a move step is  
 \begin{align}
 \label{eq:Move}
 &q(c'_{1:\kappa'}\mid c_{1:\kappa}) = \frac{P(M_{glob}\mid c_{1:\kappa})}{\kappa \, n(c_{1:\kappa}\!\setminus\! c',\, \ell)} + 
 	\frac{P(M_{loc}\mid c_{1:\kappa})\,e^{-\lambda\,|c^*-c'|}}{\kappa \sum_{\chi= \bar{c}_l}^{\bar{c}_r}e^{-\lambda\,|c^*-\chi|}}\mathbbm{1}_{\{c^*\in[\bar{c}_l, \bar{c}_r]\}}& \nonumber \\[0.2cm] 
  & \qquad \kappa'= \kappa, \quad c'_{1:\kappa'}=c_{1:\kappa}\!\setminus\!c'\cup c^*, \quad 
  c'\in c_{1:\kappa}, \quad c^*\in \mathsf{B}(c_{1:\kappa}\!\setminus\!c',\ell)&
 \end{align}

\begin{algorithm}[!htb]
	\caption{\label{alg:out}Outer algorithm - Reversible jump M-H}
	\begin{tabular}{ll}
	 \hspace*{\algorithmicindent} \textbf{Input}:& change point sequence $c_{1:\kappa}$; $\mathsf{B}(c_{1:\kappa},\ell)$;  likelihood value $P(Y_{1:T} | c_{1:\kappa})$;\\ & prior
	    value $P(c_{1:\kappa})$. \\
	\hspace*{\algorithmicindent} \textbf{Output}:& new change point sequence $\tilde{c}_{1:\tilde{\kappa}}$; $\mathsf{B}(\tilde{c}{_{1:\tilde{\kappa}}},\ell)$; likelihood value $P(Y_{1:T} | \tilde{c}_{1:\tilde{\kappa}})$; \\  & 
	 prior value $P(\tilde{c}_{1:\tilde{\kappa}})$. \\[0.2cm]
	\end{tabular}
\begin{algorithmic}
	\State Sample event $E$ from $\{B, D, M_{glob}, M_{loc}\}$ according to \eqref{eq:BDM}; \vspace{0.2cm}
	\If  {$E=B$} \vspace{0.2cm}
		\State Sample a new change point uniformly from $\mathsf{B}(c_{1:\kappa},\ell)$ and propose $c'_{1:\kappa'} = c_{1:\kappa}\cup c'$; \vspace{0.2cm}
		\State Compute $q(c'_{1:\kappa'}\mid c_{1:\kappa})$ according to \eqref{eq:Birth}; \vspace{0.2cm}
	\Else \vspace{0.2cm}
		\State Sample uniformly and remove a change point $c'$ from $c_{1:\kappa}$; \vspace{0.2cm}
		\If {$E=M_{glob}\textrm{ or }M_{loc}$} \vspace{0.2cm}
		    \If{$E = M_{glob}$} \vspace{0.2cm}
			    \State Sample a new change point $c^*$ uniformly  from $\mathsf{B}(c_{1:\kappa}\!\setminus\! c',\ell)$; \vspace{0.2cm}
			 \Else \vspace{0.2cm}
			    \State Sample a new change point $c^*$ from the interval $[\bar{c_l},\bar{c_r}]$ according to (\ref{eq:loc}); \vspace{0.2cm}
			 \EndIf
              \State Propose $c'_{1:\kappa'} = c_{1:\kappa}\!\setminus\!c' \cup c^*$ and compute $q(c'_{1:\kappa'}\mid c_{1:\kappa})$ according to \eqref{eq:Move}; \vspace{0.2cm}
		\Else \vspace{0.2cm}
			\State Propose $c'_{1:\kappa'} = c_{1:\kappa}\!\setminus\!c'$ and 
   compute $q(c'_{1:\kappa'}\mid c_{1:\kappa})$  according to \eqref{eq:Death}; \vspace{0.2cm}
		\EndIf	
	\EndIf
	\State Determine $\mathsf{B}(c'_{1:\kappa'},\ell)$ and compute $q(c_{1:\kappa}\mid c'_{1:\kappa'})$ according to \eqref{eq:Birth}, \eqref{eq:Death}, or \eqref{eq:Move}, respectively; \vspace{0.2cm}
	\State  Compute  prior $P(c'_{1:\kappa'})$ and  likelihood  $P(Y_{1:T} \mid c'_{1:\kappa'})$ for proposed configuration; \vspace{0.2cm}
	\State Sample $u$ from a Uniform$(0,1)$; \vspace{0.2cm}
	\If{$u\leq\frac{P(Y_{1:T} | c'_{1:\kappa'})P(c'_{1:\kappa'})\,q(c_{1:\kappa}\mid c'_{1:\kappa'})}{P(Y_{1:T} | c_{1:\kappa})P(c_{1:\kappa})\,q(c_{1:\kappa}\mid c'_{1:\kappa'})}$} \vspace{0.2cm}
	\State Return $c'_{1:\kappa'}$, $\mathsf{B}(c'_{1:\kappa'},\ell)$, $P(Y_{1:T} | c'_{1:\kappa'})$, $P(c'_{1:\kappa'})$. \vspace{0.2cm}
     \Else \vspace{0.2cm}
     \State Return $c_{1:\kappa}$, $\mathsf{B}(c_{1:\kappa},\ell)$, $P(Y_{1:T} | c_{1:\kappa})$, $P(c_{1:\kappa})$.
	\EndIf
\end{algorithmic}
\end{algorithm}

Note that we prefer to write the joint kernel for both the global and local move, as the same configuration $c'_{1:\kappa}$ might be reached by both  type of moves. This needs to be accounted for  when computing the M-H acceptance probability to ensure  detailed balance.

  Finally, we choose the event probabilities as 
  \begin{align}
  \label{eq:BDM}
  \begin{aligned}
  	P(B\mid c_{1:\kappa})=\begin{cases}
  	    1, &\text{ if } \kappa=0\\
  		0, &\text{ if } n(c_{1:\kappa},\ell)=0\\
  		q_B, &\text{ otherwise}
  	\end{cases}\qquad 
  	P(D\mid c_{1:\kappa})=\begin{cases}
  	0, &\text{ if } \kappa=0\\
  	q'_{D}, &\text{ if } n(c_{1:\kappa},\ell)=0\\
  	q_{D}, &\text{ otherwise} 
  \end{cases}\\[0.3cm]
\textrm{and }\ \  P(M_{glob}\mid c_{1:\kappa}) = P(M_{loc}\mid c_{1:\kappa}) = [1 - P(B\mid c_{1:\kappa}) - P(D\mid c_{1:\kappa}) ] / 2
  \end{aligned}
  \end{align}

The Metropolis-Hastings acceptance probability is equal to
\begin{equation*}
    1 \wedge \frac{P(Y_{1:T} | c'_{1:\kappa'})P(c'_{1:\kappa'})\,q(c_{1:\kappa}|c'_{1:\kappa'})}{P(Y_{1:T} | c_{1:\kappa})P(c_{1:\kappa})\,q(c'_{1:\kappa'}|c_{1:\kappa})}
\end{equation*}
where $P(Y_{1:T} \mid c_{1:\kappa}) $ is the marginal likelihood of the data given the change points, i.e.,
\begin{equation*}
	P(Y_{1:T} \mid c_{1:\kappa}) = \int P(Y_{1:T}\mid G_{1:T})P(G_{1:T}|c_{1:\kappa}) dG_{1:T}
\end{equation*}
with 
\begin{equation*}
 P(Y_{1:T}\mid G_{1:T}) = \int  P(Y_{1:T}\mid G_{1:T}, \Omega_{1:T} ) P(\Omega_{1:T} \mid G_{1:T}) d \Omega_{1:T}
 \end{equation*} 
Since $P(Y_{1:T} \mid c_{1:\kappa}) $ is not available in closed form, it needs to be estimated. {Algorithm~\ref{alg:out} contains the pseudo-code for the outer part of the algorithm described in this section.} In the next section we describe the SMC algorithm to approximate the marginal likelihood.

\subsection{Inner Component: Particle Filter}
\begin{algorithm}[!htb]
	\caption{\label{alg:inner} Inner Algorithm - Particle Filter with Tempering}
	\begin{tabular}{ll}
		\hspace*{\algorithmicindent} \textbf{Input}:&  Data $Y_{1:T}$; change points $c_{1:\kappa}$; number of particles $N$; ESS threshold $\epsilon$; \\
		 & number $S_j$ of temperatures for graph $j$, 
  $\phi_{0,j}\equiv 0$, $\phi_{0,S_j+1}\equiv 1$,  $0\le j\le \kappa$; \\ 
  &  number of mutation steps $M\ge 1$; temperatures $\{\phi_{1,j},\ldots,\phi_{S_j,j}\}$; \\
  & M-H kernel $\bar{P}_{j,s}({G}_{c_{j}}|G_{c_{j}},G_{c_j-1})$, $0\le j\le \kappa$, $1\le s\le S_j+1$;\\
  &$\bar{P}_{j,s}^{M}$ denotes $M$ iterations of such a kernel;  \\
  & hyper-parameters $\omega, \, z, \,d, \, D$.\\
		\hspace*{\algorithmicindent} \textbf{Output}:& \textcolor{black}{ Unbiased estimate $\hat{P}(Y_{1:T}|c_{1:\kappa})>0$ and proposed sequence $(G_{c_0},\ldots,G_{c_\kappa}$).} \\ 
	\end{tabular}
	\begin{center}\emph{(Actions over $n$ are understood to be repeated for $1\le n\le N$.) \vspace{0.2cm} }\end{center}
	\vspace{-\baselineskip}
	\begin{algorithmic}
	    \State Set $\hat{P} = 1$; \vspace{0.2cm}
		\For {$j$ in $0:\kappa$} \vspace{0.2cm}
		\If{$j = 0$ } \vspace{0.2cm}
		\State Sample $G^{0,n}_{c_0}\stackrel{iid}{\sim}P(G_{c_0})$ and set $w_{0}^{0,n}=1$;  \vspace{0.2cm}
		\Else \vspace{0.2cm}
		\State Initialise particles $G^{0,n}_{c_{j}}\stackrel{ind.}{\sim}  P(G_{c_{j}}| G^{S_{j-1},n}_{c_{j-1}})$; \vspace{0.2cm}
		\State Initialise weights 
		$w_{j}^{0,n} = w_{j-1}^{S_{j-1}+1,n}$; \vspace{0.2cm}
		
		\EndIf
	
		\For{$s$ in $1:S_j+1$} \vspace{0.2cm}
		\State Calculate weights $w_{j}^{s,n} = w_{j}^{s-1,n}\cdot \left[P(Y_{c_{j}:c_{j+1}-1}|G^{s-1,n}_{c_{j}})\right]^{\phi_{j,s}-\phi_{j,s-1}}$;\vspace{0.2cm}
		\If{$ESS(w_{j}^{s,1:N})<\epsilon\,N$} \vspace{0.2cm}
		\State $\hat{P} \gets \hat{P}\cdot \frac{1}{N}\sum_{n=1}^{N}w_{j}^{s,n}$;  \vspace{0.2cm}
		\State Resample  $\{G^{s-1,n}_{c_{j}}, G^{S_{j-1},n}_{c_j-1}\}$ according to the weights $\{w_{j}^{s,n}\}$; \vspace{0.2cm}
		\State Mutate particles, i.e., sample $G^{s,n}_{c_{j}}\stackrel{ind.}{\sim} \bar{P}^{M}_{j,s}(G_{c_{j}}|G^{s-1,n}_{c_{j}},G^{S_{j-1},n}_{c_{j-1}})$; 
		\vspace{0.1cm}
		\State Set $w_{j}^{s,n}=1$; \vspace{0.2cm}
		\Else  \vspace{0.2cm}
		\State set $G^{s,n}_{c_{j}} = G^{s-1,n}_{c_{j}}$;\vspace{0.2cm}
		\EndIf
		\EndFor
		\EndFor
            \State \textcolor{black}{Sample a graph 
            $G_{c_\kappa}$ from  $\{G_{c_\kappa}^{S_{\kappa}+1,n},w_{\kappa}^{S_{\kappa+1},n}\}_{n}$ and retrieve its genealogy, $(G_{c_0},\ldots,G_{c_\kappa}$), amongst particles $\{G_{c_0}^{S_{0}+1,n}\}_n$,\ldots, $\{G_{c_{\kappa-1}}^{S_{\kappa-1}+1,n}\}_n$. } \vspace{0.2cm}
		\State Return $\hat{P}(Y_{1:T|G_{1:T}})= \hat{P}$ {and the proposed sequence $(G_{c_0},\ldots,G_{c_\kappa}$)}.
	\end{algorithmic}
\end{algorithm}

As already mentioned, the inner component of the algorithm is used to compute $P(Y\mid c_{1:\kappa})$, i.e., the likelihood values required in the acceptance probability of the outer M-H, and to provide proposed samples of the graphs to be accepted or rejected by the outer algorithm.

To compute the marginal likelihood given the change point sequence, a standard bootstrap particle filter with multinomial resampling carried out at each change  point, samples $N\ge 1$ particles  $\{G^{(n)}_{c_{j}}\}_{n=1}^{N}$, for $0\le j\le \kappa$, 
from the joint distribution
\begin{equation*}
\prod_{n=1}^{N}P(G^{(n)}_{c_0})\times 
\prod_{j=1}^{\kappa}\bigg\{ \prod_{n=1}^{N} \Big(\,\sum_{l=1}^{N}\,\frac{w_{j-1}^{(l)}}{\sum_{m=1}^{N} w_{j-1}^{(m)} }\,P(G_{c_{j}}^{(n)} \mid G_{c_{j-1}}^{(l)})\,\Big) \bigg\}
\end{equation*}
where the unnormalised weights are defined as
\begin{equation*}
w_{j}^{(n)} = P(Y_{c_{j}:c_{j+1}-1} \mid G_{c_{j}}^{(n)}), \qquad 1\le n\le N, \quad 0\le j \le \kappa
\end{equation*}
The unbiased estimate $\hat P(Y\mid c_{1:\kappa})$ of $P(Y\mid c_{1:\kappa})$ could then be obtained as
\begin{equation}\label{eq:likest}
	\hat P(Y_{1:T} \mid c_{1:\kappa}) = 
	\prod_{j=0}^{\kappa}
	\Big(\frac{1}{N}\sum_{n=1}^N w_{j}^{(n)}\Big)
\end{equation}
However, it is often the case that further algorithmic advances must complement the standard particle filter to control the variance of the estimate (\ref{eq:likest}). It is well-understood that such variability is critically linked to the performance of the overall PMCMC algorithm. See e.g.~\cite{pitt:12, douc:15,sher:15} where, in various model settings, standard deviations 
centred around $1$ are proposed for the estimate of the logarithm of the normalising constant, with exponential decay in performance for PMCMC reported when the standard deviation exceeds a (not too high) threshold.

A standard approach to reduce standard deviation for given number of particles, is via the application of tempering, i.e.~introduction of  
a sequence of temperatures together with corresponding mutation steps. Such approach has been shown, in cases, to reduce the required number of particles for a target error from exponential to quadratic in the number $T$ of log-likelihood terms, see e.g.~\cite{besk:14, ruza:21}. Application of tempering will indeed  be critical for the class of models we are considering in this work, as shown in Section~\ref{ssec:SMCsimul}. 
The temperatures are determined on-the-fly, thus avoiding the introduction of additional tuning parameters in the algorithm.
The complete algorithm can be understood as a particle filter applied on a Feynman-Kac model \citep{del:04} that we extend to include additional Markov iterations and potentials. 
The overall approach is summarised in Algorithm~\ref{alg:inner}.
{We stress that the particle filter that includes the tempering and mutation steps will still provide unbiased estimates of $p(Y_{1:T}|c_{1:\kappa})$, and the induced overall PMCMC method will still target the exact posterior 
$P(c_{1:\kappa},G_{c_{0}:c_{\kappa}}|Y_{1:T})$, see e.g.~the original paper on PMCMC \citep{andr:10} for a detailed justification.}

\subsubsection{Preliminary run -- Determination of temperatures}
\label{ss:tuning}
Within Algorithm \ref{alg:inner}, the sequence of temperatures is treated as given.
In practice, at each iteration of the outer algorithm, the temperatures are determined by a separate preliminary and independent execution of the particle filter, that identifies and stores the temperatures, that are later used within Algorithm~\ref{alg:inner}.
Similar ideas have been used in the SMC literature, see e.g.~\cite{jasra:11}.
That is, we first determine the temperatures according to a target effective sample size (ESS), 
and, then, apply the particle filter in Algorithm \ref{alg:inner}, with the obtained  temperatures, to produce a robust unbiased estimator of the likelihood needed to compute the acceptance probability in Algorithm~\ref{alg:out}.

\begin{algorithm}[!ht]
	\caption{Inner Algorithm - Temperature Tuning (Preliminary particle filter)}
 \label{alg:temp}
	\begin{tabular}{ll}
		\hspace*{\algorithmicindent} \textbf{Input}:&  Data $Y_{1:T}$; change points $c_{1:k}$; number of particles $N$; \\ & hyper-parameters $w, \, z, \,d, \, D$; ESS threshold $\epsilon$; mutation steps $M\ge 1$; \\
  & M-H kernel $\bar{P}_{j,s}({G}_{c_{j}}|G_{c_{j}},G_{c_j-1})$ that preserves the law \\ & $\left[P(Y_{c_{j}:c_{j+1}-1}|G_{c_{j}})\right]^{\phi_{j,s}}\cdot  P(G_{c_{j}}|G_{c_j-1})$, \\ &$\bar{P}_{j,s}^{M}$ denotes $M$ iterations of such a kernel.  \\
		\hspace*{\algorithmicindent} \textbf{Output}:& Temperatures $\{\phi_{1,j},\ldots,\phi_{S_j,j}\}$, $S_j\ge 0$, $\phi_{0,j}\equiv 0$, $\phi_{0,S_j+1}\equiv 1$,  $0\le j\le \kappa$. \\
	\end{tabular}
		\begin{center}\emph{(Actions over $n$ are understood to be repeated for $1\le n\le N$.) \vspace{0.2cm} }\end{center}
	\begin{algorithmic}
		\For {$j$ in $0:\kappa$} \vspace{0.2cm} 
		\If{$j = 0$ } \vspace{0.2cm} 
		\State Sample $G^{0,n}_{c_0}\stackrel{iid}{\sim}P(G_{c_0})$ and set $w_{0}^{0,n}=1$;  \vspace{0.2cm}\vspace{0.2cm} 
		\Else \vspace{0.2cm} 
		\State Initialise particles $G^{0,n}_{c_{j}}\stackrel{ind.}{\sim}  P(G_{c_{j}}\mid G^{S_{j-1},n}_{c_{j-1}})$; \vspace{0.2cm}
		\State Initialise weights 
		$w_{j}^{0,n} = w_{j-1}^{S_{j-1}+1,n}$; \vspace{0.2cm}
		\EndIf
		\State $s \gets 1$; $\phi_{0,j}\gets0$; $\phi_{s,j}\gets1$; \vspace{0.3cm} 
		\While{$\phi_{s,j}\neq \emptyset$} \vspace{0.3cm} 
		\State Find $\phi_{s,j}\in (\phi_{s-1,j}, 1]$ so that $ESS(\phi_{s,j}) \geq  \epsilon N$; \vspace{0.2cm} 
            \If{$\phi_{s,j} = \emptyset$ } \vspace{0.2cm}
            \State $S_j=s-1$; \vspace{0.2cm}
            \Else \vspace{0.2cm} 
		\State Set $w_j^{s,n} = \left[P(Y_{c_{j}:c_{j+1}-1}\mid G^{s-1,n}_{c_j} )\right]^{\phi_{s,j} - \phi_{s-1,j}}$;  \vspace{0.2cm} 
		\State Resample  $\{G^{s-1,n}_{c_{j}:c_{j+1}-1}, G^{S_{j-1},n}_{c_j-1}\}$ according to the weights $\{w_{j}^{s,n}\}$; \vspace{0.2cm}
  \State Set $w_{j}^{s,n}=1$; \vspace{0.2cm}
  \State Mutate particles, i.e., sample $G^{s,n}_{c_{j}}\stackrel{ind.}{\sim} \bar{P}^{M}_{j,s}(G_{c_{j}}\mid G^{s-1,n}_{c_{j}},G^{S_{j-1},n}_{c_{j-1}})$; \vspace{-0.2cm}
		\State $s \gets s + 1$ \vspace{0.2cm} 
  \EndIf
		\EndWhile \vspace{0.2cm}
  \State Set $w_j^{S_{j}+1,n} = \left[P(Y_{c_{j}:c_{j+1}-1}\mid G^{s-1,n}_{c_j} )\right]^{1 - \phi_{s-1,j}}$;  \vspace{0.2cm} \vspace{0.2cm} 
		\EndFor \vspace{-0.2cm} 
  \State Return $\{\phi_{1,j},\ldots,\phi_{S_j,j}\}$, $S_j\ge 0$, $0\le j\le \kappa$.
	\end{algorithmic}
\end{algorithm}
We describe here how to compute the temperatures $\{\phi_{1,j},\ldots,\phi_{S_j,j}\}$ used in Algorithm~\ref{alg:inner},
for $S_j\ge 0$, where $\phi_{0,j}\equiv 0$, $\phi_{0,S_j+1}\equiv 1$, and $0\le j\le \kappa$. Note that $S_j$, the number of temperatures, can vary across graphs at different change points, i.e., it depends on $j$. 
Within this subsection, particles and weights $G^{s,n}_{c_j}$, $w^{s,n}_{j}$  refer to such a preliminary execution of the particle filter.
The temperatures are selected on-the-fly, based on the target ESS, denoted by $ESS_0$, with $ESS_0 = \epsilon\,N$, $\epsilon\in(0,1)$.  Consider the current collection of particles and weights, $G^{s-1,n}_{c_j}$ and $w^{s-1,n}_j$, generated while filtering data points $Y_{c_j:c_{j+1}-1}$, and the corresponding  likelihood factor $\left[P(Y_{c_j:c_{j+1}-1}\mid G_{c_j})\right]^{\phi_{s-1,j}}$ up to the present step. Then, the next temperature $\phi_s$ is determined so that the ESS equals  the target, i.e., $ESS(\phi_s)= ESS_0$. More precisely, define the next set of weights as function of the next temperature
\[
w^{s,n}_j(\phi) = \left[P(Y_{c_j:c_{j+1}-1}\mid G^{s-1,n}_{c_j})\right]^{\phi - \phi_{s-1,j}}
\]
and consider
\[
ESS(\phi):= \frac{\left(\sum_{n = 1}^Nw^n_{s,j}(\phi) \right)^2}{\sum_{n = 1}^N \left(w^{n}_{s,j}(\phi)\right)^2} = \epsilon\,N
\]
whose solution -- assuming it exists within $(\phi_{s-1,j},1]$ -- provides the next temperature $\phi_{s,j}$. 
The solution is obtained with a simple fast bisection method.
If $ESS(1) \geq \epsilon N$, we simply select $\phi=1$. With this procedure we obtain all temperatures related to data $Y_{c_j:c_{j+1}-1}$ and we can then proceed to the next filtering step. We set $\epsilon$ to $1/2$ to obtain a minimum ESS of $N/2$, which is a common choice \citep[see, e.g.,][p.133]{chopin2020introduction}. See Algorithm \ref{alg:temp} for a detailed description.

\subsubsection{Determination of mutation kernel $\bar{P}$}
The mutation kernel $\bar{P}$ is used within the algorithm to jitter particles, and move them towards the centre of the support of each filtering distribution  under consideration during a full application of the particle filter. The addition of mutation steps has been shown to be, in many cases, critical, both in theoretical and experimental works, see e.g.~\cite{besk:14, ruza:21} and \cite{llop:18, van:21}, respectively. In Section~\ref{ssec:SMCsimul}, we illustrate such impact for the specific model at hand through a simulation study.

For the overall algorithm to ensure a correct particle filter on an extended space, the user-specified mutation kernel $\bar{P}_{j,s}(G_{c_{j}}\mid G_{c_{j}},G_{c_j})$ must have invariant distribution  
\begin{align*}
\label{eq:invariant}
G_{c_{j}} \mapsto \left[P(Y_{c_j:c_{j+1}-1}\mid G_{c_{j}})\right]^{\phi_{j,s}}
\times P(G_{c_{j}}\mid G_{c_{j}-1})
\end{align*}
where we use the convention that $G_{c_0-1}=\emptyset$ in which case the rightmost term becomes the prior defined by \eqref{eq:G}.
This is readily achieved via a M-H step. That is, for each current segment $c_j:c_{j+1}-1$, temperature $\phi_{j,s}$, graph $G_{c_j}$, with adjacency matrix $A_{c_j}=A_{c_j}[h,k]$, and given $G_{c_j-1}$, we 
define a proposed graph $G'_{c_j}$, with adjacency matrix $A'_{c_j}=A'_{c_j}[h,k]$, using the symmetric transition 
\begin{equation}
\label{eq:mutationprop}
		A_{c_j}[h,k]'\,\mid \,A_{c_j}[h,k] \overset{ind}{\sim}
	\mid A_{c_j}[h,k]  -	\text{Bernoulli}\Big( \tfrac{2s_0}{p-1}\Big) \mid 
\end{equation}
for algorithmic tuning parameter $s_0\in (0,(p-1)/2)$.
Thus, under the proposal in \eqref{eq:mutationprop}, the expected number of flips in the edges in $s_0\cdot p$.
The acceptance probability for the mutation step is
\begin{equation*}
\label{eq:acc_prob1}
1 \,\wedge\, \frac{\left[P(Y_{c_{j}:c_{j+1}-1}\mid  G'_{c_{j}})\right]^{\phi_{j,s}}\times P( G'_{c_{j}}\mid  G_{c_{j-1}})}{\left[P(Y_{c_{j}:c_{j+1}-1}\mid  G_{c_{j}})\right]^{\phi_{j,s}}\times P( G_{c_{j}}\mid  G_{c_{j-1}})}
\end{equation*}

\subsubsection{Likelihood given the graph structure}\label{ss:SMC}

	An important quantity required within the particle filter is the marginal likelihood
	\[
	P(Y_{c_{j}:c_{j+1}-1}\mid G_{c_{j}}) = \int_{M^{+}(G_{c_j})} P(Y_{c_{j}:c_{j+1}-1}\mid \Omega_{c_{j}}) P(\Omega_{c_{j}}\mid  G_{c_{j}})\,d \Omega_{c_{j}}
	\]
	Since the  $G$-Wishart law is conjugate, we can  integrate out the precision matrices $\Omega_{1:T}$. That is, we have \citep{atay:05}
	\begin{align*}
		P(Y_{c_{j}:c_{j+1}-1}\mid G_{c_{j}}) = \frac{1}{(2\pi)^{(c_{j+1}-c_{j})p/2}}
		\frac{I_{G_{c_{j}}}(d + (c_{j+1}-c_{j}),D+H_{j})}{I_{G_{c_{j}}}(d,D)}
	\end{align*}
	where, for $j=0,\dots, \kappa$,
	\begin{align*}
		H_j = \sum_{i=c_{j}}^{c_{j+1}-1}Y_i\,Y_i^{\top}
	\end{align*}
{Notice that, while computing the likelihood of the graphs, we marginalize over $\Omega_t$ and, thus, the particles (from the inner algorithm) consist of only the graphs. However, after running the particle filter and thanks to the conjugacy properties of the $G$-Wishart law, a straightforward independent sampler can be used to get both marginal and conditional posterior of the precision matrices, where with conditional posterior distribution we mean the posterior distribution conditional on the point estimates of the graphs.}
    The normalising constant  of the $G$-Wishart prior can be factorised  \citep{rove:02, uhler2018exact},  i.e., 
	for a given graph $G$,   
	\begin{align}
 \label{eq:factorise}
		I_{G}(d,D) = \frac{\prod_{m=1}^{r} I_{G_{P_m}}(d, D_{P_{m}})}
		{\prod_{m=2}^{r} I_{G_{S_m}}(d, D_{L_{m}})}
	\end{align}
	where $P_1; L_2, P_2;\ldots ;P_r,L_r,$ is a perfect sequence of prime components and corresponding minimal separators of $G$ \citep[see, e.g., Chapter~2 of][for details]{laur:96} and $D_{P_m}$ is the submatrix of $D$ corresponding to the
rows and columns in $P_m$. In the case of a \emph{decomposable} graph $G$, all prime components are complete graphs. For complete graphs the $G$-Wishart distribution coincides with the Hyper-Wishart distribution \citep{dawid1993hyper},  for which an analytical expression for the normalising constant is available: 
	\begin{align*}
 \label{eq:factor}
		I_{G_{P_m}}(d,D_{P_m}) = \frac{2^{(d+p_m-1)p_m/2}\Gamma_{p_m}(\frac{d+p_m-1}{2})}{|D_{P_m}|^{(d+p_m-1)}}
	\end{align*}
Here $\Gamma_{d}(\cdot)$ is the multivariate Gamma function of dimension $d$ and $p_m$ is the dimension of $D_{P_m}$. Note that, by construction, the minimal separators are complete sub-graphs of $G$, thus the terms in the denominator in (\ref{eq:factorise}) are analytically computable. 
For a general, \emph{non-decomposable} graph $G$, 
	\cite{rove:02, dell:03, atay:05, carvalho2007simulation}  propose Monte-Carlo methods
	for the approximation of $I_{G}(d,D)$. Herein, to compute the normalizing constant, we employ the method of \cite{atay:05} implemented in the function \textsf{gnorm} of the \textsf{R} package \textsf{BDgraph} \citep{BDgraph}. When dealing with large number of nodes, the implementation of more sophisticated algorithms, such as the exchange algorithm by \cite{murray2006mcmcfor}, is advisable \citep[see, e.g.,][for an application to GGMs]{cheng2012hierarchical,van2022unbiased}.

\subsubsection{Evaluation of SMC approximation}
\label{ssec:SMCsimul}
The inner SMC algorithm provides unbiased estimates of the marginal likelihood conditionally on the change points. The variability of such estimates depends on the number of particles $N$, and the effect of the tempering and mutation steps, with the number of the latter, $M$, specified by the user. 
Thus, a trade-off is posed between accuracy of estimates and computational time.  

To assess the effect of the number of particles and the mutation step, and, in general, obtain insights into the performance of the SMC component, we perform a series of simulation studies. We simulate data for $p=10$ nodes and $T=200$ observation instances. We then fix the change point sequence to its known true value, and carry out 30 independent executions of the SMC algorithm, for each different combination of $N \in \{200, 500, 750\}$ and $M \in \{0,5,10,20\}$. Recall that the mutation steps are performed only when the ESS falls below the threshold $\epsilon\, N$, where $\epsilon$ is here fixed to be $N/2$. 
We consider two data generating mechanisms. The first (Scenario A) has no change points and the $p$ variates are mutually independent (see Figure B.1.1 (a) in the Supplement for the corresponding graph). In Scenario~B we set a change point at $t = 70$, and the two graphs (before and after the change point) encode some non-trivial dependence. The full graph structure of Scenario~B is described later in Section~\ref{sec:simmodel3} and displayed in Figure B.1.2 of the Supplement. 
 Figure A.1 of the Supplement and Figure~\ref{MN_scenario2} show the box-plots of the estimates of the log-likelihood, the standard deviation, and the running time of the inner SMC algorithm coded in \textsf{R} (and run with an Intel Xeon W-1250 processor), under scenarios A and B, respectively.

Under Scenario A, the variability of the estimates is 
limited, as expected, for all pairs $(N,M)$ since data are simulated under the assumption of independence with no change points. However, in real scenarios, as the analysis of financial markets, this is highly unlikely to be the case and the computational machinery here developed is essential. 
Figure~\ref{MN_scenario2} shows similar box-plots obtained under the more realistic and challenging Scenario B. Here,  we obtain higher variability with values of standard deviation ranging from  2.986 to 29.297. These results highlight the importance of the tempering and mutation steps
 for the overall algorithmic performance (even accounting for the increased computational time), and in particular for recovering the complex dependence structures. Their role is essential in reducing variability of the estimates of the normalising constant. 
Lastly, notice that computational time is increased compared to Scenario A as a consequence of: (i) the presence of a change point; (ii) the computational complexity of the Monte Carlo iterations used to compute $I_{G}(d, D)$, which increases when particles concentrate on less sparse 
graphs; (iii) and the increased number of times the ESS threshold is reached. 

\begin{figure}[tb]
\centering
\includegraphics[width=\textwidth]{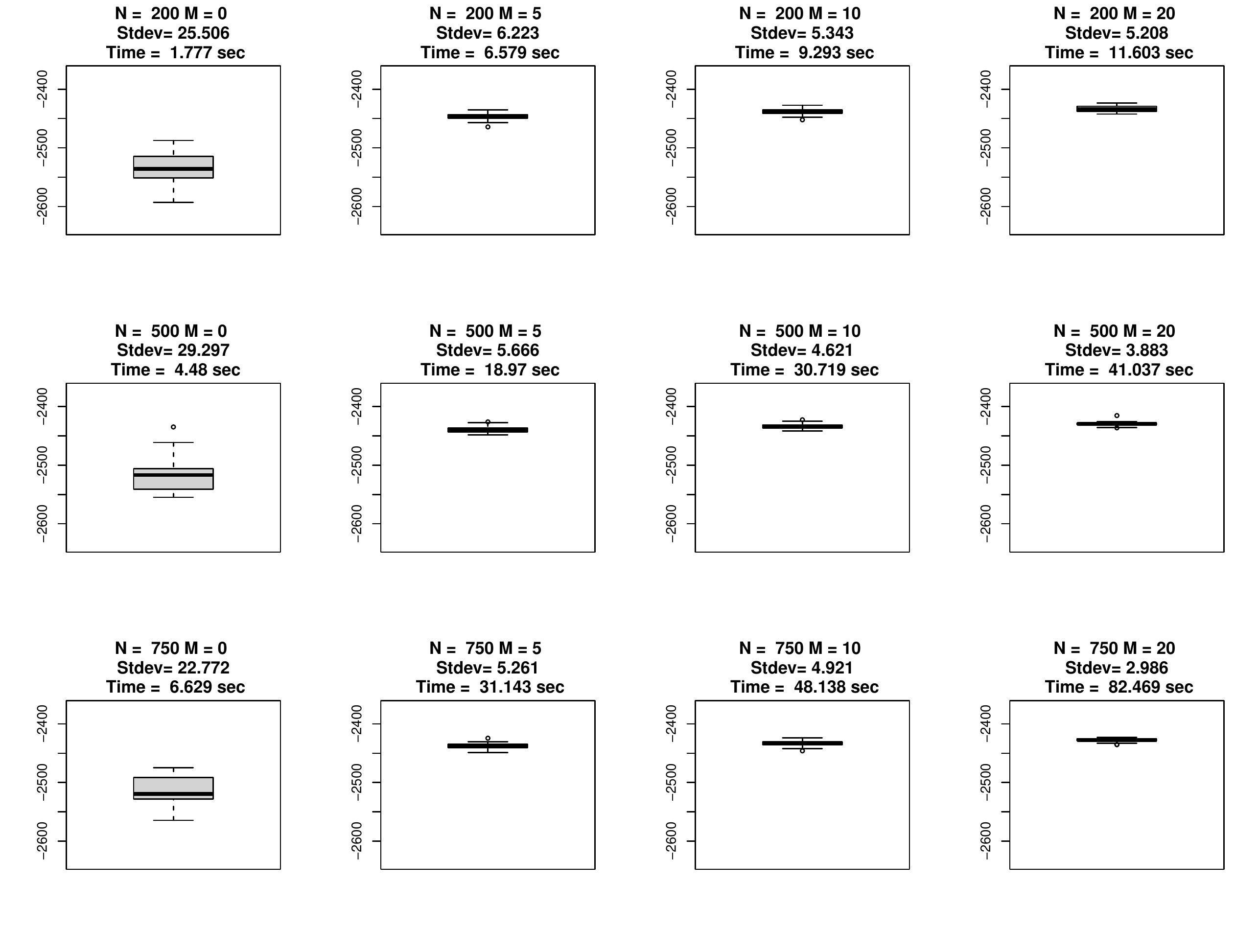}
\caption{\label{MN_scenario2}Log-likelihood estimates for Scenario~B obtained with the particle filter by fixing the change points to the truth. Distinct box-plots correspond to different numbers of particles $N$ and/or mutations steps~$M$. 
For each pair $(N,M)$, we run the algorithm 30 times and obtain the log-likelihood estimates. 
Each box-plot shows the distribution of such estimates. The variability of the estimates is rather limited for all pairs $(N,M)$, provided that $M\geq 5$.}
\end{figure}

\section{Model performance on simulated data}\label{sec:simmodel}
{WE investigate model performance through a simulation study which includes  five different scenarios. For each scenario, we simulate data for $T=200$ observation times. There are no change points in Scenarios 1 and 2, there is one change point in Scenario 3, and there are three change points in Scenario 4.Finally, in Scenario 5 the dependence structure, as captured by the precision matrix,  presents smooth changes and thus our model is misspecified. The number of nodes is $p=10$ in Scenarios 1, 2, 3, and 5, and $p=20$ in Scenario 4. In terms of the abruptness of changes in the precision matrix, Scenario 4 presents highly-abrupt changes, Scenario 3 presents mildly-abrupt changes, and Scenario 5 presents smooth changes. For a detailed description, see  Sections \ref{sec:simmodel4}, \ref{sec:simmodel3} and Section B.4 of the Supplement.}

{
To carry out posterior inference, we run the algorithm to estimate the change point sequence, using $N=200$ particles, $M=10$ mutation steps, and performing 10,000 iterations of the outer component, of which the first 2,000 are discarded as burn-in. 
Then, we re-run only the inner SMC with $N=1,000$ particles and $M=20$ mutation steps to obtain the graph estimates conditionally on the maximum-a-posteriori (MAP) estimate of the change point sequence obtained in the first step. 
When the true sequence of change points is the null set (Scenarios 1 and 2), we initialise the MCMC chain at $(c_1 = 51,\, c_2 = 101,\, c_3 = 151)$, whereas when the graph and/or the precision matrix changes (Scenario 3, 4, and 5), we initialise the chain to the state of zero change points. In real applications, when the change points are unknown, we suggest initialising the chain to no change points.  The adopted initialization for these simulations better tests the convergence speed of the algorithm.}

{The inference
results for no change points show the expected good performance of the model in terms of both identification of  change points and recovery of the dependence structure. The posterior concentrates on the true state of no change points, with posterior probability of no-change points  not falling below 0.98 in all replicates. The area under the curve (AUC) for edge detection is approximately 1. We refer  to section B of the Supplement for a more detailed presentation of the simulation studies for Scenarios 1, 2, and 5. The following Sections {describe} Scenarios 3 and 4.} 

\subsection{Simulation results for one change point}\label{sec:simmodel3}

The third simulation scenario (Scenario 3) is obtained by setting one change point at $t=70$ and generating two precision matrices as in \cite{peterson2015bayesian} and \cite{molinari2022bayesian}. In particular, 
we first define the  $\Omega_{c_0}$ and then derive $\Omega_{c_1}$ as perturbation of $\Omega_{c_0}$, which defines a mildly-abrupt change. Firstly, $\Omega_{c_0}$ is obtained by setting diagonal elements equal to 1, the first off-diagonal elements to 0.5, i.e., $\Omega_{c_0}[h,h+1] = \Omega_{c_0}[h+1, h] = 0.5$, for $h=1,\ldots,9$, and the remaining elements to 0. To construct $\Omega_{c_1}$, we randomly remove five edges among the active ones in $G_{c_0}$ and set to 0 the corresponding entries in the precision matrix. Then we add five randomly selected edges drawn from the set of inactive edges in $G_{c_0}$. Finally, a precision entry equal to 0.2 is assigned to the new  edges. The obtained matrix  is not necessary positive-definite, and, to
this end, we compute the nearest positive-definite approximation through the \textsf{R} function \textsf{nearPD} \citep{higham2002computing}, available in the \textsf{R} package \textsf{Matrix} \citep{Matrix}.  The resulting graphs are shown in Figure B.1.2 of the Supplement. We note that the computation of the nearest positive-definite matrix may result in a strong shrinkage of the non-zero elements in the precision matrix, which may cause unrealistic high values in the correspondent covariance matrix. However, this is not the case in our simulation scenario (see Figure B.1.4 and B.1.5 in the Supplement, where the simulated covariance matrices and data are displayed). We consider 20 replicates of Scenario 3.

The hyperparameter $\omega$ in \eqref{eq:G} is determined using an approach inspired by  empirical Bayes techniques, so that a priori the expected number of edges for the graphs is equal to the number of edges detected by estimating one unique graph using all the time points. To this end, we estimate the graph using an adaptive lasso approach, which is a modification of the estimation procedure proposed by \cite{meinshausen2006high} inspired by the adaptive lasso of \citep{zou2006adaptive}, as implemented in  the \textsf{R} package \textsf{GGMselect} \citep{GGMselect}. For the hyperparameter $z$ in \eqref{eq:GG}, we opt for $z =0.1$, so that a priori we expect only one  edge to change at each change point, favouring graph similarity. This choice also allows us to better understand model performance and hyperparameter sensitivity, as in our simulations we force 10 edges to change across the change point, an event to which our prior associates a probability lower than $4\cdot 10^{-8}$. The hyperparameter $p_0$, which controls the a priori number of the change points, is set to $p_0=0.1$ to favour sparsity. The hyperparameters of the G-Wishart distribution are set to the common values of $\delta = 3$ and $D = \text{Id}_p$. In Section C.5 of the Supplement we provide  hyperparameter sensitivity analysis carried on the real dataset. The change point detection procedure appears to be unsensitive to the choice of the hyperparameters and the graph recovery performance is limitedly affected by the choice of $z$.

\begin{table}[tb]
\newcolumntype{b}{>{\columncolor{Gray}}c}
\begin{center}
\resizebox{\textwidth}{!}{ 
\begin{tabular}{b| c|b|c|c|c|c|c|c|c}
\cellcolor{white}& \cellcolor{LightCyan}True  &\cellcolor{white}&\cellcolor{LightCyan}MAP&\cellcolor{LightCyan}MAP&\cellcolor{LightCyan}prob&\cellcolor{LightCyan}90\%&\cellcolor{LightCyan}95\%&\cellcolor{LightCyan}&\cellcolor{LightCyan}\\
\cellcolor{white}&\cellcolor{LightCyan}change point &\cellcolor{white}& \cellcolor{LightCyan} est. &\cellcolor{LightCyan}  prob.& \cellcolor{LightCyan}$\kappa = 1$&\cellcolor{LightCyan} C.I.&\cellcolor{LightCyan} C.I.&\cellcolor{LightCyan} Mean& \cellcolor{LightCyan} Median\\
\hline
          &&Rep. 10 & $(70)$ & 0.306& 0.971 & $[70,74]$&$[70,79]$&72.21 & 71\\
          &&Rep. 5 & $(70)$ & 0.294 & 0.987 & $[70,80]$ & $[70,80]$ & 73.34 & 72\\
          &&Rep. 3 & $(70)$ & 0.289 & 0.975 & $[69,79]$ & $[68,79]$ & 71.17 & 70 \\
          \hline 
Scenario 3&(70)\\
          \hline
          &&Rep. 11 & $(68)$ & 0.245 & 0.966 & $[68,76]$ & $[68,81]$ & 70.60 & 70\\
          &&Rep. 20 & $(73)$ & 0.179 & 0.965 & $[67,75]$ & $[66,77]$ & 71.19 & 71\\
          &&Rep. 12 & $(74)$ & 0.191 & 0.933 & $[67,83]$ & $[67,83]$ & 73.53 & 73\\
     \hline
\end{tabular}
}
\caption{\label{tab:scenario3} Scenario 3: Posterior summaries for change points of the three best and three worst replicates in terms of MAP estimate and MAP probability. MAP estimates, MAP probabilities (for the posterior over all configurations of change points), posterior probability of the number of change points being 1, and credible intervals, mean and median of the position of the change point (conditionally on having one change point). Credible intervals are obtained computing the smallest credible sets with 90\% and 95\% credibility, which are not necessary continuous intervals, and then using the minimum and the maximum time points in the credible set as extremes of the provided interval.}
\end{center}
\end{table}
\begin{figure}[tb]
\centering
\begin{subfigure}[b]{0.49\textwidth}
\includegraphics[width=\textwidth]{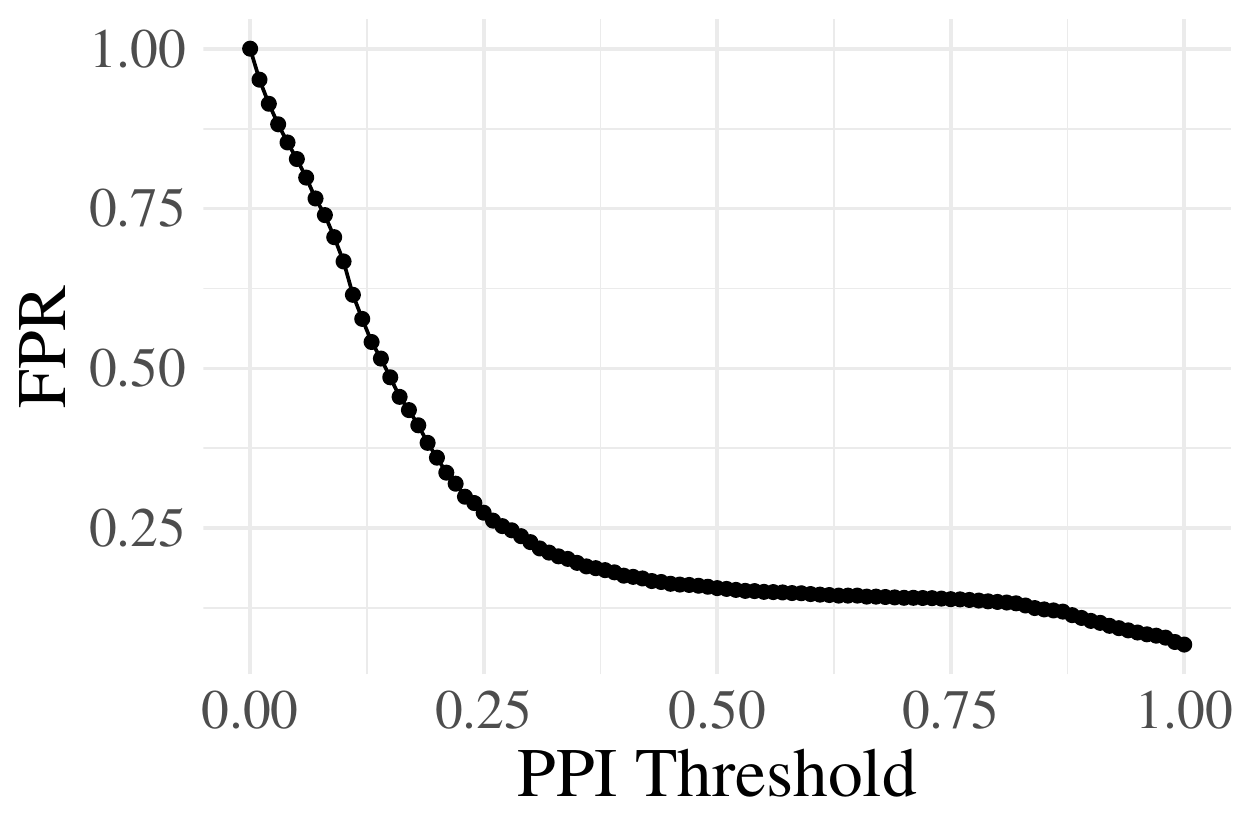}
\caption{\label{fig:FPR3}False positive rate}
\end{subfigure}
\hfill
    \begin{subfigure}[b]{0.45\textwidth}
    \includegraphics[width=\textwidth]{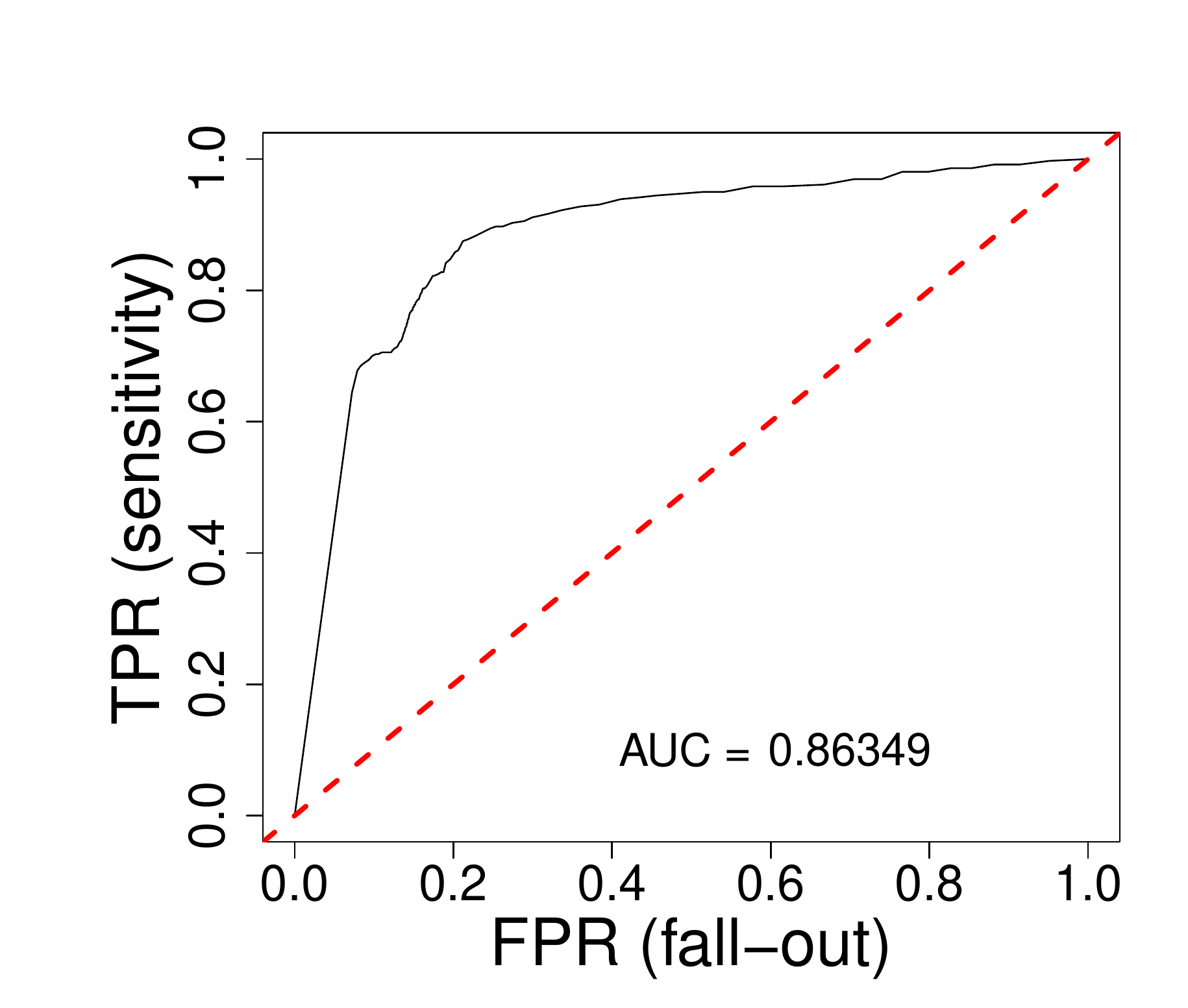}
    \caption{\label{fig:ROC3}Receiver operating characteristic curve}
    \end{subfigure}
    \caption{Panel (a): FPR versus PPI threshold, for Scenario 3, computed via 20 replicates. Panel (b): ROC curve in Scenario 3, computed via 10 replicates. }
\end{figure}

 Table~\ref{tab:scenario3} contains posterior summaries of the three best and three worst replicates based on the accuracy in the recovery of the change point configuration. Summaries for all replicates can be found in Section B.3 of the Supplement. In all replicates, the posterior distribution of the number of change points is concentrated around~1, with a posterior probability greater or equal to 0.91. Moreover, though the space of sequences that satisfy the minimum duration constraint of $\ell=12$ for $T=200$ includes more than $4\cdot 10^{12}$ sequences, in eight of the 20 replicates the MAP estimate (which minimises the $0-1$ loss function) coincides with the true state $c^{\star} = 70$, and in all replicates the MAP is contained in the interval  $[68, 74]$.  Moreover, Table~\ref{tab:scenario3} reports also the $90\%$ and $95\%$ credible intervals, the mean, and the median for the position of the change point, confirming that the posterior is concentrated around the true state in all replicates. 
 Figure~\ref{fig:FPR3} shows the combined false positive rate (FPR) of edge detection for the two graphs as a function of the threshold used for the posterior probability of inclusion (PPI). FPRs show a reasonable pattern, and, for the 0.5 threshold, the FPR is 0.156.  Figure~\ref{fig:ROC3} displays the combined ROC curve, with an AUC  approximately equal to 0.86. {The graph estimates used to evaluate the FPRs are obtained conditionally on the estimated change point sequence, even when it does not coincide with the true one. Indeed, the MAP always identifies a single change point, so, we compare the estimated graphs before/after the estimated change point, with the true graph before/after the true change point for fairness of results.}

{Lastly, we compare the results with those obtained by applying the group-fused graphical lasso (GFGL), introduced by \cite{gibberd2017regularized} \citep[see also][]{gibberd2017multiple}, and the LOcal Group Graphical Lasso Estimation (\textsf{loggle}) of \cite{yang2015fused}. Similarly to our proposal, GFGL consists of a piece-wise constant graphical model. Differently from our model, it is estimated employing a penalized likelihood approach, where two penalties act favouring both sparsity of the graph structure (i.e., shrinkage penalty) and sparsity in the number of change points (i.e., smoothing penalty). 
 The \textsf{loggle} approach is also based on a penalised likelihood approach, but it assumes that the graph topology is gradually changing over time and, thus, cannot be used to detect change points. We compare our approach with an ``oracle version" of \textsf{loggle}, where we estimate the graphs knowing the position of the change points.} 
 
  {From a theoretical point of view, both GFGL and \textsf{loggle} prohibit principled uncertainty quantification on the number and location of change points and  graph structure. Contrarily, our strategy allows for straightforward  uncertainty quantification, which is one of teh main advantages of the Bayesian framework. However, as it is well-known,  Bayesian posterior inference typically comes at the expense of the computational time needed to estimate the model. So, in this scenario, GFGL and \textsf{loggle} produce estimates in a few seconds or minutes and our dynamic Gaussian graphical model requires hours to be estimated (see D.2 in the Supplement for details on  computational time).}  
{ Results from the GFGL model are obtained   for different values of the hyperparameters $\lambda_1$ and $\lambda_2$. The hyperparameters $\lambda_1$ and $\lambda_2$ control respectively the shrinkage and the smoothness of the solution; for more details see \cite{gibberd2017regularized}. Detailed output summaries for the GFGL are presented in Section B.3 of the Supplement. As already noticed by \cite{gibberd2017regularized}, GFGL's results can be highly sensitive to the choice of the hyperparameters in terms of both detected change points and recovered graph structure. In our experiment, the number of change points estimated by GFGL varies from one to seven, depending on the simulation replicate and on the choice of hyperparameters $\lambda_1$ and $\lambda_2$. The location of the change points also varies largely across the different simulation replicates. Contrary, our model identifies the correct number of change points and their approximate position in all replicates. } {The ``oracle version" of the \textsf{loggle} model gives a FPR of 0.242 and a TPR of 0.825. Contrary, with our approach, if we fix the FPR to 0.242 the corresponding TPR is
0.895, fixing the TPR to 0.825 leads to a FPR of 0.181. (see Table B.3.2 in the Supplement).  }

\subsection{Simulation results for more changes points and nodes}\label{sec:simmodel4}
{The fourth simulation scenario (Scenario 4) is obtained simulating data for $T=200$ time points and $p=20$ variables/nodes. The data generating mechanism presents three change points located at $t =60$, $t=100$ and $t=150$ and, thus, four different graph structures. The true graph structures are displayed in Figure B.1.3 of the Supplement and obtained fixing the first graph $G_{1}$, which presents 11 {activated} edges, and subsequently randomly changing the graph structure across change points. In correspondence of each change point, any {active}/non-active edge is deactived/activated with probability 0.4. The four precision matrices are then generated sampling from a $G$-Wishart distribution, independently conditionally on graph structure. Note that the abruptness in the changes in the dependence structure is higher in this scenario than in Scenario 3, since the precision matrix is generated independently at each change point and not obtained as a perturbation of the previous precision matrix as in the previous section. The hyperparameters are chosen as described in the previous section.} 

{The posterior distribution for the number of change points assigns probability one to  the correct value of 3 (see Figure B.4.1 in the Supplement). The posterior expectation  for the first change point location is 60.68, for the second change point is 98.11, and for the third one is 149.94. More details on posterior inference results are provided in Section B.4 of the Supplement.
Figure~\ref{fig:FPR4} shows the combined  FR of edge detection for the four graphs as function of the threshold used for the PPI. Again, FPRs are appears reasonable for any PPI threshold, quickly decaying to zero, and, for the 0.5 threshold, the FPR is 0.035. Figure~\ref{fig:ROC4} displays the ROC curve, with an AUC  approximately equal to 0.89.
In the Supplement, we report also results obtained with the GFGL model of \cite{gibberd2017regularized}, which, similarly to what already observed in the previous section, shows high variability of the estimates depending on the value of the hyperparameters. Moreover, in this scenario, we note that the GFGL model leads  also to poor graph recovery even when the correct change points are detected (see Figures B.4.3 and B.4.4 of the Supplement). The ``oracle version" of the \textsf{loggle} model produces a FPR of 0.113 and a TPR of 0.426, presenting a significant worse performance than  our approach. For example,   for a 0.5 PPI threshold we obtain a FPR of 0.035 and a TPR of 0.611, 
if we fix the FPR to 0.113 the corresponding  TPR is 0.764, fixing the  TPR to 0.426  leads to a FPR of 0.021. See Table~\ref{tab:logglescen4}. }
\begin{figure}[tb]
\centering
\begin{subfigure}[b]{0.49\textwidth}
\includegraphics[width=\textwidth]{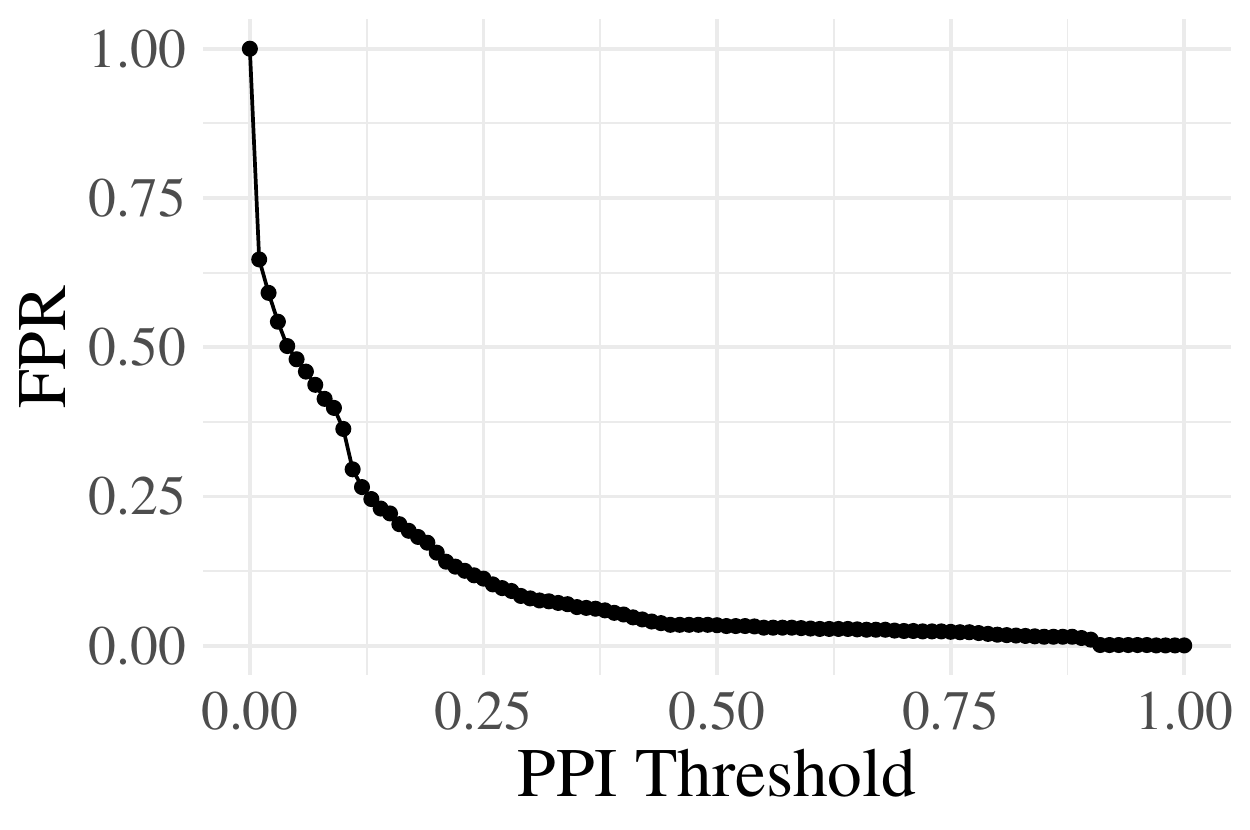}
\caption{\label{fig:FPR4}False positive rate}
\end{subfigure}
\hfill
    \begin{subfigure}[b]{0.45\textwidth}
    \includegraphics[width=\textwidth]{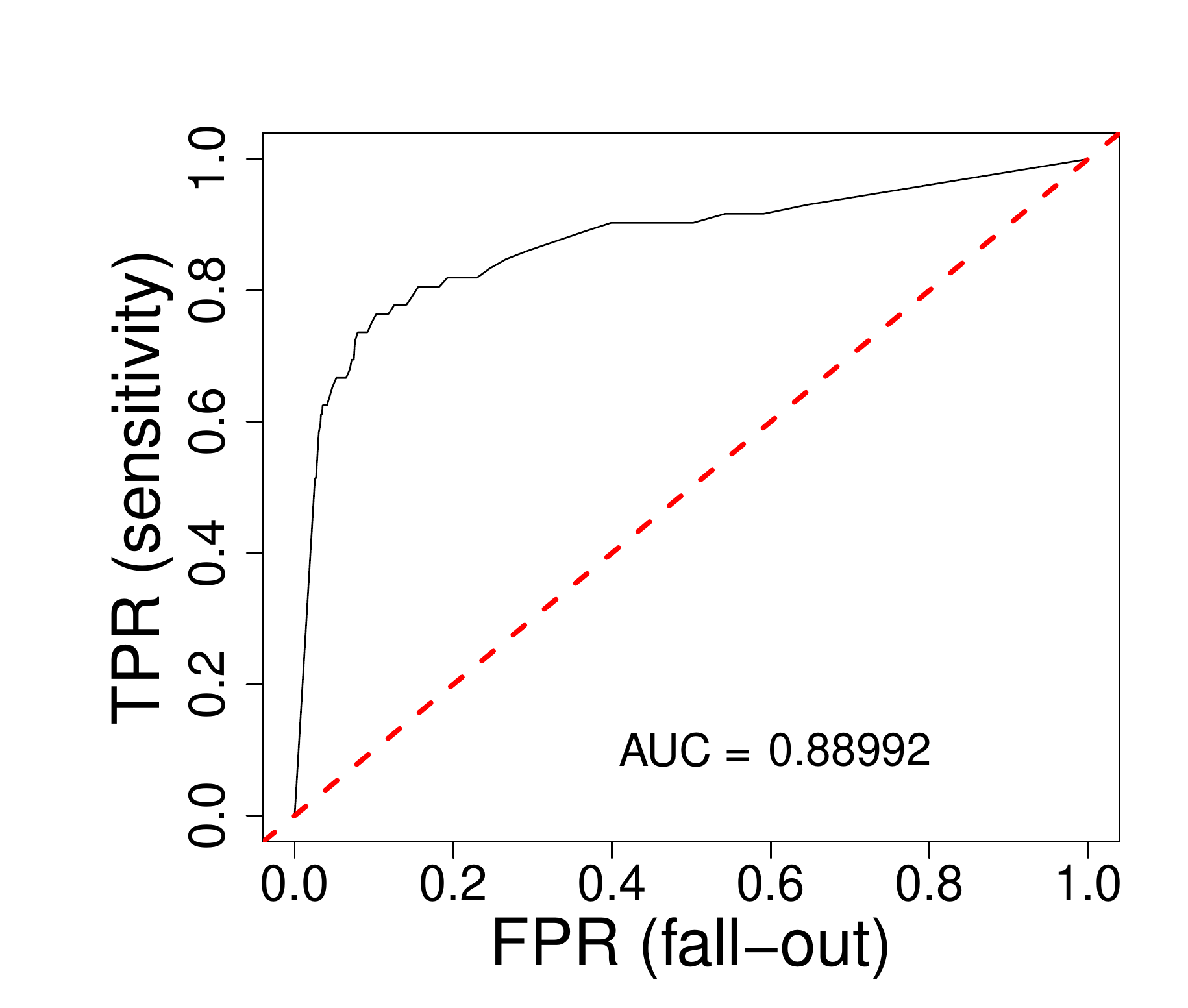}
    \caption{\label{fig:ROC4}Receiver operating characteristic curve}
    \end{subfigure}
    \caption{Panel (a): FPR versus PPI threshold, for Scenario 4. Panel (b): ROC curve in Scenario 4. }
\end{figure}

\begin{table}[!h]
\begin{center}
\begin{tabular}{c|c|c|c|c}
&loggle&\multicolumn{3}{c}{Bayesian dynamic GGM}\\
&&PPI thres. 0.5&PPI thres. 0.25&PPI thres. 0.78\\
\hline
FPR&0.113&0.035&0.113&0.021\\
TPR&0.426&0.611&0.764&0.426\\
\end{tabular}
\end{center}
\caption{\label{tab:logglescen4} Scenario 4: Comparison on graph recovery between the results obtained with the ``oracle version" of the \textsf{loggle} model and our model.}
\end{table}

\section{Industry returns during COVID-19 pandemic}
\label{sec:real}

We apply the model to detect changes in the dependence structure of the nine industry portfolios' weekly returns described in Section~\ref{sec:intro}. We consider weekly data over a time horizon of three years: from January 2019 to December 2021 so that $T=157$. {When choosing  which type of returns to include in the analysis, i.e. daily, weekly, or monthly, we are faced with a  a trade-off: higher frequency data may show lower degree of dependence, making harder to detect structure changes in the dependence structure \cite[see, for instance,][]{ab2018dependence}; on the other hand, lower frequency data provide a less detailed representation of markets' trends. For this reason, we consider weekly returns to attain a more detailed level of information compared to monthly data and a potentially stronger signal on correlations compared to daily returns.} Logarithmic weekly returns are computed starting from the daily returns available from Kenneth R. French's Data Library at \url{https://mba.tuck.dartmouth.edu/pages/faculty/ken.french/data_library.html},  where the industry classification used to associate each stock to one of the nine portfolios is defined as follows. Stocks listed in the New York Stock Exchange (NYSE), the American Stock Exchange (AMEX), and National Association of Securities Dealers Automatic Quotation System (NASDAQ) are assigned to an industry at the end of June of year $t$ based on their four-digit standard industrial classification (SIC) code at that time. Then, returns are computed from July of year $t$ to June of year $t+1$. The corresponding standardized time series are represented in Figure C.1 of the Supplement. 

\begin{figure}[!t]
\centering
\includegraphics[width=0.85\textwidth]{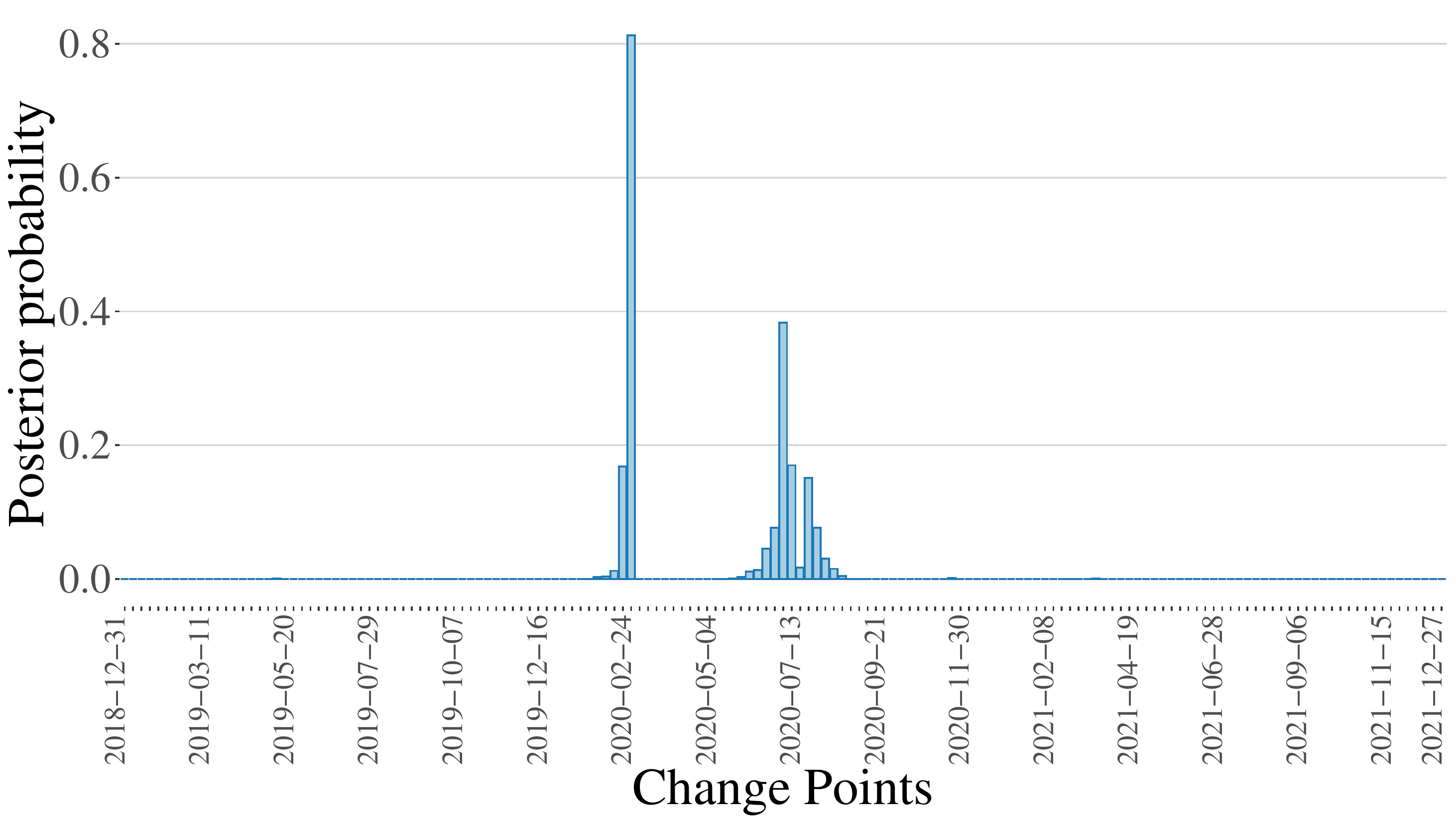}
\caption{\label{real_where}Marginal posterior probability of every time point to be a change point.}
\end{figure}

To estimate the dependence structure for the weekly returns, we firstly run the algorithm to estimate the change point sequence, using 200 particles, 10 mutation steps, performing 32,000 iterations of the outer component, of which the first 2,000 are discarded as burn-in, and  thinning every ten iterations.   
Secondly, we re-run only the particle filter with 1,000 particles and 20 mutation steps to sample the graphs from their posterior distribution conditionally on the MAP estimate of the change point configuration. The algorithm is  initialised assuming  no change points and hyperparameters are set as described in Section~\ref{sec:simmodel3}.

\begin{table}[!b]
\newcolumntype{a}{>{\columncolor{Gray}}l}
\centering
\begin{tabular}{ac|ac|ac|ac}
\rowcolor{LightCyan}change & post & change & post & change & post & change & post \\
\rowcolor{LightCyan}points & prob & points & prob & points & prob & points & prob \\
\hline
\hline
(57 79) 	 &0.0010&	(59 82)	 &0.0073&	(60 99)	 &0.0003&	(61 81)	 &0.0137\\
(57 82)	 &0.0010&	(59 84)	 &0.0017&	(61 116)	 &0.0010&	(61 82)	 &\textbf{0.1094}\\
(57 84)	 &0.0007&	(60 78)	 &0.0020&	(61 73)	 &0.0007&	(61 83)	 &0.0650\\
(57 85)	 &0.0003&	(60 79)	 &0.0073&	(61 74)	 &0.0033&	(61 84)	 &0.0150\\
(58 79)	 &0.0003&	(60 80)	 &\textbf{0.1010}&	(61 75)	 &0.0103&	(61 85)	 &0.0127\\
(58 82)	 &0.0033&	(60 81)	 &0.0033&	(61 76)	 &0.0133&	(61 86)	 &0.0043\\
(59 77)	 &0.0013&	(60 82)	 &0.0297&	(61 77)	 &0.0447&	(61 99)	 &0.0010\\
(59 78)	 &0.0007&	(60 83)	 &0.0100&	(61 78)	 &\textbf{0.0737}&	(19 61 83)	 &0.0010\\
(59 79)	 &0.0010&	(60 84)	 &0.0127&	\textcolor{blue}{(61 79)}	 &\textcolor{blue}{\textbf{0.3735}}&	(39 61 83)	 &0.0003\\
(59 81)	 &0.0003&	(60 85)	 &0.0020&	(61 80)	 &\textbf{0.0694}&	(61 80 113)	 &0.0003
\end{tabular}
\caption{\label{tab:postcp} Posterior distribution of change point configuration. In \textbf{bold} we highlight probabilities greater than 0.05 and in \textcolor{blue}{\textbf{blue}} the MAP estimate.}
\end{table}

Figure~\ref{real_where} shows, for each time point, the marginal posterior probability of being a change point and Table~\ref{tab:postcp} reports the joint posterior distribution of the configurations of change points, which have been accepted by the algorithm. The posterior distribution on the number of change points assigns  probability  0.9984 to two change points and the remaining mass 0.0016 to three change points. {The posterior distribution  for the number and location of change points are highly concentrated around the posterior mode, showing a low level of uncertainty.} Our analysis highlights a first structural change at $t=61$, i.e., during the week starting on February 24, 2020, in correspondence of what appears to be the market's reaction to the first significant world-wide increase in Coronavirus confirmed cases and deaths outside  China over the previous weekend. In particular, during the weekend February 21-23, 2020, Italy, the first and  hardest-hit country in Europe in 2020, reported the first local cases of COVID-19 \citep[see, e.g.,][]{just2020stock}. During the week of February 24, 2020, the Dow Jones and S\&P 500 fell by 11\% and 12\%, respectively, marking the biggest weekly declines to occur since the financial crisis of 2008. {The identification of a first change point in correspondence of a major shock is coherent with the known stylized fact that, during crisis, dependence among investments typically increases diminishing diversification benefits \citep[see, for instance,][]{kotkatvuori2013stock}.}
A second change point is detected at $t = 79$, i.e., during the last week of June 2020, interpretable as a subsequent and partial re-stabilization of the financial markets after the initial and most uncertain period of the pandemic. {The credible intervals, obtained computing the smallest credible sets with 95\% credibility and then using the minimum and the maximum time points in the set as extremes of the interval, are $[60,61]$ and $[76,83]$ for the first and second change point respectively.}

\begin{figure}[!t]
\begin{center}
\resizebox{\columnwidth}{!}{%
\begin{tikzpicture}
\thispagestyle{empty}
\small
\tikzstyle{mystyle}=[circle,minimum size=4mm,draw=black, draw opacity=0.2, fill=green, fill opacity=0.2]

    \draw [line width=0.5mm,|-|, draw = blue, draw opacity=0.2] (-2 , 2.2) -- (2 , 2.2);
    \node [rectangle split,rectangle split parts=2] (graph1) at (0,2.8) {\textbf{From December 31, 2018 } \nodepart{second} \textbf{to February 21, 2020}}; 
    \node[mystyle,draw=black,label=center:Manuf] (Manuf) at (0,0) {};
    \node[mystyle,draw=black,label=center:Enrgy] (Enrgy) at (-1.5,1.5) {};
    \node[mystyle,draw=black, label=center:Utils] (Utils) at (1.4,1.8) {};
    \node[mystyle,draw=black, label=center:NoDur] (NoDur) at (1.6,0.3) {};
    \node[mystyle,draw=black, label=center:Shops] (Shops) at (1,-2.8) {};
    \node[mystyle,draw=black, label=center:Durbl] (Durbl) at (-1.35,-1.5) {};
    \node[mystyle,draw=black, label=center:Hlth] (Hlth) at (-1.2,-2.8) {};
    \node[mystyle,draw=black, label=center:HiTec] (HiTec) at (-0.5,-2) {};
    \node[mystyle,draw=black, label=center:Telcm] (Telcm) at (0.5,-1.5) {};

    \path [-, text = blue] (Manuf) edge node {0.96} (Enrgy);
    \path [-, text = blue] (Manuf) edge node {1 } (Durbl);
    \path [-, text = blue] (Manuf) edge node {1 }  (HiTec);
    
    \path [-, text = blue]  (NoDur) edge node {1} (Utils);
    \path [-,  text = blue]   (NoDur) edge node {1} (Telcm);
    
    \path[-, text = blue] (Shops) edge node {0.99} (Telcm);

    \path [-, text = blue] (HiTec) edge node {1 } (Shops);
    \path [-, text = blue] (HiTec) edge node {0.99 } (Hlth);
    
    \draw [line width=0.5mm,|-|, draw = blue, draw opacity=0.2] (3 , 2.2) -- (7 , 2.2);
    \node [rectangle split,rectangle split parts=2] (graph3) at (5,2.8) {\textbf{From February 24, 2020 } \nodepart{second} \textbf{to June 26, 2020}}; 
    \node[mystyle,draw=black,label=center:Manuf] (Manuf3) at (5,0) {};
    \node[mystyle,draw=black,label=center:Enrgy] (Enrgy3) at (3.5,1.5) {};
    \node[mystyle,draw=black, label=center:Utils] (Utils3) at (6.4,1.8) {};
    \node[mystyle,draw=black, label=center:NoDur] (NoDur3) at (6.6,0.3) {};
    \node[mystyle,draw=black, label=center:Shops] (Shops3) at (3.8,-2.8) {};
    \node[mystyle,draw=black, label=center:Durbl] (Durbl3) at (3.65,-1.5) {};
    \node[mystyle,draw=black, label=center:Hlth] (Hlth3) at (6,-2.8) {};
    \node[mystyle,draw=black, label=center:HiTec] (HiTec3) at (4.5,-2) {};
    \node[mystyle,draw=black, label=center:Telcm] (Telcm3) at (5.5,-1.5) {};
    
    \path [-, text = blue] (Manuf3) edge node {0.82} (Enrgy3);
    \path [-, text = blue] (Manuf3) edge node {0.97} (Durbl3);
    \path [-, text = blue] (Manuf3) edge node {0.93} (NoDur3);
    \path [-, text = blue] (Manuf3) edge node {0.84} (Utils3);
    
    \path [-, text = blue] (NoDur3) edge node {0.94} (Telcm3);
    \path [-, text = blue] (NoDur3) edge node {1 } (Hlth3);
    
    \path [-, text = blue] (HiTec3) edge node {1 } (Hlth3);
    \path [-, text = blue] (Shops3) edge node {1 } (Hlth3);
    
    \path [-, text = blue] (Utils3) edge node {0.87} (Enrgy3);
    
    \path [-, text = blue] (HiTec3) edge node {1} (Durbl3);
    \path [-, text = blue] (HiTec3) edge node {0.98} (Shops3);

     \draw [line width=0.5mm,|-|, draw = blue, draw opacity=0.2] (8 , 2.2) -- (12 , 2.2);
     \node [rectangle split,rectangle split parts=2] (graph5) at (10,2.8) {\textbf{From Jun 29, 2020 } \nodepart{second} \textbf{to December 31, 2021}};  
    \node[mystyle,draw=black,label=center:Manuf] (Manuf5) at (10,0) {};
    \node[mystyle,draw=black,label=center:Enrgy] (Enrgy5) at (8.5,1.5) {};
    \node[mystyle,draw=black, label=center:Utils] (Utils5) at (11.4,1.8) {};
    \node[mystyle,draw=black, label=center:NoDur] (NoDur5) at (11.6,0.3) {};
    \node[mystyle,draw=black, label=center:Shops] (Shops5) at (8.8,-2.8) {};
    \node[mystyle,draw=black, label=center:Durbl] (Durbl5) at (8.65,-1.5) {};
    \node[mystyle,draw=black, label=center:Hlth] (Hlth5) at (11,-2.8) {};
    \node[mystyle,draw=black, label=center:HiTec] (HiTec5) at (9.5,-2) {};
    \node[mystyle,draw=black, label=center:Telcm] (Telcm5) at (10.5,-1.5) {};
    
    \path [-, text = blue] (Manuf5) edge node {0.97} (Enrgy5);
    \path [-, text = blue] (Manuf5) edge node {1} (NoDur5);
    \path [-, text = blue] (Manuf5) edge node {0.97} (Utils5);
    \path [-, text = blue] (Manuf5) edge node {1} (Shops5);
    
    \path [-, text = blue] (NoDur5) edge node {0.88} (Utils5);
    \path [-, text = blue] (NoDur5) edge node {1} (Telcm5);
    \path [-, text = blue] (NoDur5) edge node {1 } (Hlth5);
    
    \path [-, text = blue] (Durbl5) edge node {0.98 } (Shops5);

    \path [-, text = blue] (HiTec5) edge node { 0.96} (Shops5);
    \path [-, text = blue] (HiTec5) edge node { 0.94} (Hlth5);
    
    \path [-, text = blue, bend left] (Shops5) edge node {0.96} (Enrgy5);
    \path [-, text = blue] (Utils5) edge node { 1} (Enrgy5);
    
\end{tikzpicture} }
\end{center}
\caption{\label{fig:real_graphs}{Posterior estimates of the graphs  and PPI for the selected edges. Threshold of inclusion is set to achieve an expected posterior specificity of at least 95\%.}}
\end{figure}
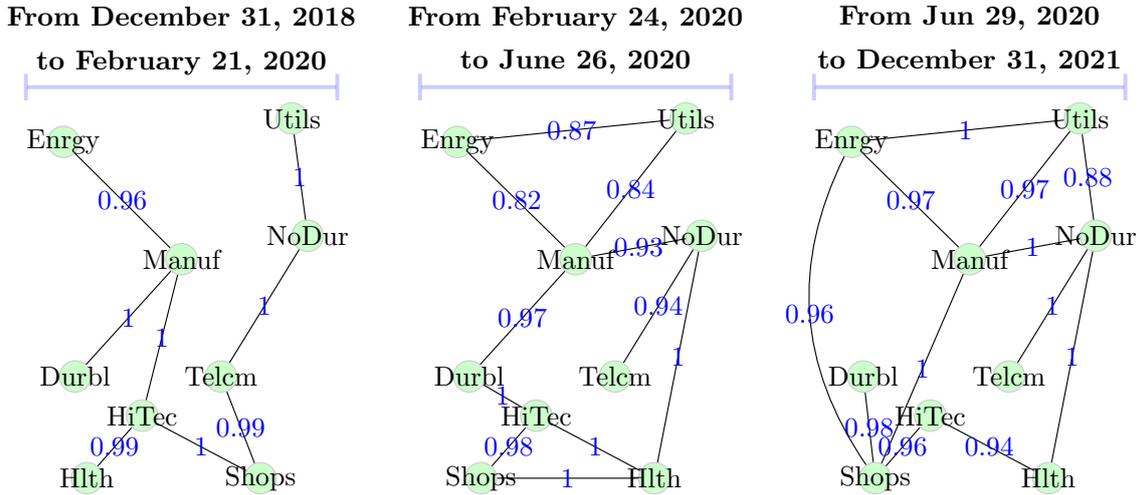

Conditionally on the MAP change point configuration, estimates of the three graphs are provided in Figure~\ref{fig:real_graphs}, while the estimated variance and covariance matrices are displayed in Figure C.2.1 of the Supplement. The graphs are obtained based on the marginal PPI of the edges in order to control the corresponding Bayesian false discovery rate \citep{newton2004detecting}. In particular, we set the threshold of inclusion based on the PPI to 0.8 in order to guarantee an expected rate of false detection not higher than 0.05, i.e., a specificity of at least 0.95 \citep[for more details, see,][]{leday2019fast,williams2021bayesian}. In Section~C.1 of the Supplement we report the values of degree centrality, betweenness centrality \citep{freeman1977set}, local clustering and global clustering coefficients \citep{watts1998collective} for the estimated graphs, which give insights into the role of each node. 

A clearly noticeable feature from Figure~\ref{fig:real_graphs} is the increase in the number of edges from the first change point (8 edges) to the followings (11 and 12 edges, respectively), reflected also in the global measures of clustering of the graphs which varies from 0 in the pre COVID-19 period, to 0.43 during the first COVID-19 outbreak, to 0.24 to the post COVID-19 outbreak (see Table C.1.3 in the Supplement). Such increase in the connectivity of the graph is coherent with the hypothesis of the COVID-19 outbreak acting as a common non-measurable risk factor driving the returns in the market. As already noticed in the preliminary analysis summarised by Figure~\ref{fig:moving_window}, the role of hub of the manufacturing portfolio (i.e., machinery, trucks, planes, chemicals, office furniture, paper) over the three years is confirmed. The corresponding degree centrality (i.e., number of vertices in the neighbourhood of the manufacturing portfolio) is the highest in all three graphs (see Table C.1.1 of the Supplement for more details). However, contrary to the conclusions of the initial exploratory analysis, consumer non-durable (i.e., food, tobacco, textiles, apparel, leather, toys) returns appear to play a less central role before the COVID-19 outbreak. Such evidence that was absent in the explanatory analysis is discovered mainly thanks to the automatic detection of the change points, that allow us to determine the most appropriate time window to estimate the graph and capture differences in structure. Moreover, we identify  another hub in the Shops portfolio (i.e., wholesale, retail, and some services, as laundries and repair shops) in the time-interval after the last change point. 

In terms of volatility, in all three periods  the consumer durables (i.e., cars, TVs, furniture, household appliances) and energy (i.e, oil, gas, and coal extraction and products) are characterised by the highest volatility. Moreover, the analysis confirms that the three periods (pre COVID-19 outbreak, during first COVID-19 global outbreak, and after) coincide with small, high, and medium volatility markets, as already evident from the time-series plot (see Figure C.1 in the Supplement). In more detail, we note that health (i.e., healthcare, medical equipment, and drugs) is the only industry in the market whose portfolio presents a similar  volatility before and after the outbreak, while all other portfolios' returns are set to higher levels of variability as consequence of a long-run effect 
of market uncertainty. Similar conclusions can be drawn also for the pair-wise correlations (that can be easily computed from the values in Figure C.2.1 and are reported in Section C.2 of the Supplement), i.e. correlations  are higher during COVID-19 outbreak.

\begin{table}[!t]
\newcolumntype{a}{>{\columncolor{Gray}}l}
\centering
\begin{tabular}{aa|c|c|c}
\rowcolor{LightCyan}&& \# of change & & \\ 
\rowcolor{LightCyan}$\lambda_1$ & $\lambda_2$ &  points & change points & Global clustering coef.\\ 
0.25 & 60 & 0 & () & (0.72)\\
0.35 & 60 & 2 & (61, 80) & (0, 1, 0.33)\\
0.50 & 60 & 2 & (61, 80) & (0, 0, 0)\\
0.25 & 55 & 2 & (61, 80) & (0.72, 0.84, 0.81)\\
0.25 & 20 & 3 & (61, 80, 98)&(0, 1, 0.92, 0.92)\\
0.35 & 20 & 4 & (61, 80, 98, 99)&(0, 1, 0.33, 0.33, 0)\\
0.50 & 20 & 4 & (61, 68, 77, 80)&(0, 1, 0.88, 0.87, 0)\\
0.25 & 10 & 6 & (57, 61, 80, 98, 99 ,116)&(0, 0, 1, 0.89, 0.90, 0.33, 0.33)\\
0.35 & 10 & 4 & (61, 80, 98, 116)&(0, 1, 0.33, 0.33, 0)\\
0.50 & 10 & 6 & (57, 61, 68, 69, 77, 80)&(0, 0, 1, 0.88, 0.88, 0)\\
\end{tabular}
\caption{\label{tab:GFGL} Results obtained using GFGL for different values of the hyperparameters.}
\end{table}

{Finally, we compare our results with those obtained by applying the group-fused graphical lasso (GFGL) \citep{gibberd2017regularized}. Output summaries for the GFGL are shown in Table~\ref{tab:GFGL} and additional figures can be found in Section C.3 of the Supplement. 
We estimate the GFGL model for different values of the hyperparameters $\lambda_1$ and $\lambda_2$. We recall that $\lambda_1$ and $\lambda_2$ control the shrinkage and the smoothness of the solution of the GFGL model, respectively . As already noticed by \cite{gibberd2017regularized} and in the simulation study in Section~\ref{sec:simmodel}, for GFGL inference results are highly sensitive to the hyperparameters in terms of both detected change points (see Table~\ref{tab:GFGL} and Figure C.3.1 in the Supplement) and recovered graph structure (see Figure C.3.2 in the Supplement). Here, we consider values for $\lambda_1$ in \{0.25, 0.35, 0.50\} and $\lambda_2$ in \{10, 20, 55, 60\}, which are in the range of those considered by \cite{gibberd2017regularized} in their work. The number of identified change points ranges widely from 0 to 6. However, we notice that in all estimated change point configurations, but the one with no estimated change points, GFGL always includes $t=61$ and at $t=80$ as  change points, which is consistent with the change points identified by our approach. Moreover, even though the graph structure estimated by the GFGL largely varies depending on the hyperparameters,  in all settings where change points are detected, the graph structure connectivity increases during the COVID-19 outbreak in February 2020 and diminishes after June 2020  (cf.~the global clustering coefficients reported in Table~\ref{tab:GFGL} and Figure C.3.2 in the Supplement). This result is again consistent with teh results obtained with our approach. Increased graph connectivity during the COVID-19 outbreak is also found applying \textsf{loggle}; see section C.3 of the Supplement.}

\section{Discussion and conclusion}
\label{sec:conclusion}
In this work, we study the impact of the COVID-19 pandemic on the US stock market, with a specific focus on changes in dependence structure across stocks related to different industries. To do so, we consider weekly returns recorded for three years starting in January 2019. We identify two structural changes. The first change is in correspondence with the last week of February 2020, a date that for most countries coincided with the beginning of the pandemic. That same week financial markets recorded the weekly biggest losses since the financial crisis of 2008. The second change point is detected after approximately four months, when there is a reduction in market uncertainty, but the dependence structure as well as the volatility are not back to pre COVID-19 levels.  Comparing the dependence structure across the three periods (i.e., before February 24th, between February 24th and  June 26th, after June 26th) we provide many insights on the impact of  the pandemic on the stock market and highlight whose effects {appear} to be persistent up to the end of 2021, the last year considered.
 
The main methodological contribution of this work is the development of a dynamic GGM, which allows for abrupt changes in the dependence structure of the random variables represented by the nodes of the graph. Our model builds on existing literature on GGMs, as well as random change points. Our model construction allows us to control sparsity in the number of change points and/or in the graph structure. We have designed a tailored SMC algorithm, arguing for its use in such a complex setup over other M-H based alternatives and demonstrating its performance on simulated data  and on  our motivating application.

Our work opens up several avenues for future research. 

\begin{itemize}
\item[(a)] Scalability. We have not made use of the full SMC machinery. We briefly discuss two directions for increase in model dimension, along the number of nodes, $p$, and along the length of time instances for observations $T$. In terms of the size of the graph, recent works have developed effective proposals, informed by the observations, for MCMC methods on graph posteriors, instead of previously used 
random-walk-type blind moves. See, e.g., \cite{van:22}, and the references therein, for approaches based on Langevin-type analogues for discrete spaces, with parallelisation employed within the specification of the proposal. Such approaches have been seen to be effective for node sizes of $p=\mathcal{O}(10^2)$. In terms of the length $T$, recent advances on modelling involving change points and accompanying SMC methodologies, can permit for recasting models so that change points also become part of the hidden Markov process \citep[see, e.g.][]{yild:13}. At the same time, SMC methods based on state-of-art particle Gibbs approaches that incorporate backward steps to improve mixing over the update of the unobserved Markovian states, are shown to provide pseudomarginal methodologies of superior mixing compared to standard PMCMC \citep{lind:14}. Such new algorithms are supported by strong theoretical results. Indicatively, the number of particles can now be allowed to remain constant, $N=\mathcal{O}(1)$, as a function of $T$, when PMCMC requires $N=\mathcal{O}(T)$. Thus, costs for the overall SMC algorithm can be brought down to $\mathcal{O}(NT)$, from the previous $\mathcal{O}(T^2)$, for big $T$ -- with $\mathcal{O}(NT)$ not taking under consideration the option of parallelisation across particles. 
\item[(b)] Smooth changes. In this work, we have considered abrupt  changes in edge inclusion probabilities. Alternatively, we could model edge inclusion probabilities as a function of time, for example, using autoregressive-type models. In this setup, shrinkage priors could be specified to link the probability of edge inclusion at time $t$ to the same probability at time $t-1$ \cite[see, for instance,][]{molinari2022bayesian2}. This approach is amenable to many generalizations, such as the inclusion of covariates. Moreover, the probability of edge inclusion at time $t$ could be a function of the probabilities of edge inclusions at time $t-1$ of a neighbourhood of each node. 
\item[(c)] Graph sub-structures. Here, we have presented changes between graphs as captured by edge flips before and after a change point. Edge detection is very sensitive to the number of nodes as well as sample size. It has been argued \citep[][and references therein]{boom2022bayesian} that in many applications a more robust approach is to shift the focus of inference to graph sub-structures such as hubs and communities,  with the goal of capturing the evolution over time of such macro-structures which better describe the underlying phenomenon. 
\item[(d)] More general response types. The model can be easily extended to accommodate different type of responses, such as binary and count data. An easy solution would be the representation of such data in terms of latent {variables} \citep{albert1993bayesian, chib1998analysis}. Moreover, it is straightforward to include time-homogeneous and time-varying covariates to model the mean of the time series, as well as  a trend and seasonal component.  
\end{itemize}

\textbf{Acknowledgements.} This work was supported by the Singapore Ministry of Education Academic Research Fund Tier 2 under Grant MOE2019-T2-2-100. The authors are grateful to the Editor, an Associate Editor and five anonymous referees for insightful comments and suggestions, which led to a substantial improvement of the manuscript. B. Franzolini is also affiliated to the Bocconi Institute for Data Science and Analytics (BIDSA). 

\bibliographystyle{chicago}
\bibliography{references.bib} 

\newpage
\begin{center}
\textbf{\Large{Supplement to \emph{``Change point detection in  
dynamic Gaussian graphical models: the impact of COVID-19 pandemic on US stock market"}}}\\
\vspace{0.8cm}
Beatrice Franzolini, Alexandros Beskos, Maria De Iorio, Warrick Poklewski
Koziell, and Karolina Grzeszkiewicz
\end{center}

\vspace{2cm}

\section*{ A. \enskip Evaluation of SMC approximations under scenario A}
\renewcommand{\thefigure}{A.1}
\begin{figure}[H]
\centering
\includegraphics[width=\textwidth]{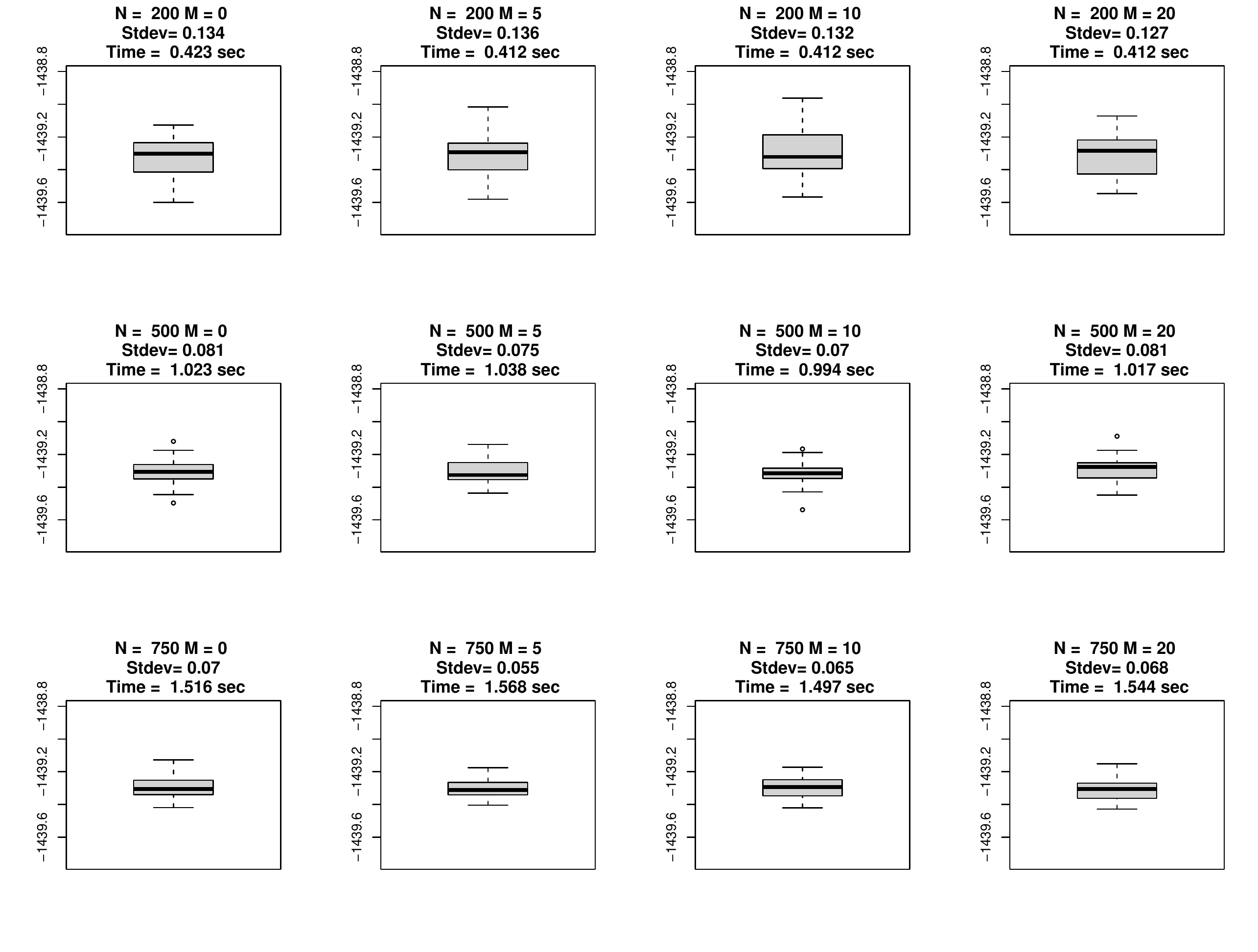}
\caption{Log-likelihood estimates for Scenario~A obtained with the particle filter by fixing the change points to the truth. Distinct box-plots correspond to different numbers of particles $N$ and/or mutations steps~$M$. 
For each pair $(N,M)$, we run the algorithm 30 times and obtain the log-likelihood estimates. 
Each box-plot shows the distribution of such estimates.  The variability of the estimates is small for all considered pairs $(N,M)$.}
\end{figure}

\newpage
\section*{B. \enskip Simulation studies: additional figures, tables, and results}
\subsection*{B.1 True data generating processes for simulation studies}
\renewcommand{\thefigure}{B.1.1}
\begin{figure}[H]
\centering
    \begin{subfigure}[b]{0.49\textwidth}
    \includegraphics[trim={5cm 0 0 0},clip,width=\textwidth]{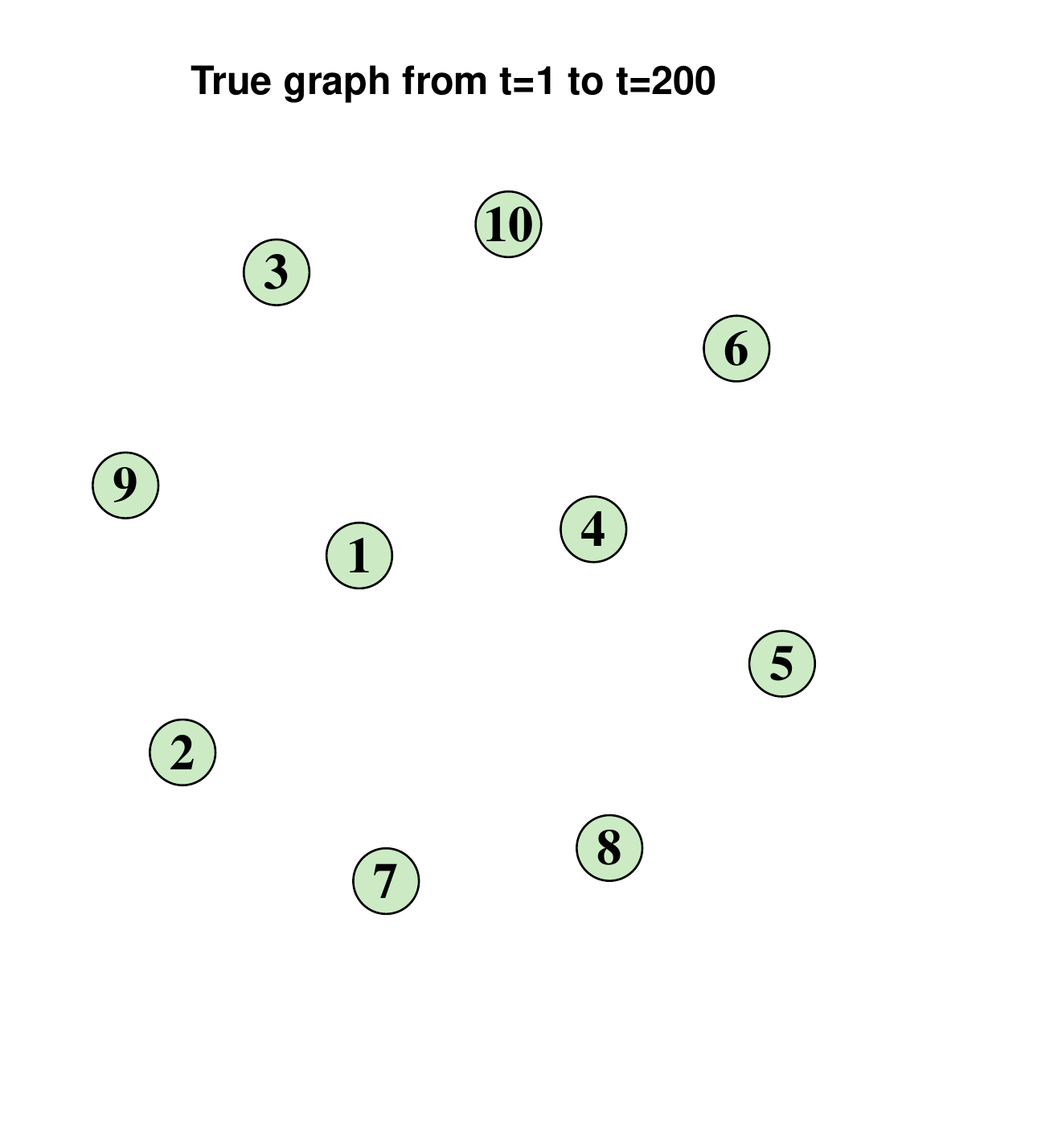}
    \caption{No change points and independence}
    \end{subfigure}
    \hfill
    \begin{subfigure}[b]{0.49\textwidth}
    \includegraphics[trim={5cm 0 0 0},clip,width=\textwidth]{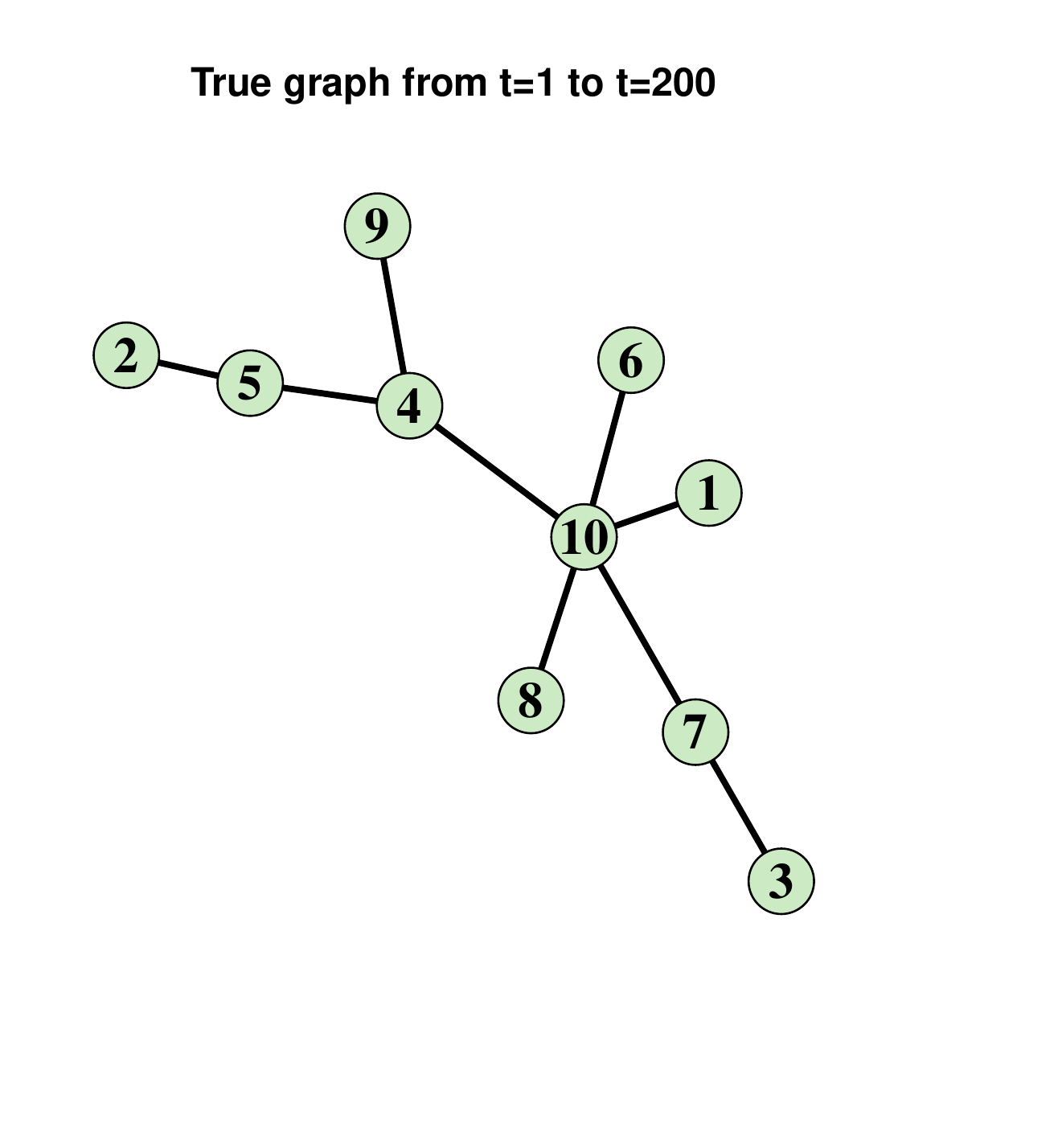}
    \caption{No change points and dependence}
    \end{subfigure}
    \caption{Graph structure of Scenarios 1 (panel a), 2 (panel b) and 5 (panel b) used as data generating mechanism for assessing model's performance.}
\end{figure}

\vspace{1 cm}

\renewcommand{\thefigure}{B.1.2}
\begin{figure}[H]
\centering
\includegraphics[trim={3.5cm 3cm 0 0},clip,width=\textwidth]{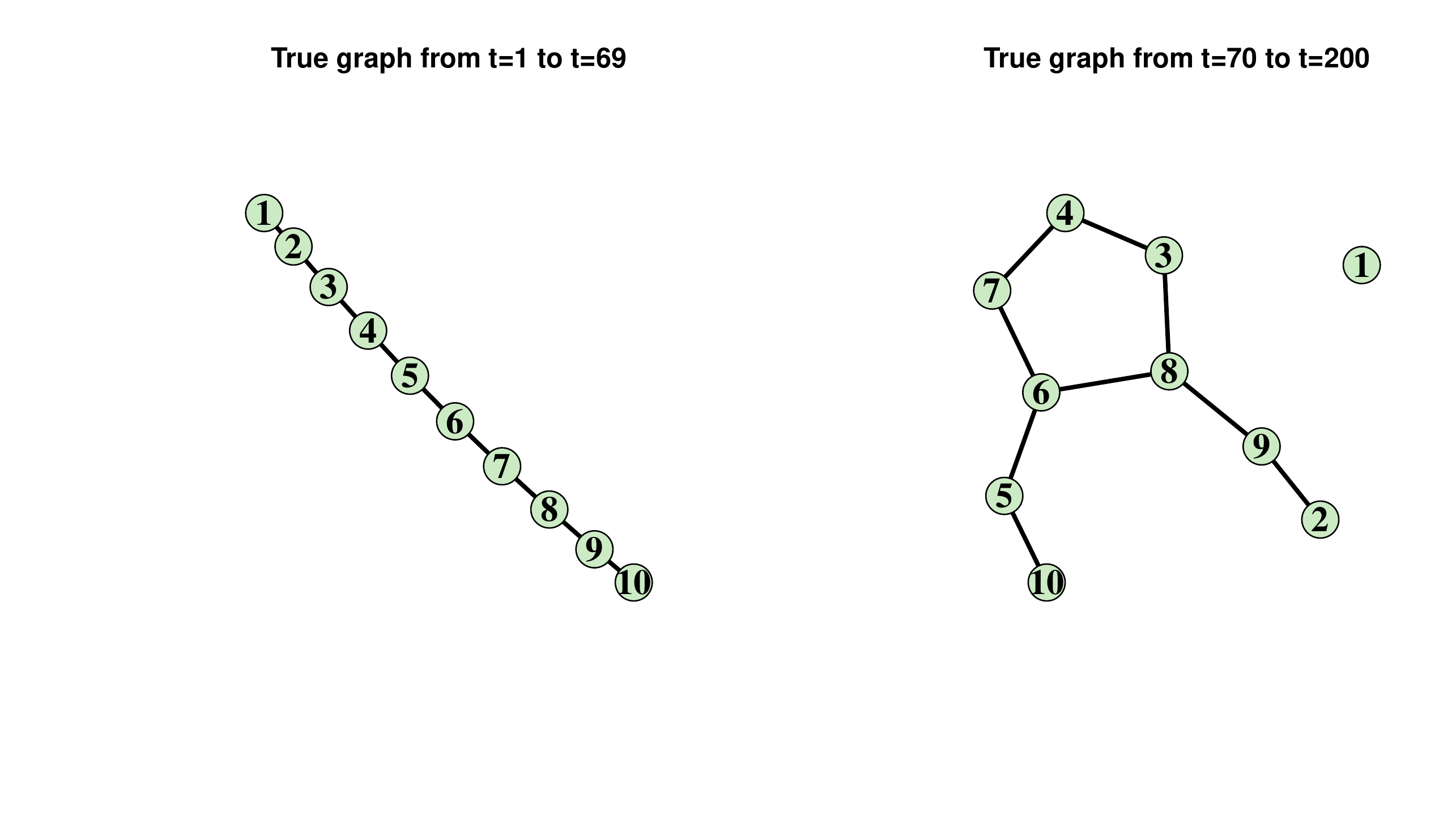}
\caption{Scenario 3: graph structure used as data generating mechanism for assessing model's performance.}
\end{figure}

\renewcommand{\thefigure}{B.1.3}
\begin{figure}[H]
\centering
\includegraphics[trim={0 2cm 0 0},clip,width=\textwidth]{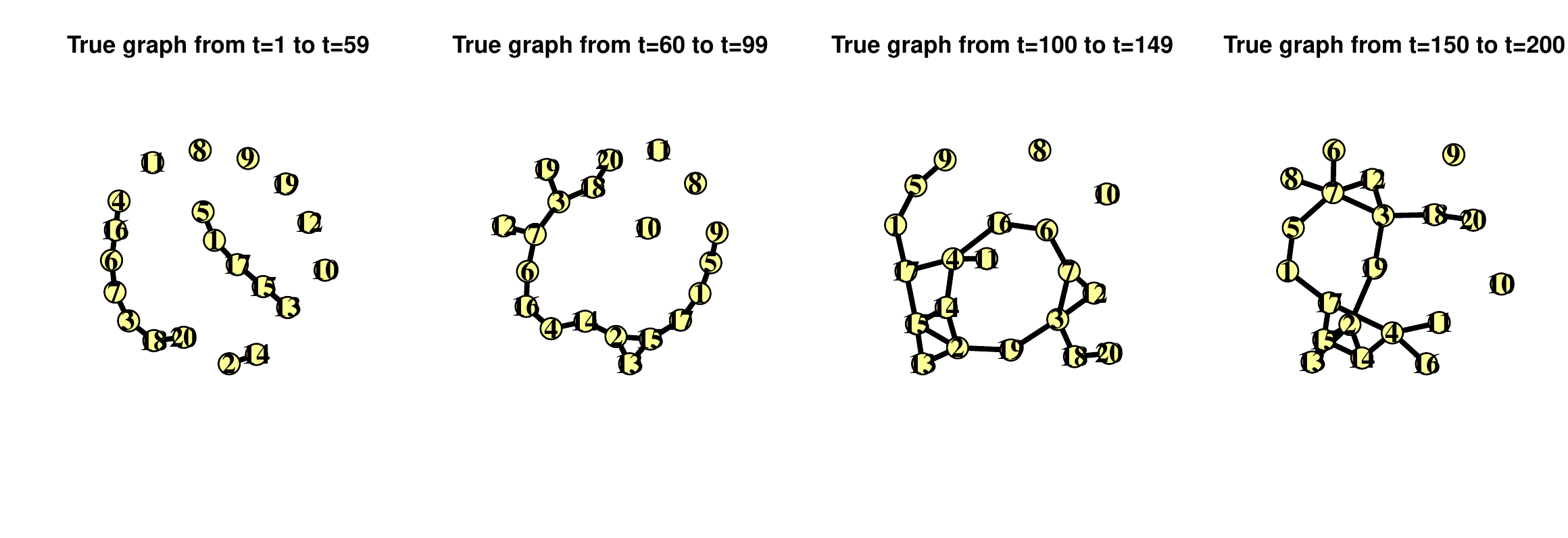}
\caption{Scenario 4: graph structure used as data generating mechanism for assessing model's performance.}
\end{figure}

\hfill
\vspace{1 cm}

\renewcommand{\thefigure}{B.1.4}
\begin{figure}[H]
\centering
\begin{subfigure}[b]{0.48\textwidth}
\includegraphics[width=\textwidth]{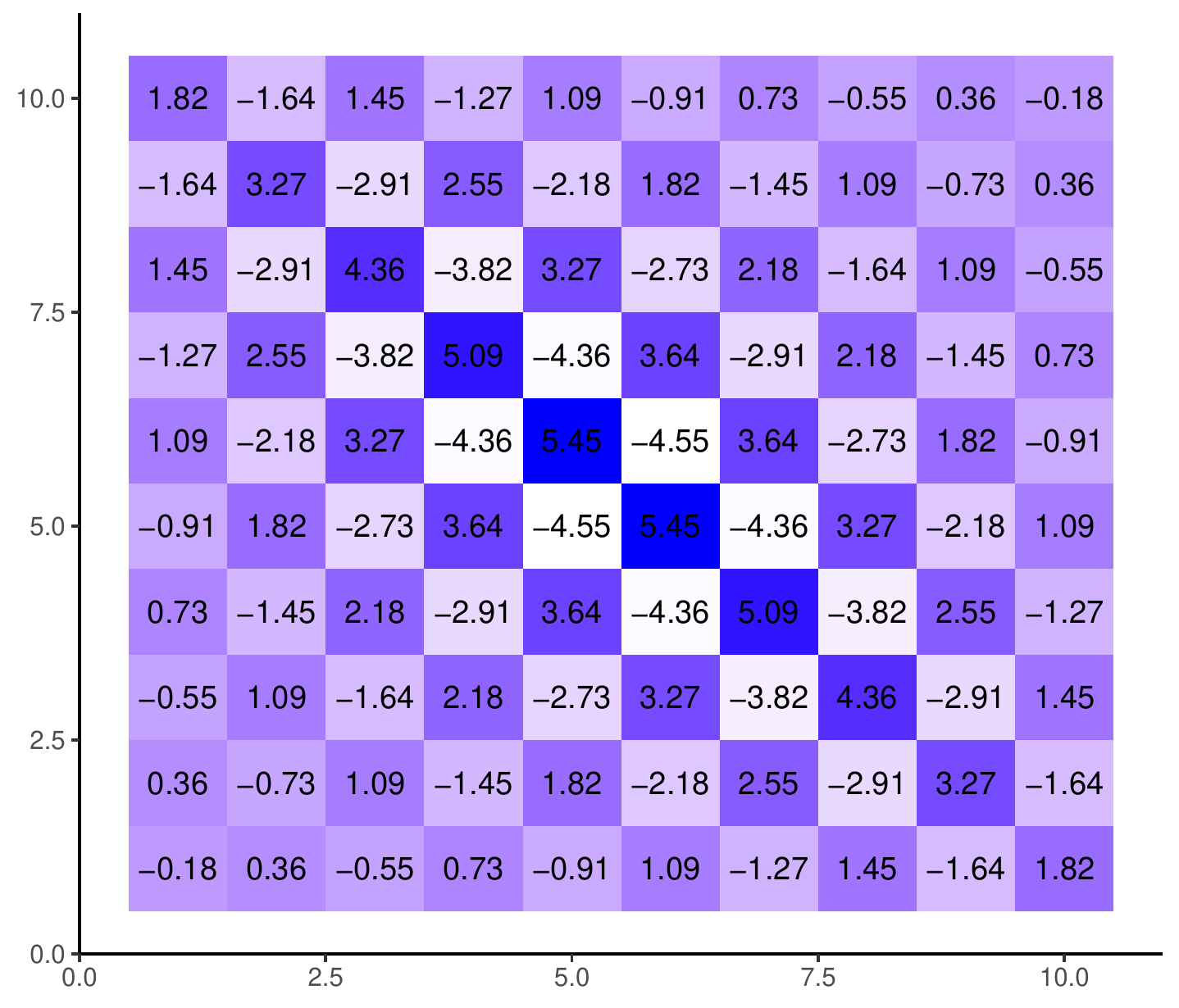}
\end{subfigure}
\begin{subfigure}[b]{0.48\textwidth}
\includegraphics[width=\textwidth]{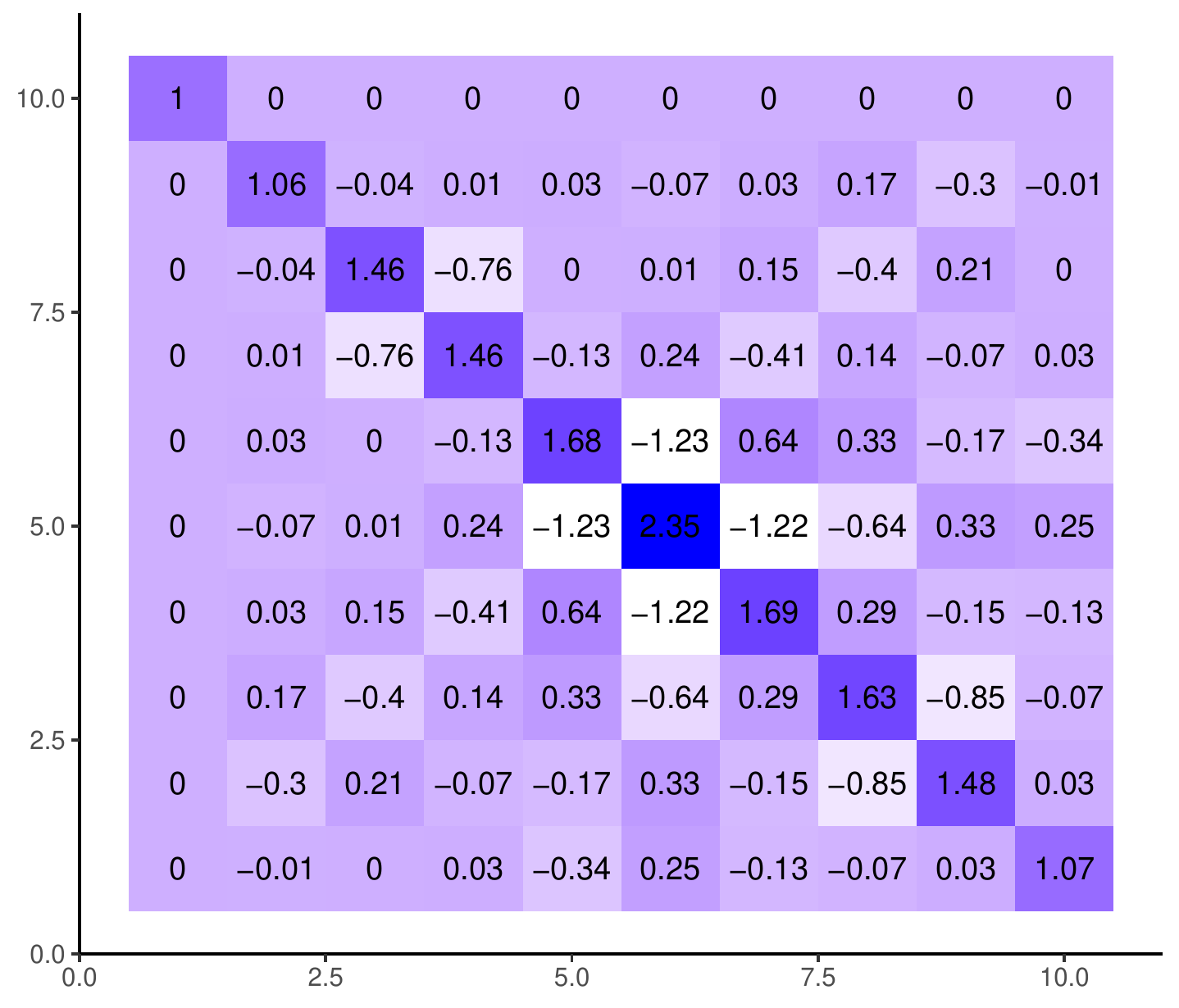}
\end{subfigure}
\caption{Scenario 3: Covariance matrices.}
\end{figure}

\hfill

\vspace{1 cm}

\renewcommand{\thefigure}{B.1.5}
\begin{figure}[H]
\centering
\includegraphics[width=0.85\textwidth]{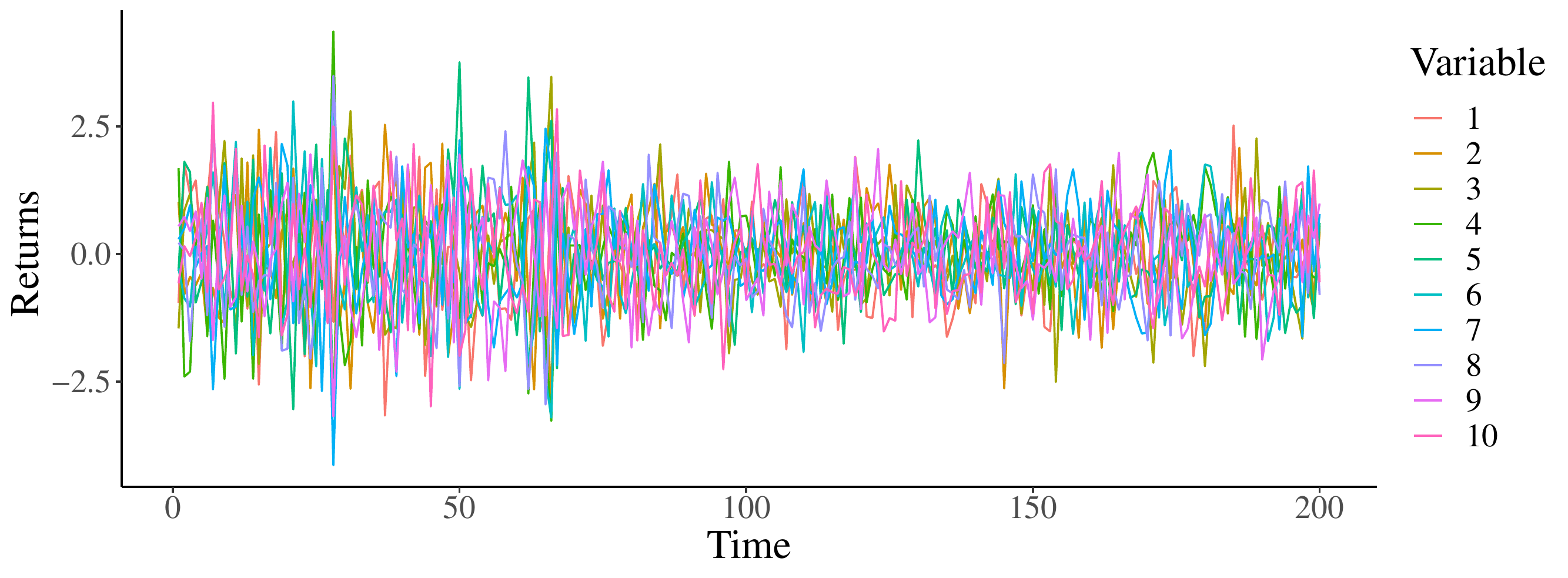}
\caption{Scenario 3: Simulated data - replicate n.1.}
\end{figure}

\clearpage

\subsection*{B.2 Scenarios 1 and 2: results}

The first two scenarios do not include change points among the $T=200$ instances. In Scenario 1 all variables are independent, while  in Scenario 2 we assume a non-trivial conditional independence structure represented by  a graph with nine edges, shown in Figure B.1.1 (b). We consider 10 simulation replicates for Scenario 1 and 2. 
The hyperparameters for Scenario 1 and 2 are chosen as described in Section 4.1 of the paper.  The inference
results for no change points show the expected good performance of the model in terms of both identification of  change points and recovery of the dependence structure. In both scenarios the posterior on the change points  concentrates on the true state of no change points, with posterior MAP probability not falling below 0.98 in all simulation replicates. 
Figure~\ref{fig:FPR12} shows the false positive rate (FPR) of edge detection as function of the threshold used for the marginal posterior probability of edge inclusion (PPI, i.e., the posterior probability associated to each edge in the graph, $\mathrm{P}(A[h,k] = 1\mid  \text{data})$). FPRs are very low for almost any PPI threshold (indicatively, for a threshold of 0.5, FPR is 0 and 0.006 for Scenarios 1 and 2, respectively). Figure~\ref{fig:ROC2} displayes  the receiver operating characteristic (ROC) curve  for  Scenario 2, with the area under the curve (AUC)  approximately~1. 

\renewcommand\thetable{B.2.1}
\begin{table}[H]
\newcolumntype{b}{>{\columncolor{Gray}}c}
\begin{center}
\begin{tabular}{b|c|c}
\cellcolor{white}& \cellcolor{LightCyan}True change &\\
\cellcolor{white}& \cellcolor{LightCyan}point configuration&\\
\hline
     Scenario 1 &  $\emptyset$ & \begin{tabular}{b|c|c}
      & \cellcolor{LightCyan}MAP estimate &\cellcolor{LightCyan} MAP prob.\\
          Replica 1 & $\emptyset$ &  1 \\
          Replica 2 & $\emptyset$ &  1 \\
          Replica 3 & $\emptyset$ &  1 \\
          Replica 4 & $\emptyset$ &  1 \\
          Replica 5 & $\emptyset$ &  1 \\
          Replica 6 & $\emptyset$ &  1 \\
          Replica 7 & $\emptyset$ &  1 \\
          Replica 8 & $\emptyset$ &  1 \\
          Replica 9 & $\emptyset$ &  1 \\
          Replica 10 & $\emptyset$ &  1 \\
     \end{tabular}\\
     \hline
     Scenario 2 & $\emptyset$ & \begin{tabular}{b|c|c}
      & \cellcolor{LightCyan}MAP estimate &\cellcolor{LightCyan} MAP prob.\\
          Replica 1 & $\emptyset$ &  0.980 \\
          Replica 2 & $\emptyset$ &  1   \\
          Replica 3 & $\emptyset$ &  0.993   \\
          Replica 4 &  $\emptyset$ &  1   \\
          Replica 5 & $\emptyset$ & 1 \\
          Replica 6 & $\emptyset$ & 1   \\
          Replica 7 &  $\emptyset$ & 0.997  \\
          Replica 8 &  $\emptyset$ & 1  \\
          Replica 9 &  $\emptyset$ & 1   \\
          Replica 10 &  $\emptyset$ & 1   \\
     \end{tabular}\\
\hline
\end{tabular}
\caption{Scenarios 1 and 2: Posterior summaries for change points. MAP estimates and MAP probabilities (for the posterior over all configurations of change points). Results are obtained simulating 10 replicates.}
\end{center}
\end{table}
\vspace{-1cm}
\renewcommand{\thefigure}{B.2.1}
\begin{figure}[H]
\centering
\begin{subfigure}[b]{0.49\textwidth}
\includegraphics[width=\textwidth]{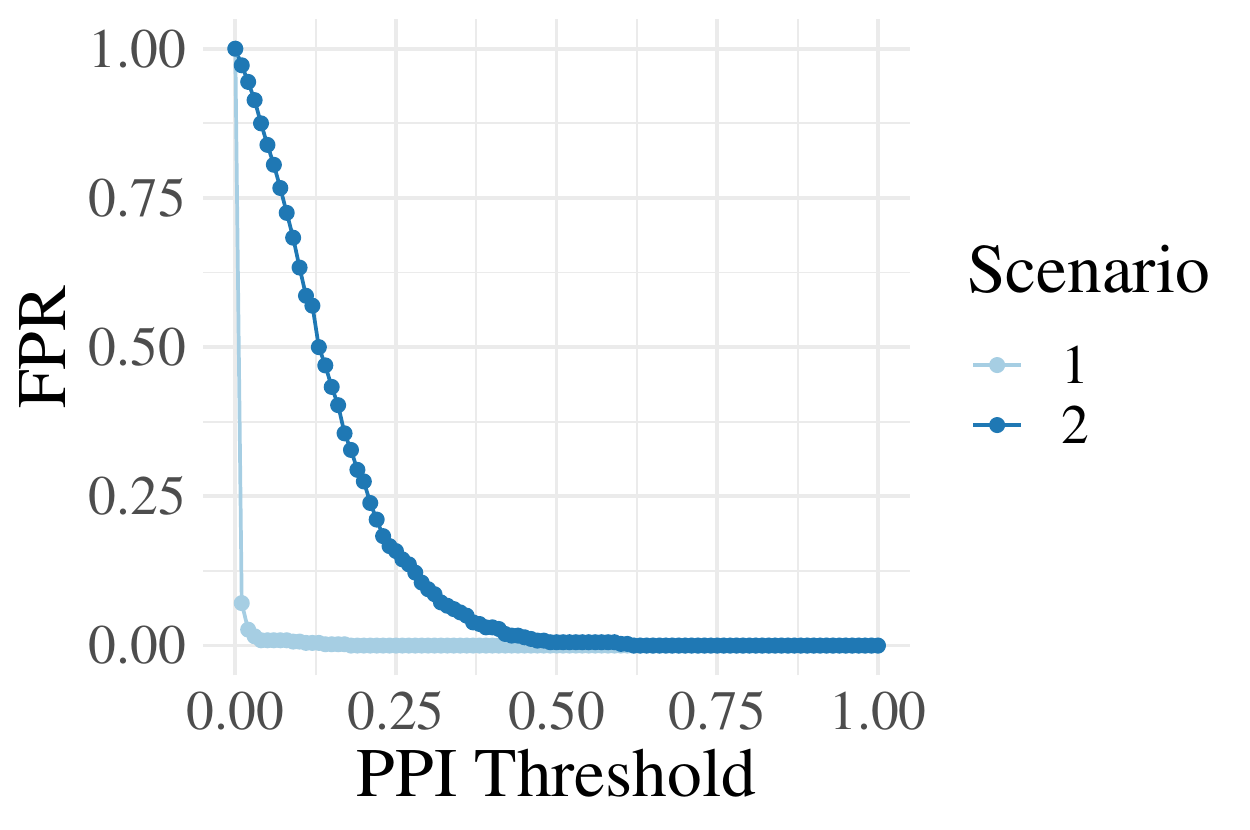}
\caption{\label{fig:FPR12}False positive rate}
\end{subfigure}
\hfill
    \begin{subfigure}[b]{0.45\textwidth}
    \includegraphics[width=\textwidth]{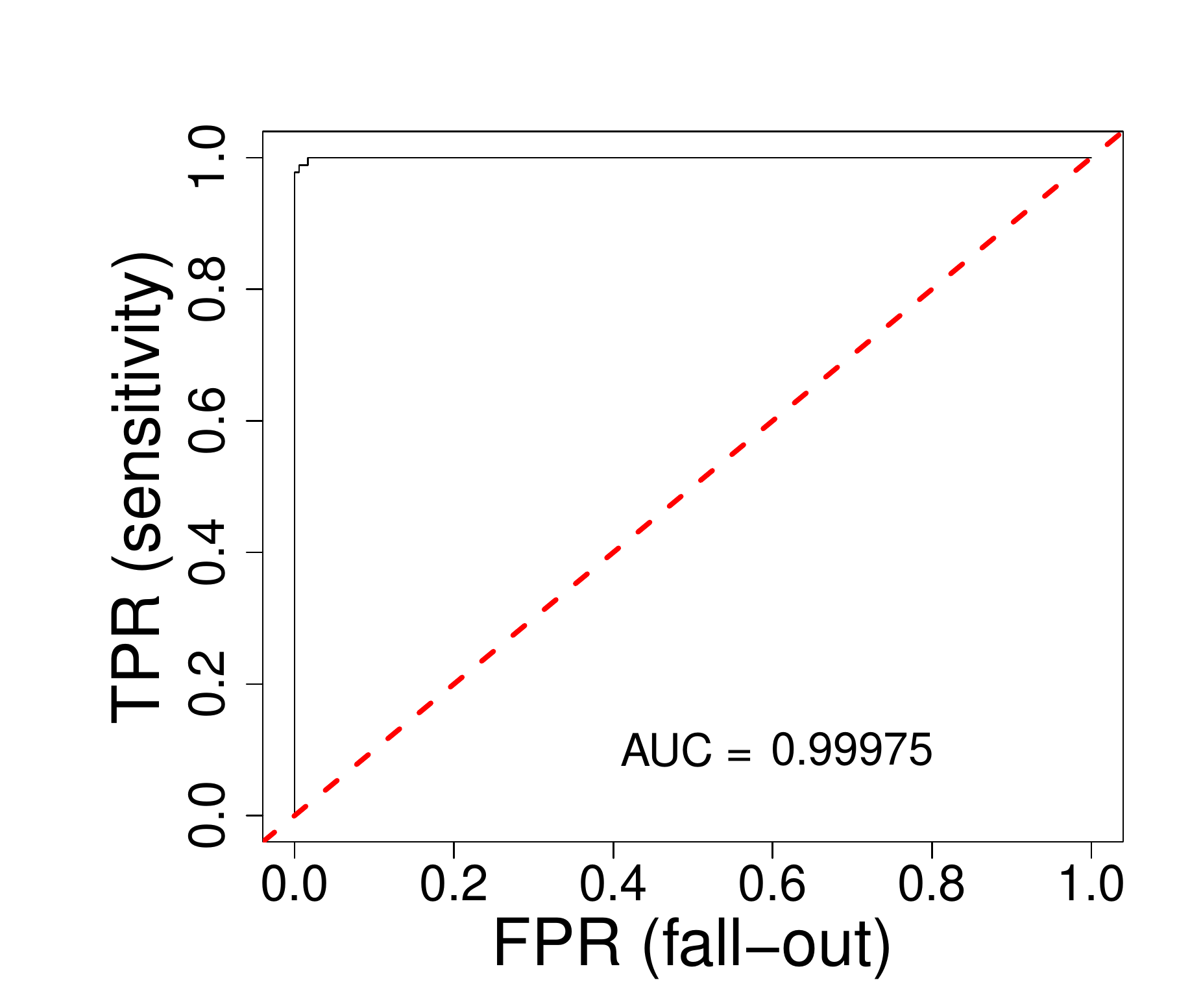}
    \caption{\label{fig:ROC2}Receiver operating characteristic curve}
    \end{subfigure}
    \caption{Scenarios 1 and 2: Panel (a): FPR versus PPI threshold, for simulation Scenarios 1 and 2, computed across 10 replicates. Panel (b): ROC curve in Scenario 2, computed across 10 replicates. }
\end{figure}

\subsection*{B.3 Scenario 3: additional results}
\renewcommand\thetable{B.3.1}
\begin{table}[!h]
\newcolumntype{b}{>{\columncolor{Gray}}c}
\begin{center}
\resizebox{\textwidth}{!}{  
\begin{tabular}{b| c|b|c|c|c}
\cellcolor{white}& \cellcolor{LightCyan}&\cellcolor{white}&\cellcolor{LightCyan}GFGL estimate&\cellcolor{LightCyan}GFGL estimate& \cellcolor{LightCyan}Bayesian\\
\cellcolor{white}&\cellcolor{LightCyan}True change &\cellcolor{white}& \cellcolor{LightCyan} $\lambda_1 =0.25$ &\cellcolor{LightCyan} $\lambda_1 =0.50$&\cellcolor{LightCyan}dynamic \\
\cellcolor{white}&\cellcolor{LightCyan}point position&\cellcolor{white}& \cellcolor{LightCyan} $\lambda_2=10$ &\cellcolor{LightCyan} $\lambda_2=60$&\cellcolor{LightCyan}GGM MAP\\
\hline
          &&Replica 1 & $(69,70)$ & $(69)$&$(70)$ \\
          &&Replica 2 & $(60,69)$ & $(60,69)$&$(69)$ \\
          &&Replica 3 & $(47,69,70,74)$ & $(69)$&$(70)$ \\
          &&Replica 4 & $(41,60,68,70)$ & $(60)$& $(70)$\\
          && Replica 5 & $(67,70)$ & $(70)$& $(70)$\\
          &&Replica 6 & $(26,53,70)$ & $(52)$& $(69)$ \\
          &&Replica 7 & $(45,67,71)$ & $(45,67)$& $(69)$ \\
          &&Replica 8 & $(44,56,59,60,61,63,65)$ & $(44,56)$& $(71)$ \\
          &&Replica 9 & $(62,69,70)$ & $(69)$& $(71)$\\
          Scenario 3 &  $(70)$ & Replica 10 & $(67,70)$  & $(70)$& $(70)$\\
          &&Replica 11 & $(25,31,42,67,68,69)$ & $(42)$& $(68)$ \\
          &&Replica 12 & $(69,70)$ & $(69)$& $(74)$\\
          &&Replica 13 & $(21,68,70)$ & $(68)$& $(69)$\\
          &&Replica 14 & $(65,66,70,87)$ & $(66)$& $(71)$\\
          && Replica 15 & $(34,54,61,66,67,68,85)$ & $\emptyset$& $(70)$ \\
          &&Replica 16 & $(30,36,66,68,69,70)$ & $(65)$& $(70)$\\
          &&Replica 17 & $(37,52,57,69,70)$ & $(49)$& $(71)$\\
          &&Replica 18 & $(53,61,70,78)$ & $(70)$&  $(71)$\\
          &&Replica 19 & $(38,61,62,69,70)$ & $(29,61)$& $(70)$\\
          &&Replica 20 & $(67,68,70,74)$ & $(67,68)$&  $(73)$\\
     \hline
\end{tabular}
}
\caption{\label{tab:scenario3comp} Scenario 3: Comparison on the estimated change points between the results obtained with the GFGL model and our model (last column).}

\end{center}
\end{table}

\renewcommand\thetable{B.3.2}
\begin{table}[!h]
\begin{center}
\begin{tabular}{c|c|c}
&loggle&Bayesian dynamic GGM with threshold 0.5\\
\hline
FPR&0.242&0.241\\
TPR&0.825&0.825\\
\end{tabular}
\end{center}
\caption{ Scenario 3: Comparison on graph recovery between the results obtained with the ``oracle version" of the \textsf{loggle} model and our model.}
\end{table}

\renewcommand\thetable{B.3.3}
\begin{table}[tb]
\newcolumntype{b}{>{\columncolor{Gray}}c}
\begin{center}
\resizebox{\textwidth}{!}{ 
\begin{tabular}{b| c|b|c|c|c|c|c|c|c}
\cellcolor{white}& \cellcolor{LightCyan}True  &\cellcolor{white}&\cellcolor{LightCyan}MAP&\cellcolor{LightCyan}MAP&\cellcolor{LightCyan}prob&\cellcolor{LightCyan}90\%&\cellcolor{LightCyan}95\%&\cellcolor{LightCyan}&\cellcolor{LightCyan}\\
\cellcolor{white}&\cellcolor{LightCyan}change point &\cellcolor{white}& \cellcolor{LightCyan} est. &\cellcolor{LightCyan}  prob.& \cellcolor{LightCyan}$\kappa = 1$&\cellcolor{LightCyan} C.I.&\cellcolor{LightCyan} C.I.&\cellcolor{LightCyan} Mean& \cellcolor{LightCyan} Median\\
\hline
          &&Rep. 1 & $(70)$ & 0.250 & 0.960 & $[68,78]$ & $[67,80]$ & 71.74 & 70 \\
          &&Rep. 2 & $(69)$ & 0.299 & 0.987 & $[65,76]$ & $[65,79]$ & 70.69 & 70\\
          &&Rep. 3 & $(70)$ & 0.289 & 0.975 & $[69,79]$ & $[68,79]$ & 71.17 & 70 \\
          &&Rep. 4 & $(70)$ & 0.230 & 0.984 & $[68,78]$ & $[60,78]$ & 71.70 & 71\\
          &&Rep. 5 & $(70)$ & 0.294 & 0.987 & $[70,80]$ & $[70,80]$ & 73.34 & 72\\
          &&Rep. 6 & $(69)$ & 0.301 & 0.985 & $[65,76]$ & $[65,79]$ & 70.56 & 70\\
          &&Rep. 7 & $(69)$ & 0.207 & 0.973 & $[68,74]$ & $[68,91]$ & 70.27 & 69\\
          &&Rep. 8 & $(71)$ & 0.123 & 0.911 & $[63,84]$ & $[60,84]$ & 70.82 & 70\\
          &&Rep. 9 & $(71)$ & 0.122 & 0.930 & $[69,83]$ & $[69,83]$ & 75.75 & 75\\
Scenario 3&(70)&Rep. 10 & $(70)$ & 0.306& 0.971 & $[70,74]$&$[70,79]$&72.21 & 71\\
          &&Rep. 11 & $(68)$ & 0.245 & 0.966 & $[68,76]$ & $[68,81]$ & 70.60 & 70\\
          &&Rep. 12 & $(74)$ & 0.191 & 0.933 & $[67,83]$ & $[67,83]$ & 73.53 & 73\\
          &&Rep. 13 & $(69)$ & 0.180 & 0.966 & $[68,74]$ & $[67,77]$ & 70.99 & 71\\
          &&Rep. 14 & $(71)$ & 0.205 & 0.988 & $[65,81]$ & $[65,81]$ & 71.63 & 71\\
          &&Rep. 15 & $(70)$ & 0.140 & 0.998 & $[66,87]$ & $[65,90]$ & 73.59 & 70\\
          &&Rep. 16 & $(70)$ & 0.140 & 0.985 & $[66,78]$ & $[66,78]$ & 71.86 & 71\\
          &&Rep. 17 & $(71)$ & 0.143 & 0.991 & $[53,75]$ & $[53,75]$ & 66.82 & 69\\
          &&Rep. 18 & $(71)$ & 0.245 & 0.954 & $[66,80]$ & $[66,80]$ & 71.78 & 71\\
          &&Rep. 19 & $(70)$ & 0.188 & 0.950 & $[69,80]$ & $[68,80]$ & 71.67 & 71\\
          &&Rep. 20 & $(73)$ & 0.179 & 0.965 & $[67,75]$ & $[66,77]$ & 71.19 & 71\\
     \hline
\end{tabular}
}
\caption{\label{tab:scenario3full} Scenario 3: Posterior summaries for change point recovery. MAP estimates, MAP probabilities (for the posterior over all configurations of change points), posterior probability of the number of change points being 1, and credible intervals, mean and median of the position of the change point (conditionally on having one change point). Credible intervals are obtained computing the smallest credible sets with 90\% and 95\% credibility, which are not necessary continuous intervals, and then using the minimum and the maximum time points in the credible set as boundaries of the  interval.}
\end{center}
\end{table}

\hfill 
\clearpage
\subsection*{B.4 Scenarios 4: additional results}
\renewcommand{\thefigure}{B.4.1}
\begin{figure}[H]
\centering
\includegraphics[width=0.65\textwidth]{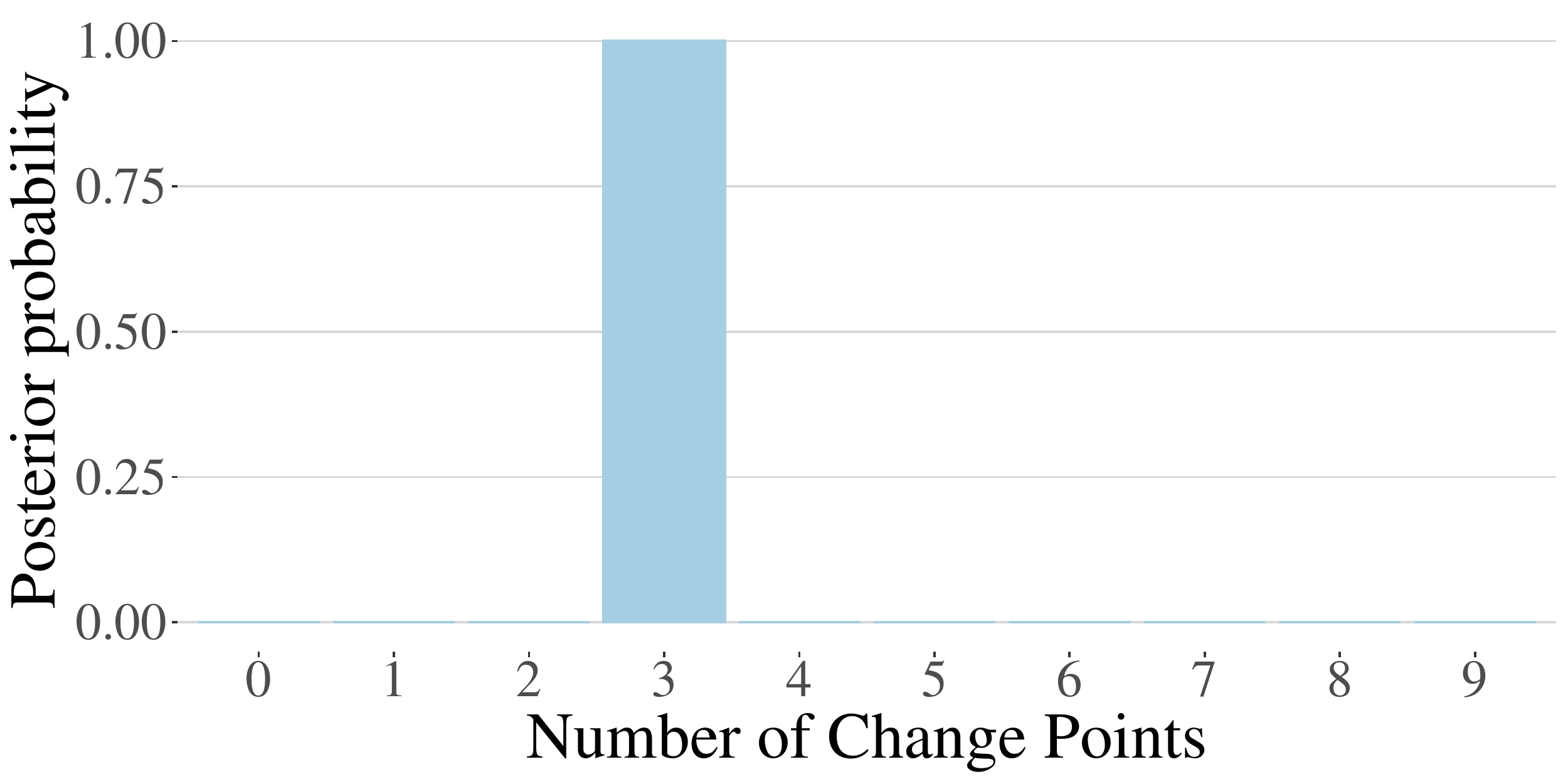}
\caption{Scenario 4: Posterior distribution on the number of change points}
\end{figure}

\renewcommand\thetable{B.4.1}
\begin{table}[!h]
\newcolumntype{a}{>{\columncolor{Gray}}l}
    \centering
    \begin{tabular}{ac}
\rowcolor{LightCyan}change & post\\
\rowcolor{LightCyan}points & prob\\
\hline
(60, 97, 148)&0.0012\\
(60, 98, 148)&0.0262\\
(60, 98, 150)&0.2909\\
(61, 98, 150)&0.1785\\
(61, 97, 150)&0.1935\\
(61, 99, 150)&0.3096\\
    \end{tabular}
    \caption{Scenario 4: Posterior distribution of the change point configuration}
\end{table}

\renewcommand{\thefigure}{B.4.2}
\begin{figure}[H]
\centering
\includegraphics[width=0.80\textwidth]{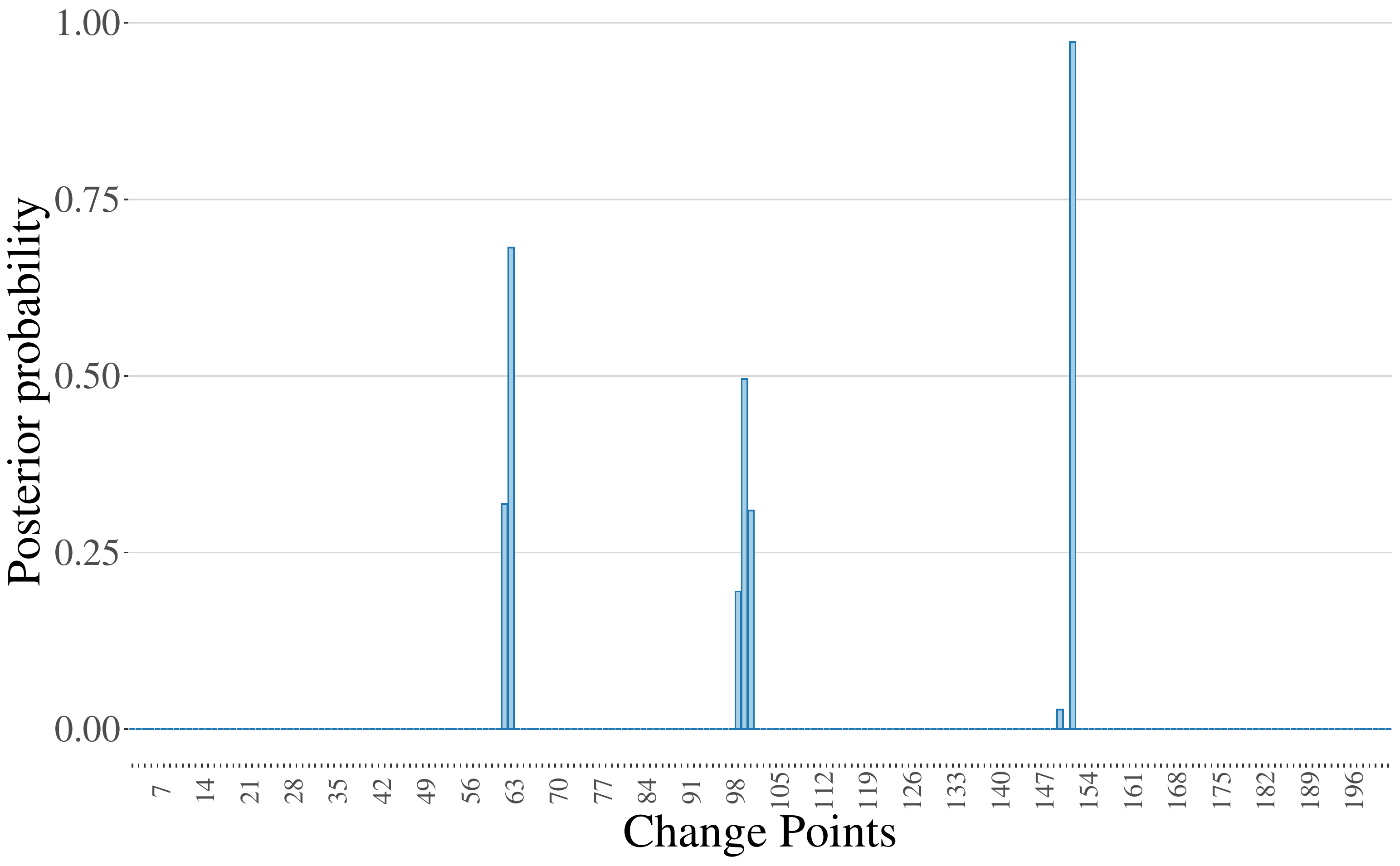}
\caption{Scenario 4: Marginal posterior probability of every time point to be a change point.}
\end{figure}

\renewcommand{\thefigure}{B.4.3}
\begin{figure}[H]
\centering
\begin{subfigure}[b]{0.3\textwidth}
\includegraphics[width=\textwidth]{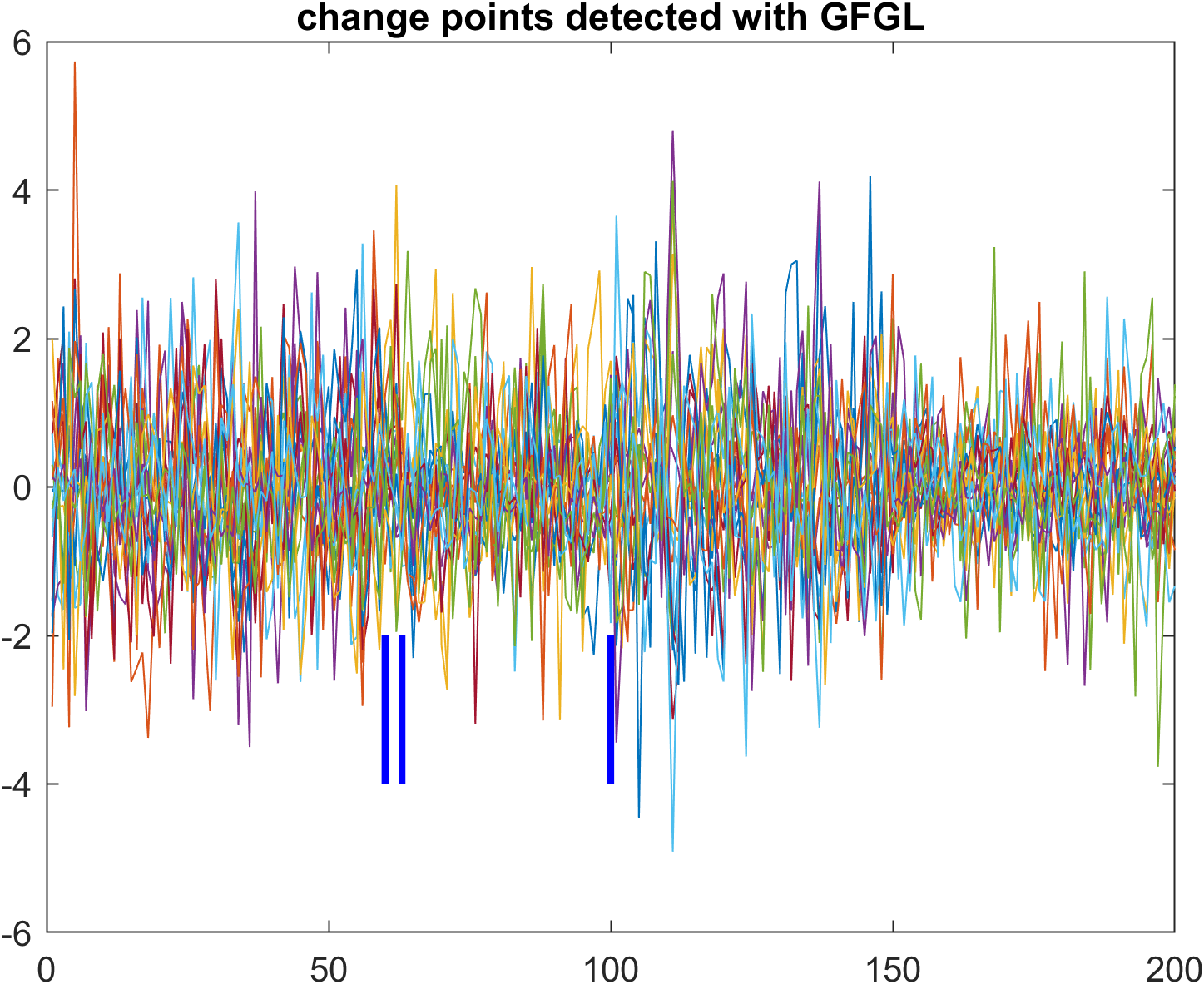}
\caption{\footnotesize The change points identified with $\lambda_1=0.10$ and $\lambda_2=60$ are $\{60,63,100\}$.}
\end{subfigure}
\hfill
\begin{subfigure}[b]{0.3\textwidth}
\includegraphics[width=\textwidth]{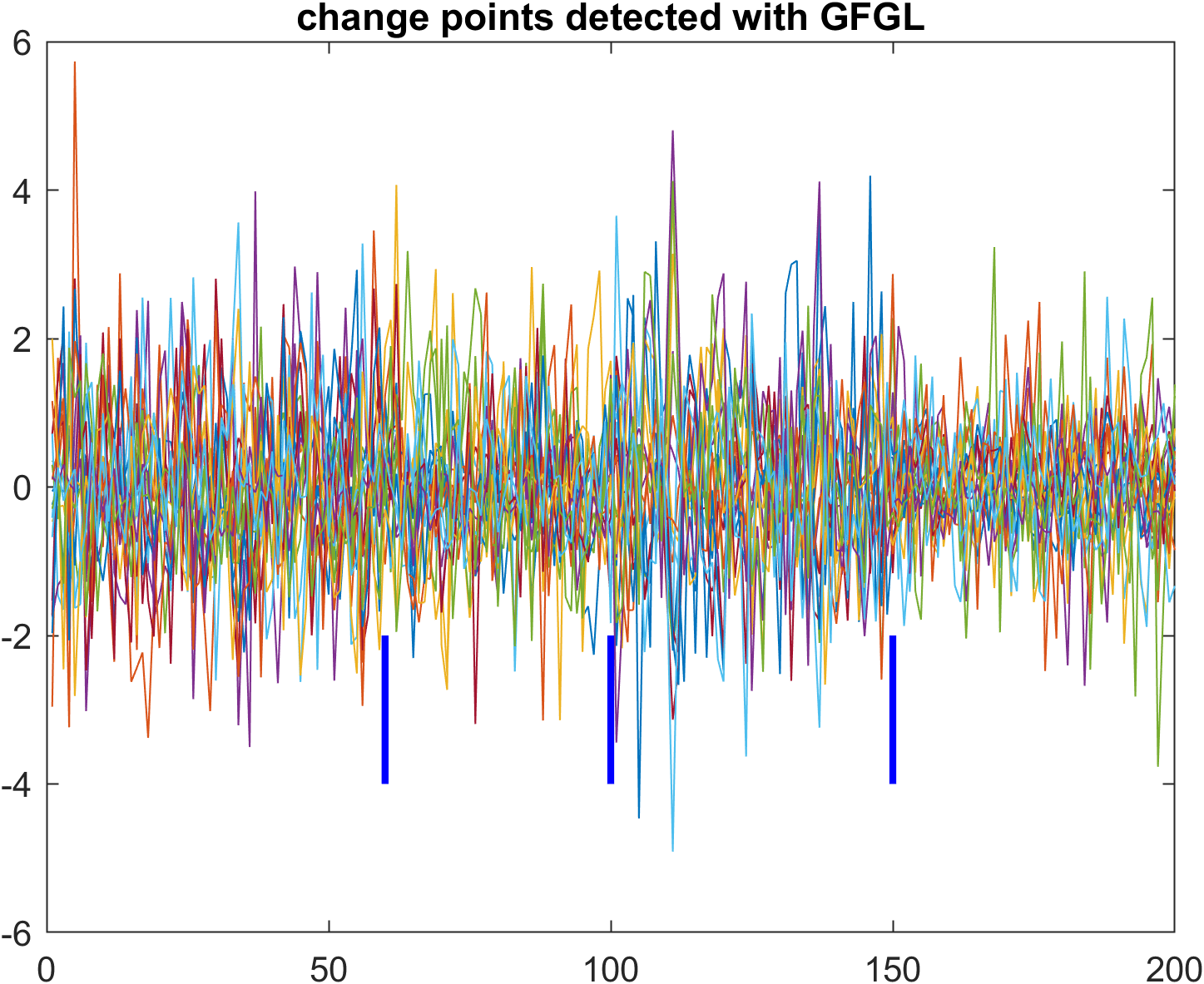}
\caption{ \footnotesize The change points identified with $\lambda_1=0.20$ and $\lambda_2=60$ are $\{60,100,150\}$.}
\end{subfigure}
\hfill
\begin{subfigure}[b]{0.3\textwidth}
\includegraphics[width=\textwidth]{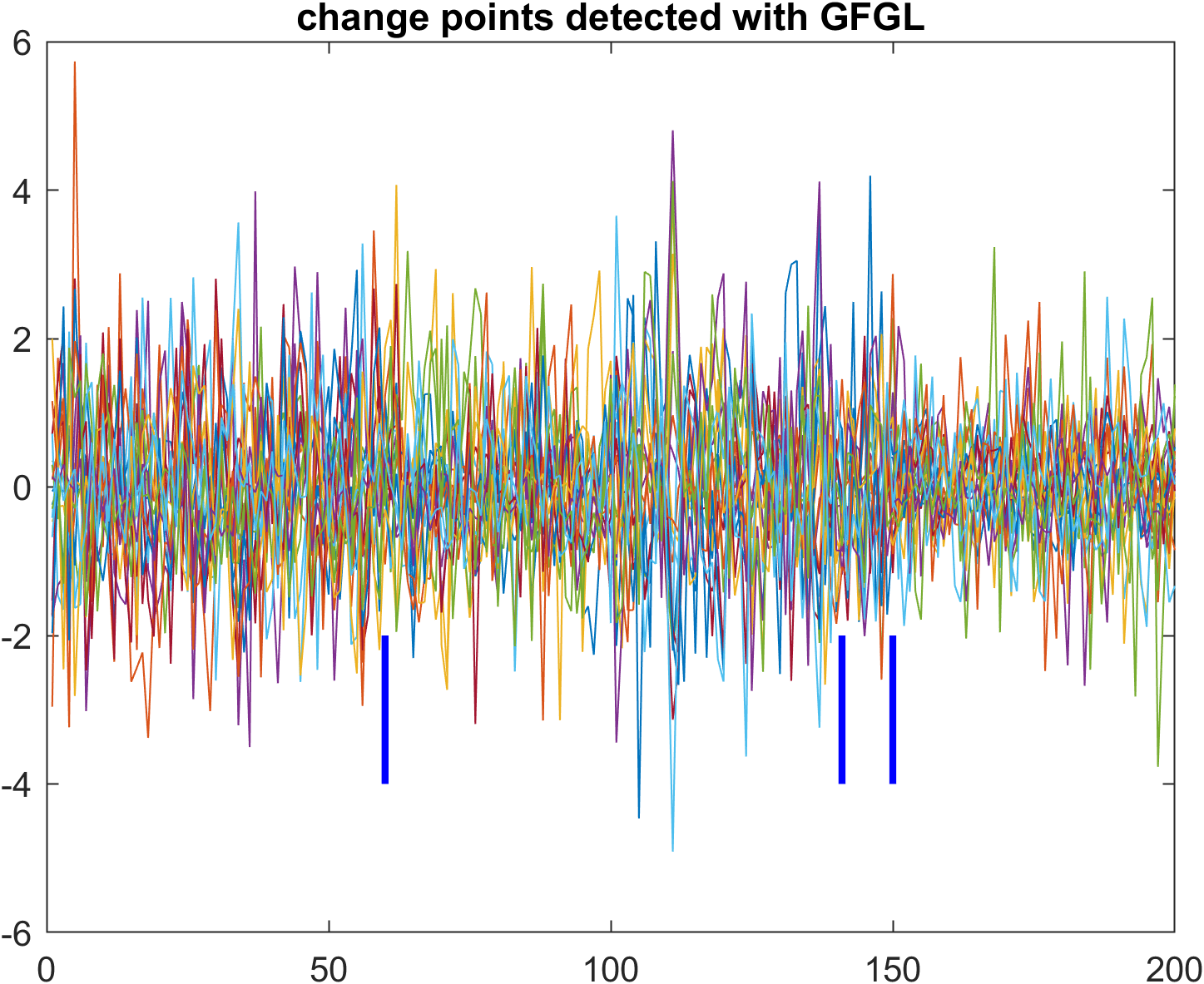}
\caption{ \footnotesize The change points identified with $\lambda_1=0.50$ and $\lambda_2=60$ are $\{60,141,150\}$.}
\end{subfigure}
\caption{Scenario 4: Change points (as blue vertical lines) detected by the GFGL model. }
\end{figure}

\renewcommand{\thefigure}{B.4.4}
\begin{figure}[H]
\centering
\begin{subfigure}[b]{0.85\textwidth}
\includegraphics[width=\textwidth]{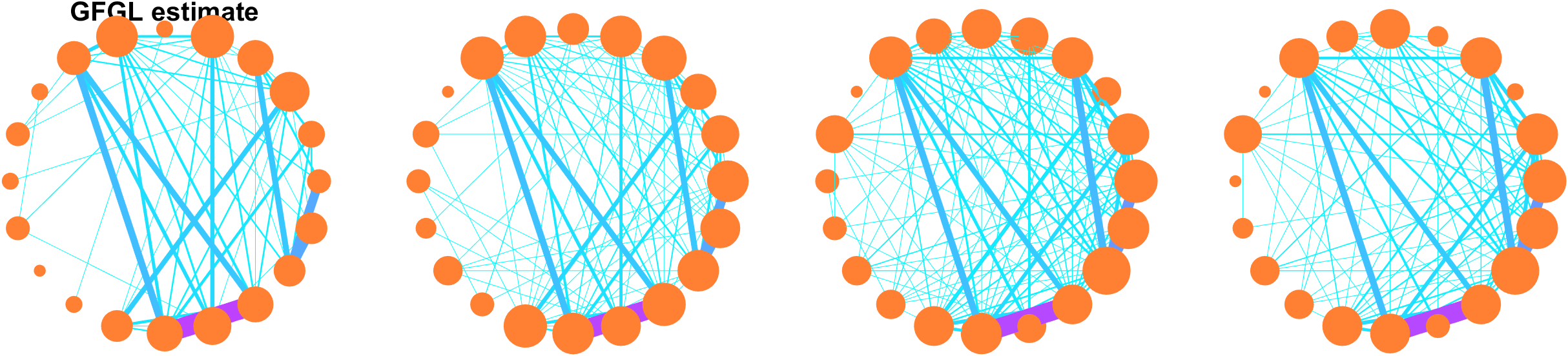}
\caption{The change points identified with $\lambda_1=0.10$ and $\lambda_2=60$ are $\{60,63,100\}$.}
\end{subfigure}
\begin{subfigure}[b]{0.85\textwidth}
\vspace{0.3cm}
\includegraphics[width=\textwidth]{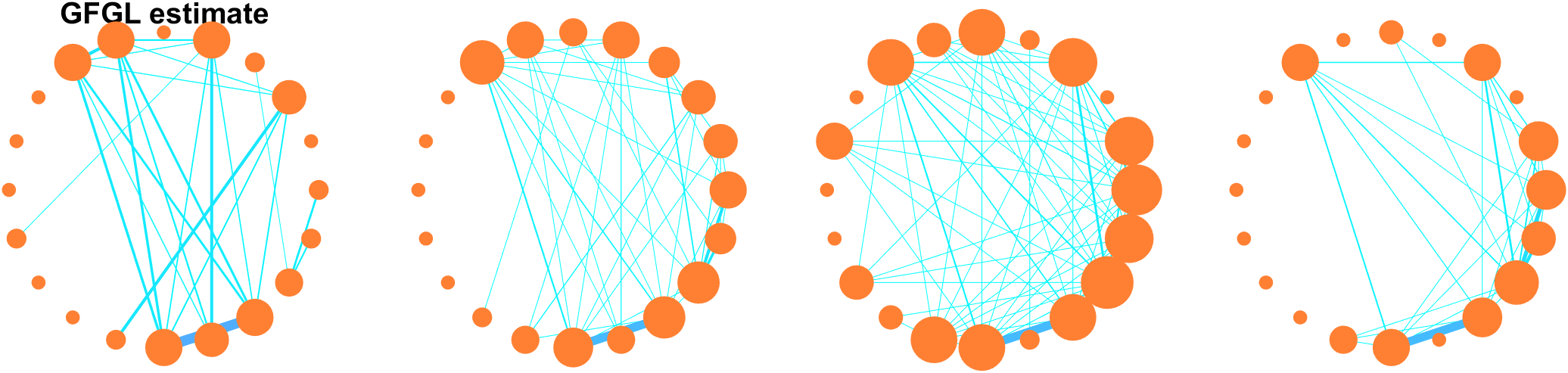}
\caption{The change points identified with $\lambda_1=0.20$ and $\lambda_2=60$ are $\{60,100,150\}$.}
\end{subfigure}
\hfill
\begin{subfigure}[b]{0.85\textwidth}
\vspace{0.3cm}
\includegraphics[width=\textwidth]{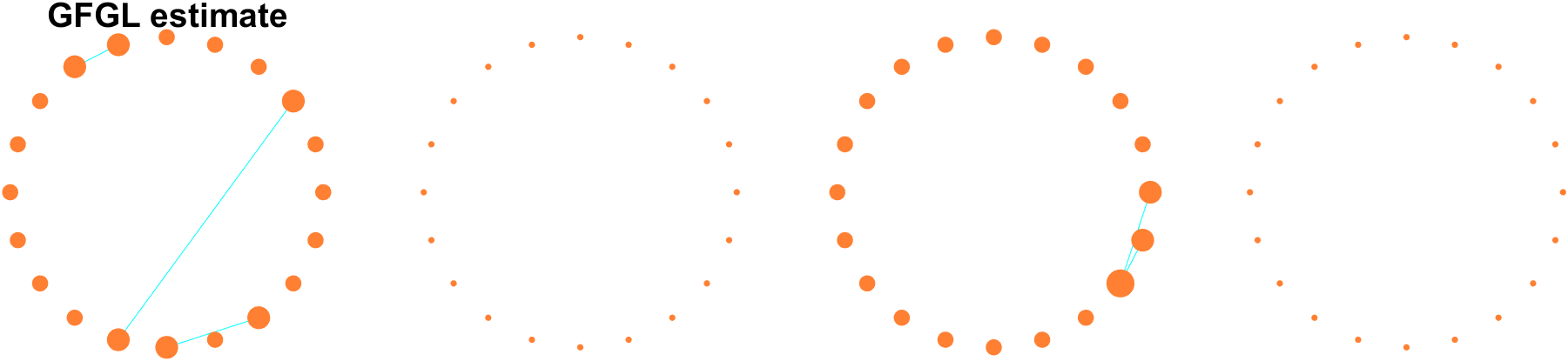}
\caption{The change points identified with $\lambda_1=0.50$ and $\lambda_2=60$ are $\{60,141,150\}$.}
\end{subfigure}
\caption{ Scenario 4: Graphs estimated by the group-fused graphical lasso model for different values of the hyperparameters.}
\end{figure}

\renewcommand{\thefigure}{B.4.5}
\begin{figure}[H]
\centering
\includegraphics[width=\textwidth]{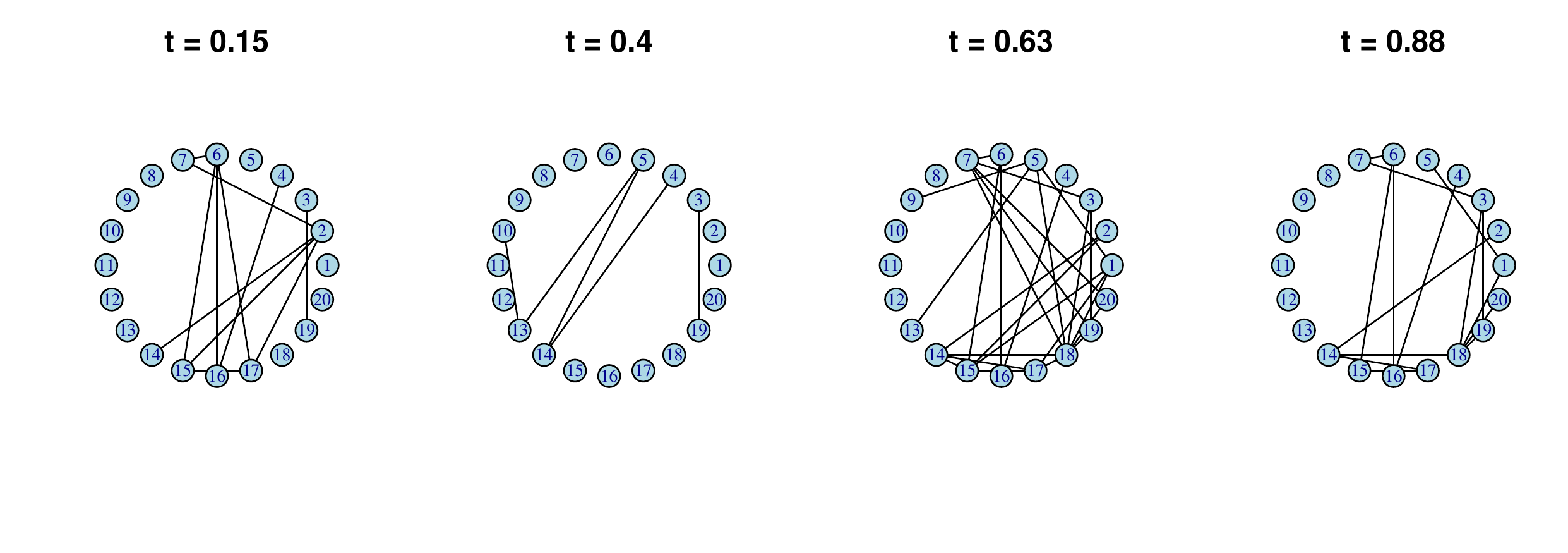}
\caption{Scenario 4: Graph estimation with the ``oracle version" of the \textsf{loggle} model.x}
\end{figure}

\subsection*{B.5 Scenario 5: results}
We consider a scenario for  our model is highly misspecified.
In this scenario, the dependence structure presents both an abrupt change point in the variance as well as smooth changes.
The true data generating mechanism is as follows. The first covariance matrix is generated starting from a graph with 21 edges.
 There is a first change point at $t=60$, at which the correlation structure remain the same as before, but standard deviations double for all variables. Then, starting from $t=100$ the dependence structure starts to smoothly change according to a multivariate GARCH model \citep[see, for instance,][]{silvennoinen2009multivariate}, i.e.
\[
Y_t \mid \Sigma_t \overset{ind}{\sim} N_p(0,\Sigma_t) \qquad \text{for }t=100,\ldots,200
\]
\[
\text{vech}(\Sigma_t)=A\enskip\text{vech}(Y_{t-1}Y'_{t-1})+B\enskip\text{vech}(\Sigma_{t-1})
\]
where $\text{vech}()$ is an operator that stacks the columns of the lower triangular part of its argument and $A$ and $B$ are $p(p + 1)/2 \times p(p + 1)/2$ dimensional matrices, that we set to be diagonal \citep[for more details on this specification of the multivariate GARCH, see][]{bollerslev1988capital}. All entries on the diagonal of $A$ are equal to 0.21 and all entries on the diagonal of $B$ are equal to 0.80. The following figures summarize the results obtained for two replicates of Scenario 5. 

\renewcommand{\thefigure}{B.5.1}
\begin{figure}[H]
\centering
\includegraphics[width=0.85\textwidth]{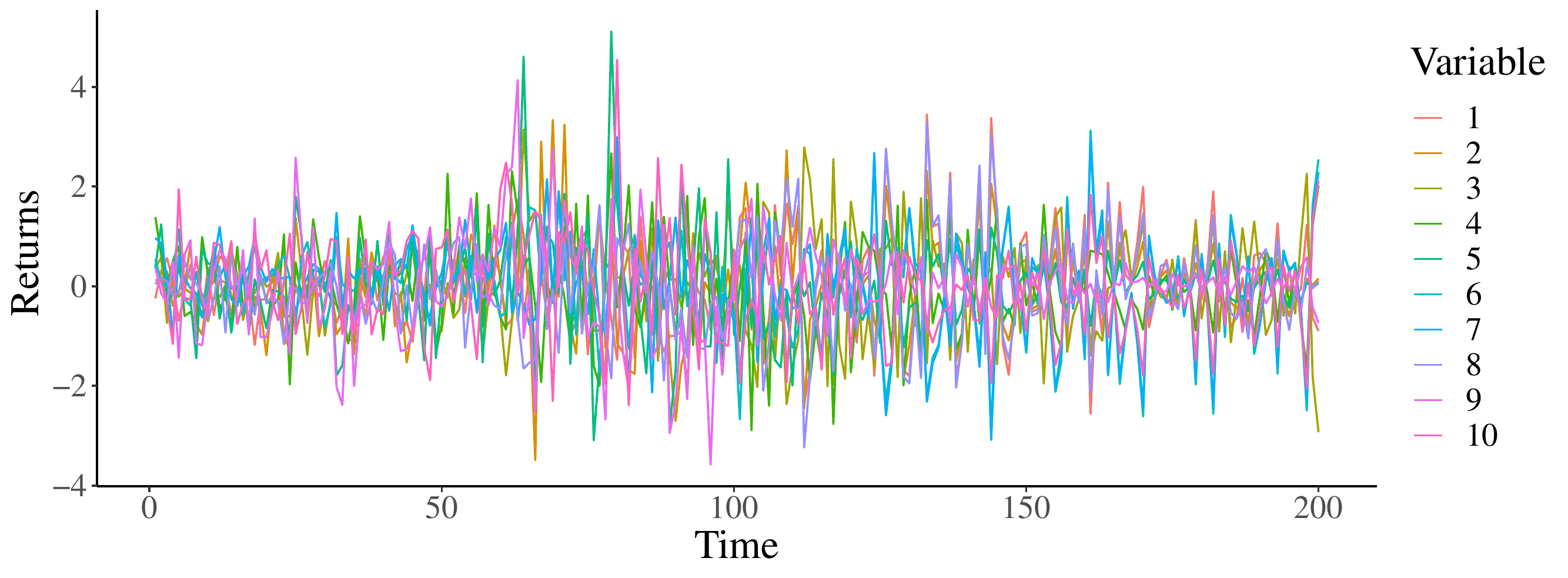}
\caption{ Scenario 5: Simulated data - replicate n.1.}
\end{figure}
\renewcommand{\thefigure}{B.5.2}
\begin{figure}[H]
\centering
\includegraphics[width=0.85\textwidth]{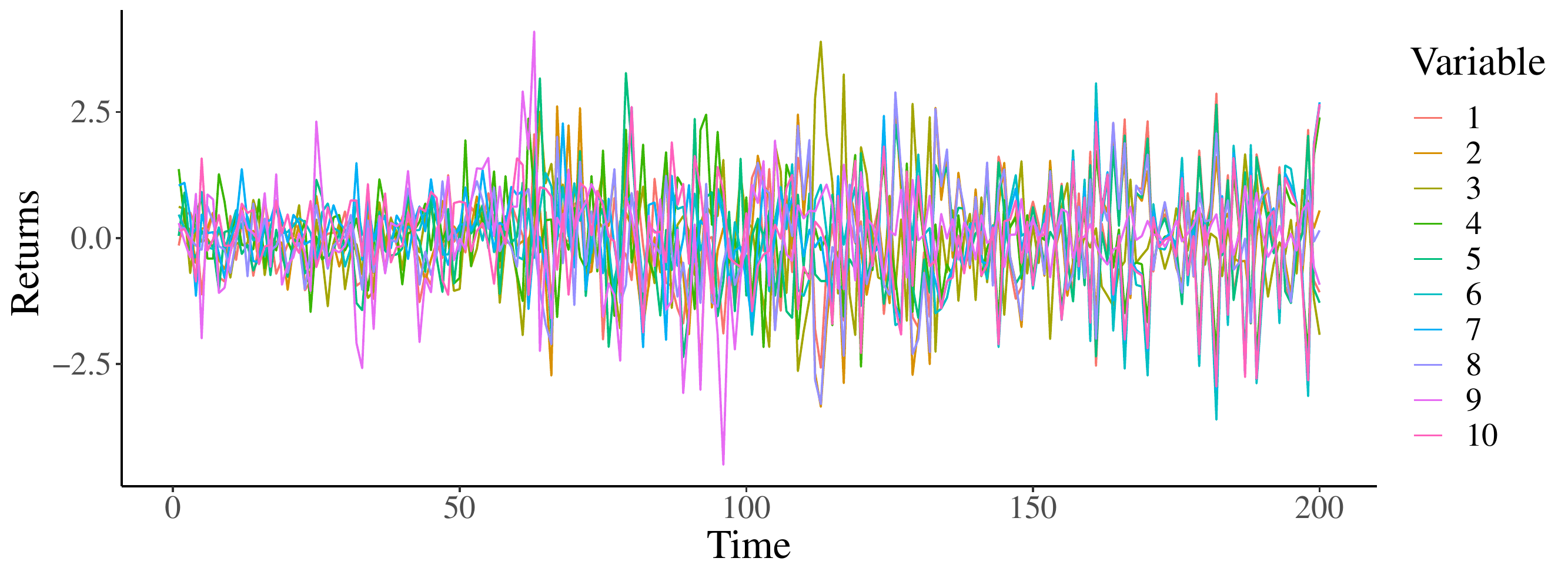}
\caption{ Scenario 5: Simulated data - replicate n.2.}
\end{figure}

\renewcommand{\thefigure}{B.5.3}
\begin{figure}[H]
\centering
\includegraphics[width=0.85\textwidth]{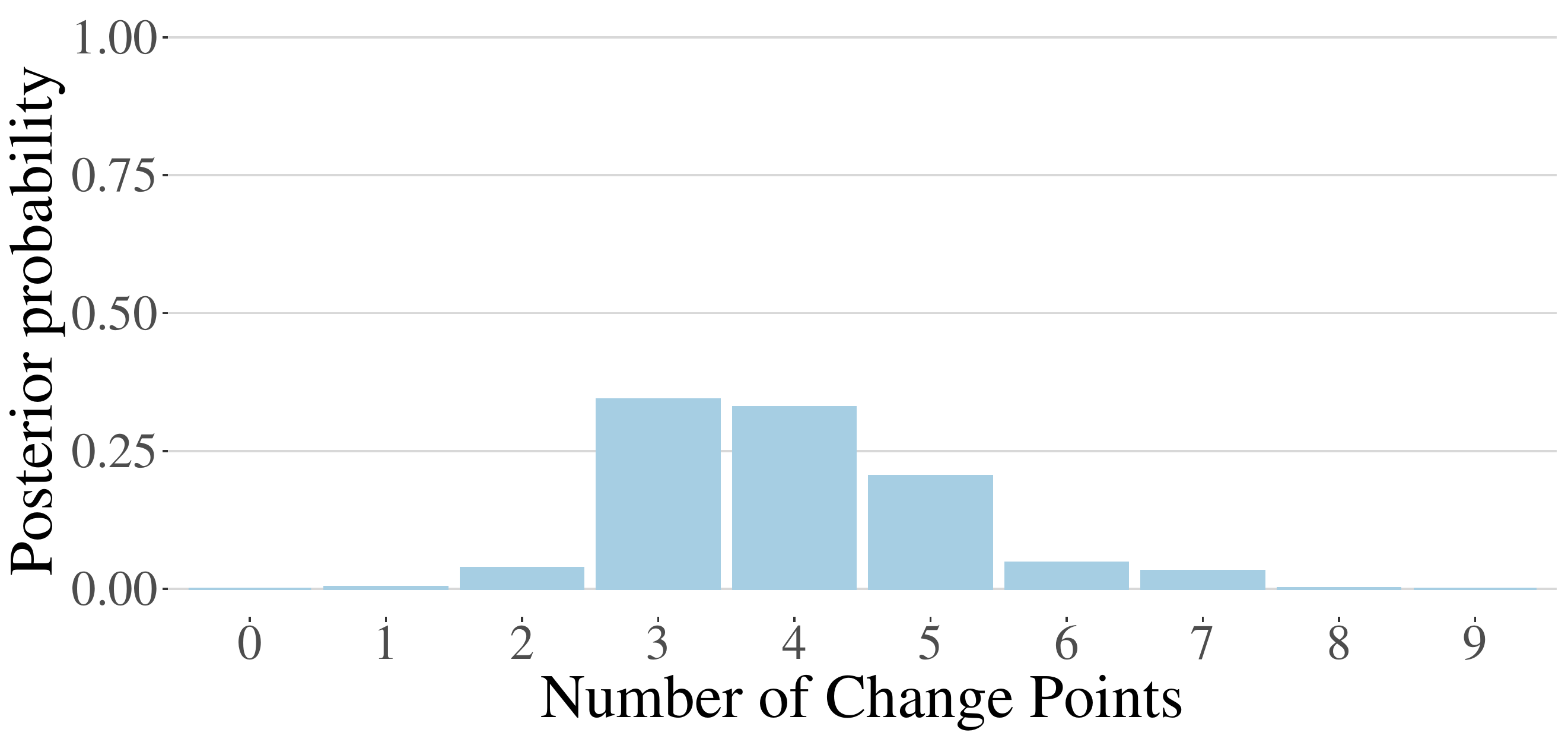}
\caption{ Scenario 5: Posterior distribution of the number of change points - replicate n.1.}
\end{figure}
\renewcommand{\thefigure}{B.5.4}
\begin{figure}[H]
\centering
\includegraphics[width=0.85\textwidth]{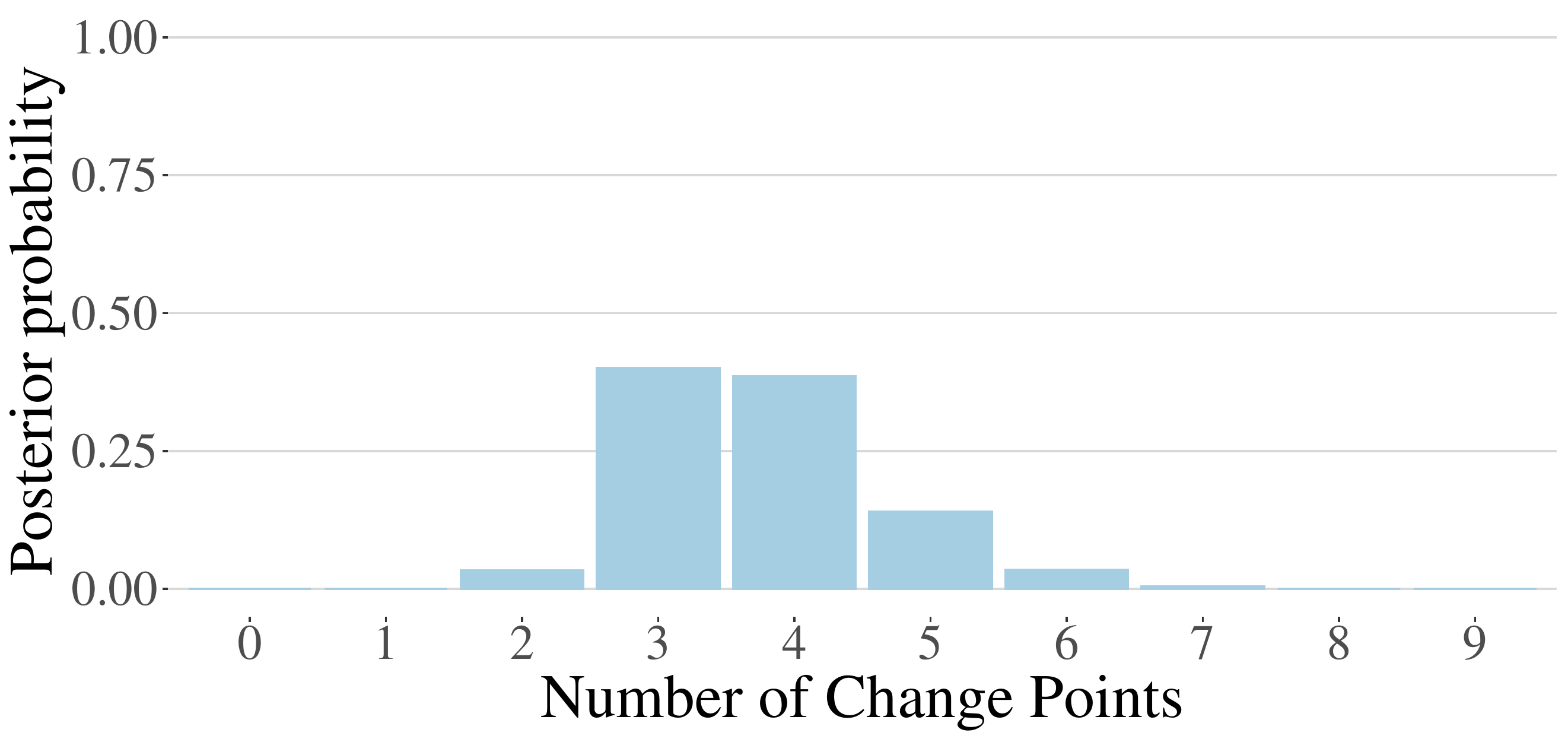}
\caption{ Scenario 5: Posterior distribution of the number of change points - replicate n.2.}
\end{figure}

\renewcommand{\thefigure}{B.5.5}
\begin{figure}[H]
\centering
\includegraphics[width=0.8\textwidth]{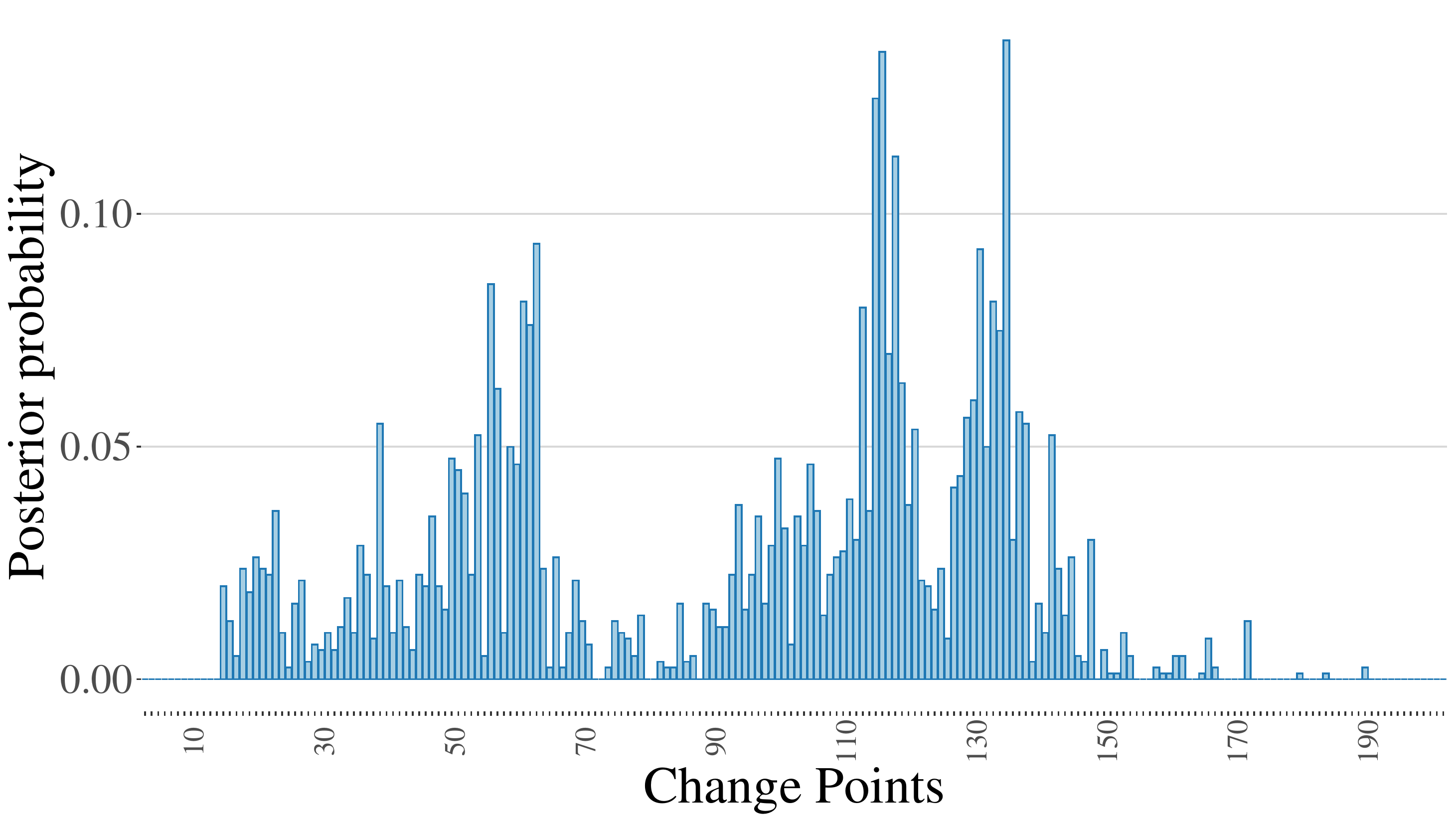}
\caption{ Scenario 5: Marginal posterior probability of every time point to be a change point - replicate n.1.}
\end{figure}
\renewcommand{\thefigure}{B.5.6}
\begin{figure}[H]
\centering
\includegraphics[width=0.8\textwidth]{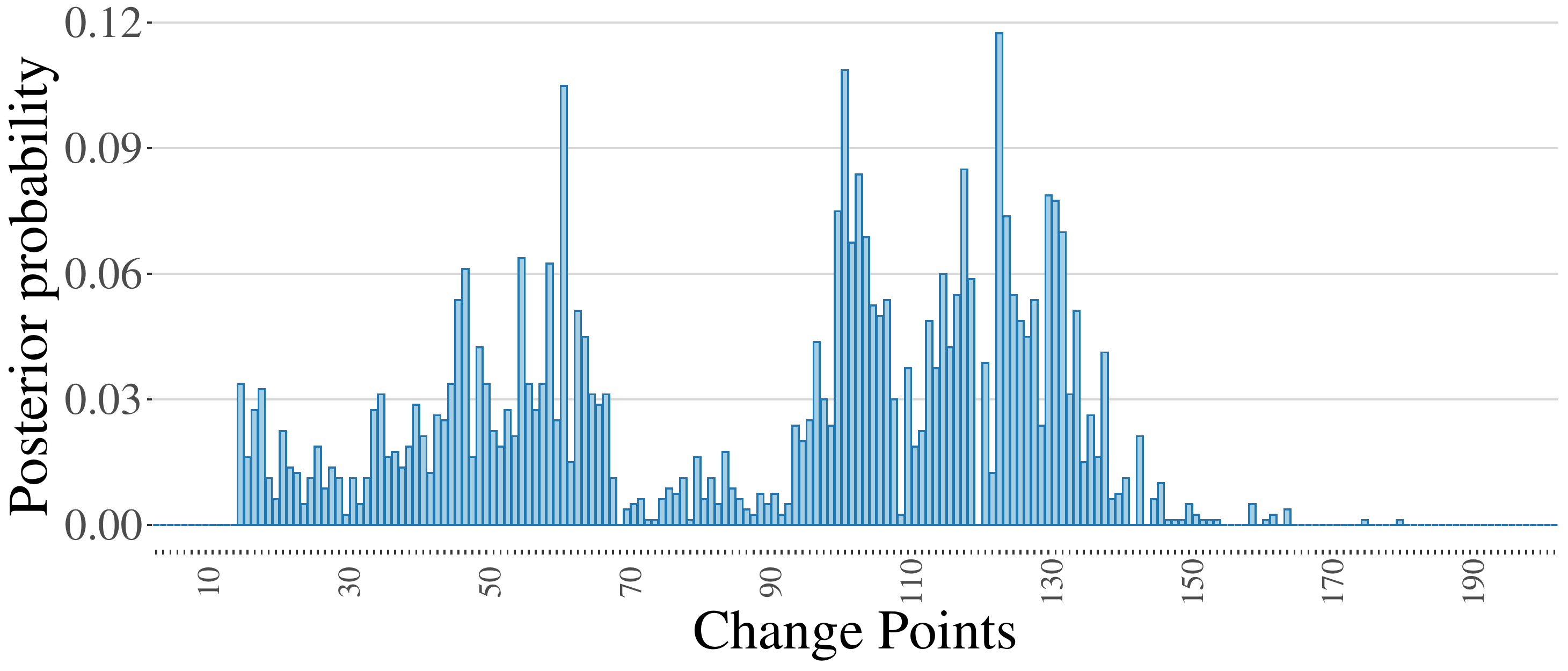}
\caption{ Scenario 5: Marginal posterior probability of every time point to be a change point - replicate n.2.}
\end{figure}

\section*{C. \enskip US stock market analysis: additional figures and tables}

\renewcommand{\thefigure}{C.1}
\begin{figure}[H]
\centering
\includegraphics[width=0.85\textwidth]{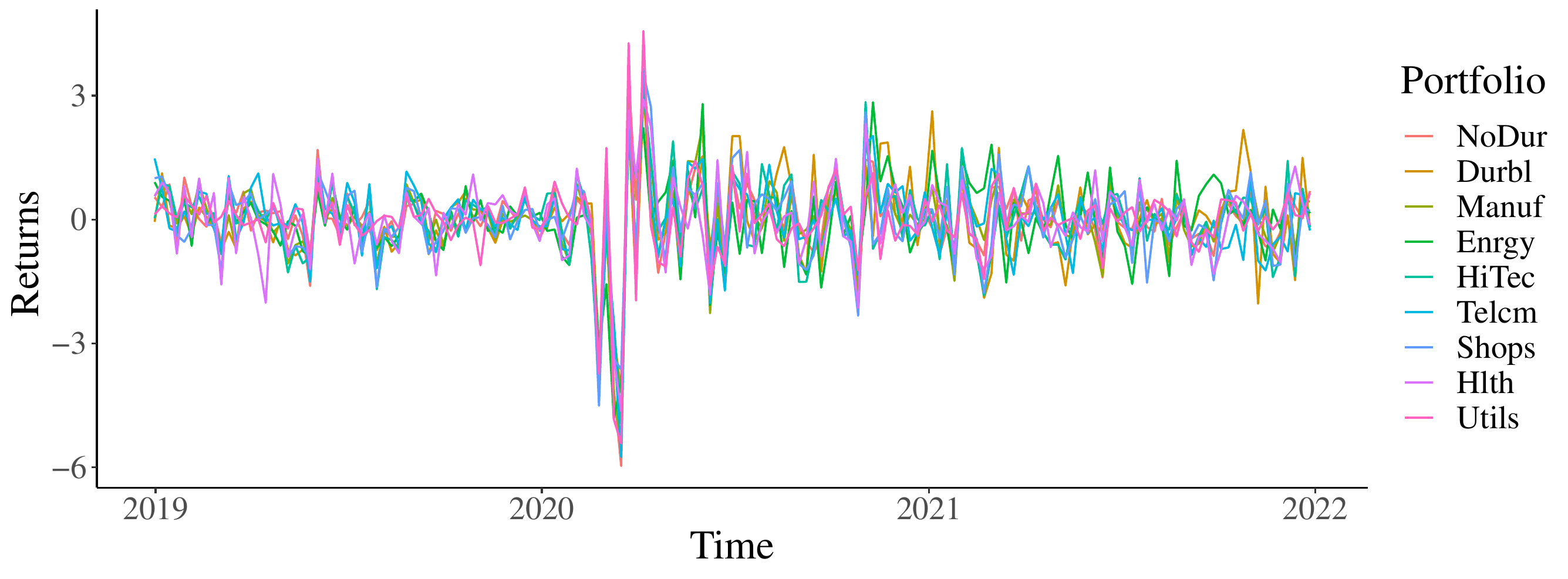}
\caption{Weekly Standardised Logarithmic Returns of nine Industry Portfolios from January 2019 to December 2021.}
\end{figure}

\subsection*{C.1 \enskip Descriptive indexes and summaries of the posterior graphs }
The following tables contain degree centrality, betweenness centrality, local clustering and global clustering coefficients for the estimated graphs.

\renewcommand\thetable{C.1.1}
\begin{table}[!h]
    \centering
    \begin{tabular}{l|c|c|c}
         Portfolios & \cellcolor{LightCyan}pre COVID-19 & \cellcolor{LightCyan}during COVID-19 & \cellcolor{LightCyan}post COVID-19  \\
         \hline \hline
         Consumer Nondurables (NoDur)&2&3&\textcolor{black}{\textbf{4}}\\
         Consumer Durables (Durbl)&1&2&1\\
         Manufacturing (Manuf)&\textcolor{black}{\textbf{3}}&\textcolor{black}{\textbf{4}}&\textcolor{black}{\textbf{4}}\\
         Energy (Enrgy)&1&2&3\\
         High Technology (HiTec)&\textcolor{black}{\textbf{3}}&3&2\\
         Telecommunications (Telcm)&2&1&1\\
         Shops (Shops)&2&2&\textcolor{black}{\textbf{4}}\\
         Health (Hlth)&1&3&2\\
         Utilities (Utils)&1&2&3\\
    \end{tabular}
    \caption{\textbf{Degree centrality}. Degree centrality of a certain node is the number of vertices in the neighbourhood of that node. In {\textbf{bold}}, we highlight the highest degree centrality for each graph.}
    \vspace{-\baselineskip}
\end{table}

\renewcommand\thetable{C.1.2}
\begin{table}[!h]
    \centering
    \begin{tabular}{l|c|c|c}
         Portfolios & \cellcolor{LightCyan}pre COVID-19 & \cellcolor{LightCyan}during COVID-19 & \cellcolor{LightCyan}post COVID-19  \\
         \hline \hline
         Consumer Nondurables (NoDur)&7&\textbf{11.5}&\textbf{10}\\
         Consumer Durables (Durbl)&0&4.5&0\\
         Manufacturing (Manuf)&\textbf{13}&\textbf{14}&\textbf{6.67}\\
         Energy (Enrgy)&0&0&1.33\\
         High Technology (HiTec)&\textbf{19}&3.5&{2.33}\\
         Telecommunications (Telcm)&12&0&0\\
         Shops (Shops)&\textbf{15}&0&\textbf{10}\\
         Health (Hlth)&0&\textbf{5.5}&2.33\\
         Utilities (Utils)&0&0&1.33\\
    \end{tabular}
    \caption{\textbf{Betweenness centrality}. Between centrality of a certain node $v$ is $c(v) = \sum_{h\neq v \neq  k}{ \sigma_{h,k}(v)}{\sigma_{h,k}}$, where $\sigma_{h,k}(v)$ is the number of geodesics (i.e., shortest paths) between nodes $h$ and $k$ going through node $j$ and $\sigma_{h,k}$ is the number of geodesics between nodes $h$ and $k$. In \textbf{bold}, we highlight the three highest values for each graph.}
    \vspace{-\baselineskip}
\end{table}

\renewcommand\thetable{C.1.3}
\begin{table}[!h]
    \centering
    \begin{tabular}{l|c|c|c}
         Portfolios & \cellcolor{LightCyan}pre COVID-19 & \cellcolor{LightCyan}during COVID-19 & \cellcolor{LightCyan}post COVID-19  \\
         \hline \hline
         Consumer Nondurables (NoDur)&0&0&0.17\\
         Consumer Durables (Durbl)&0&0&0\\
         Manufacturing (Manuf)&0&0.17&0.50\\
         Energy (Enrgy)&0&1&0.67\\
         High Technology (HiTec)&0&0.33&0\\
         Telecommunications (Telcm)&0&0&0\\
         Shops (Shops)&0&1&0.17\\
         Health (Hlth)&0&0.33&0\\
         Utilities (Utils)&0&1&0.67\\
         \hline
         \textbf{Global clustering coefficient} & 0 & 0.43 & 0.24
    \end{tabular}
    \caption{\textbf{Local clustering coefficients} Local clustering coefficients are the ratio between the number of triangular cliques, of which the node is a part, and $d \, (d-1) / 2$, where $d$ is the degree centrality of that node. Global clustering is the average of local coefficients.}
    \vspace{-\baselineskip}
\end{table}

\clearpage
\subsection*{C.2 \enskip Posterior estimates of precision, covariance, and correlation matrices. }

\renewcommand{\thefigure}{C.2.1}
\begin{figure}[H]
\centering
\begin{subfigure}[b]{0.5\textwidth}
\includegraphics[width=\textwidth]{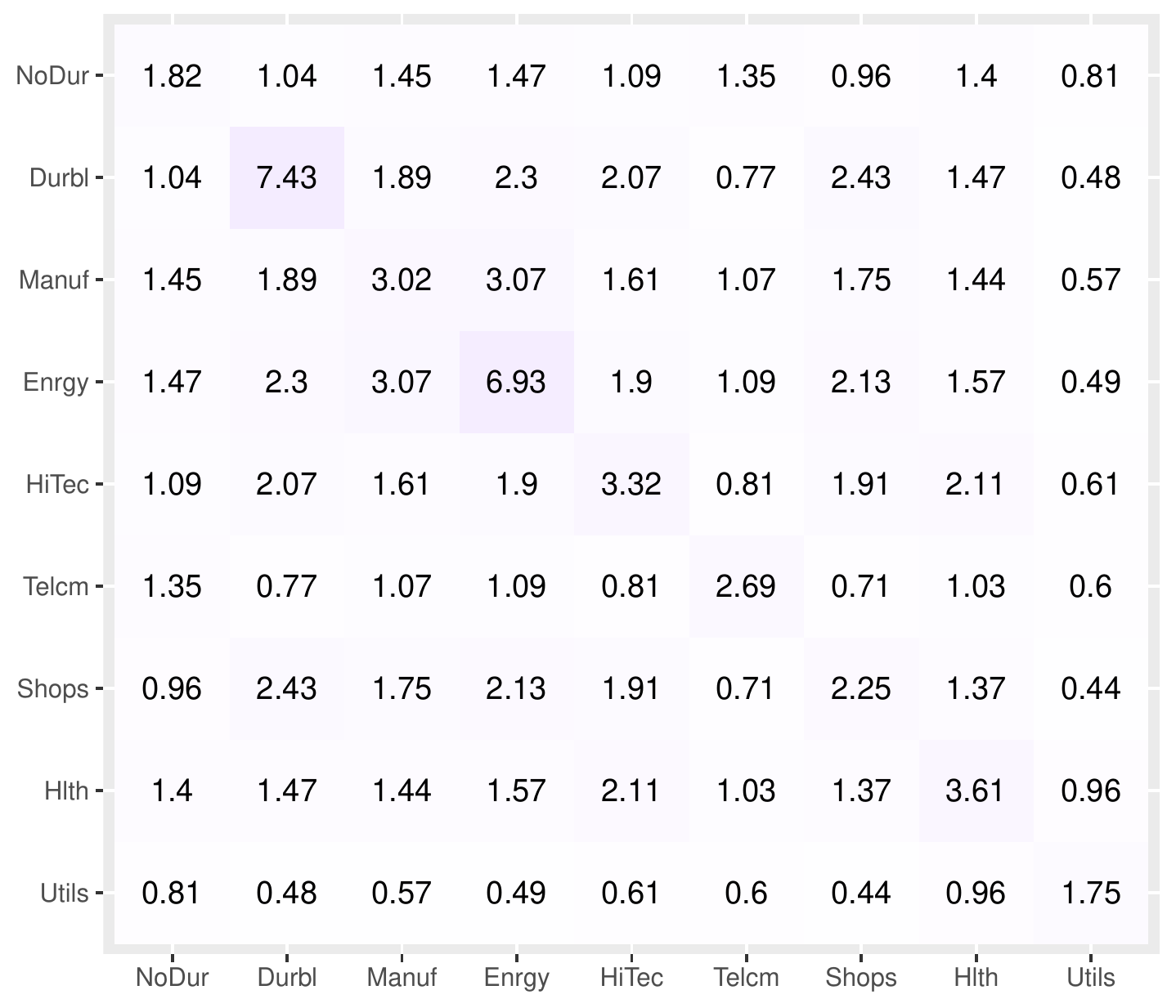}
\end{subfigure}
\begin{subfigure}[b]{0.5\textwidth}
\includegraphics[width=\textwidth]{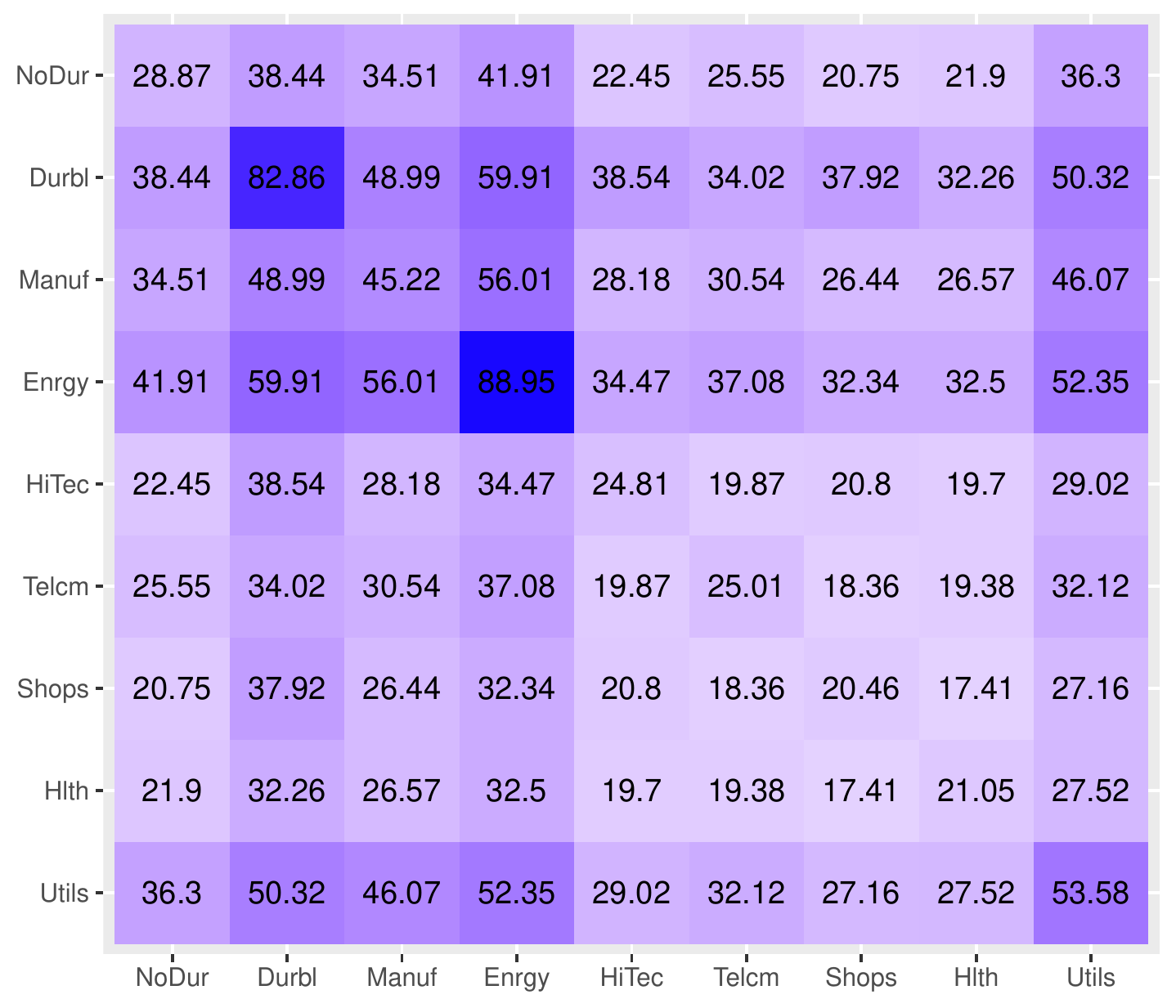}
\end{subfigure}
\begin{subfigure}[b]{0.5\textwidth}
\includegraphics[width=\textwidth]{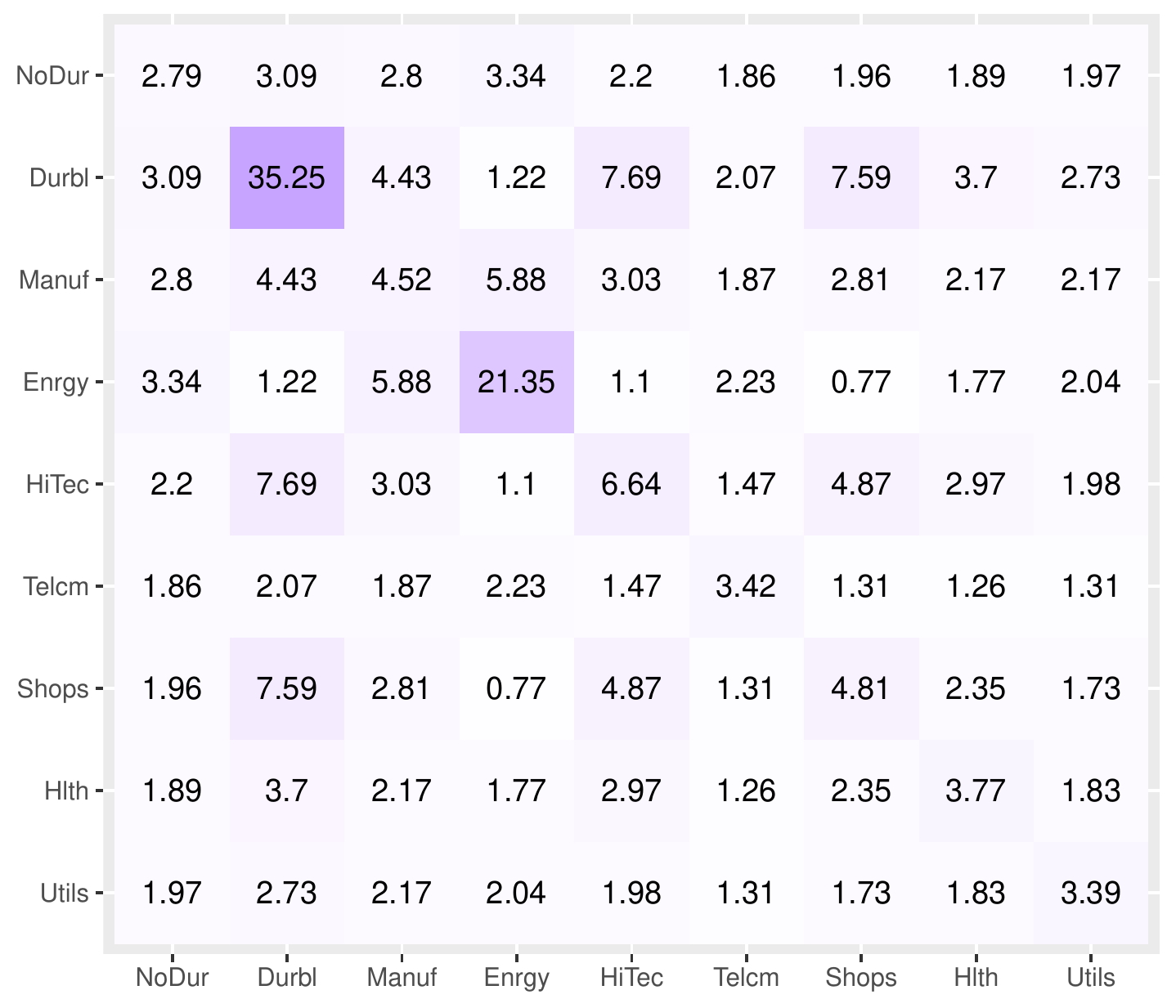}
\end{subfigure}
\caption{\label{real_sigma} Posterior estimates of the  variance and covariance matrices for the original non-standardised weekly percentage logarithmic returns. Posterior estimates are obtained computing the expected values of the entries in the matrix with respect to its posterior distribution conditionally to the graphs point estimate. }
\end{figure}

\renewcommand{\thefigure}{C.2.2}
\begin{figure}[H]
\centering
\begin{subfigure}[b]{0.5\textwidth}
\includegraphics[width=\textwidth]{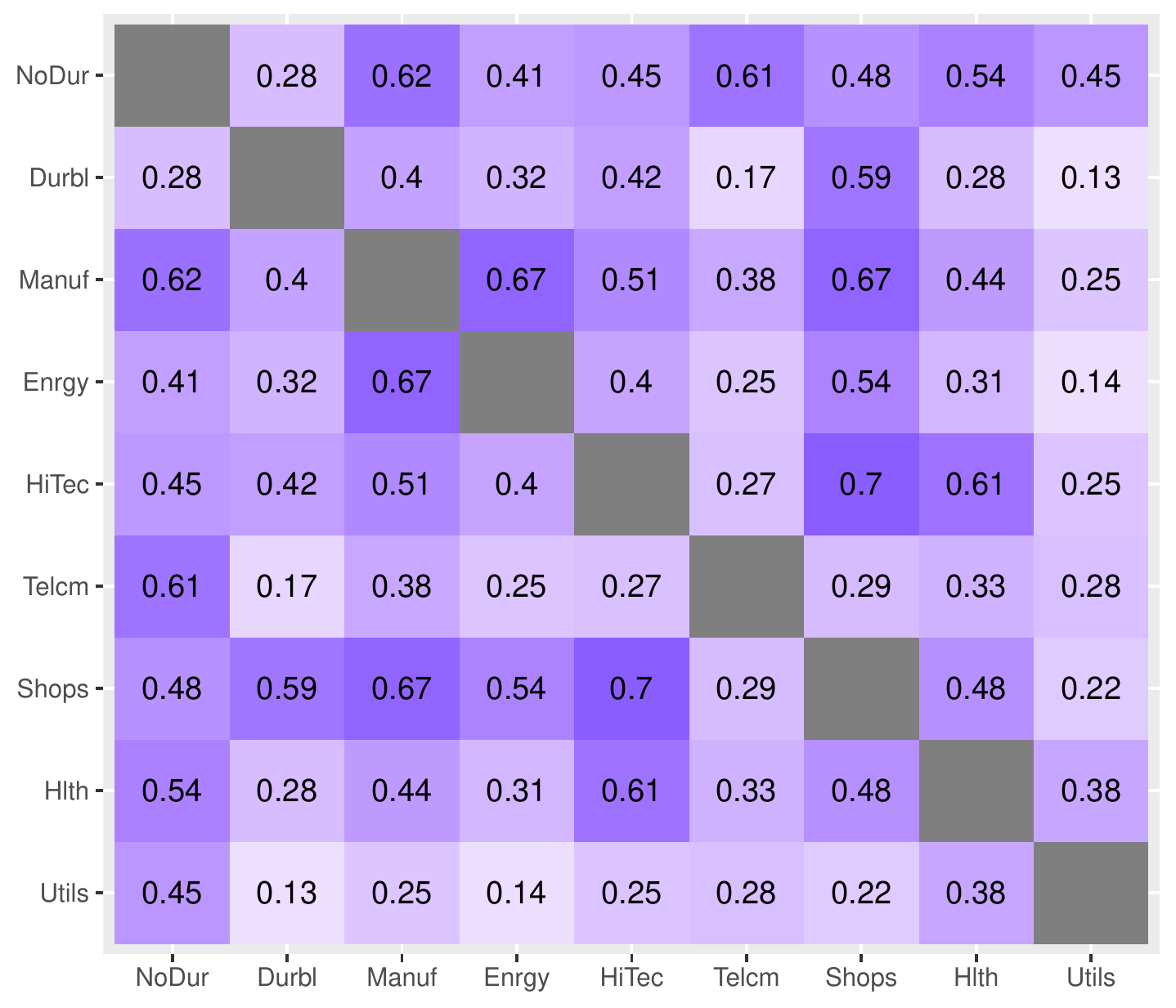}
\end{subfigure}
\begin{subfigure}[b]{0.5\textwidth}
\includegraphics[width=\textwidth]{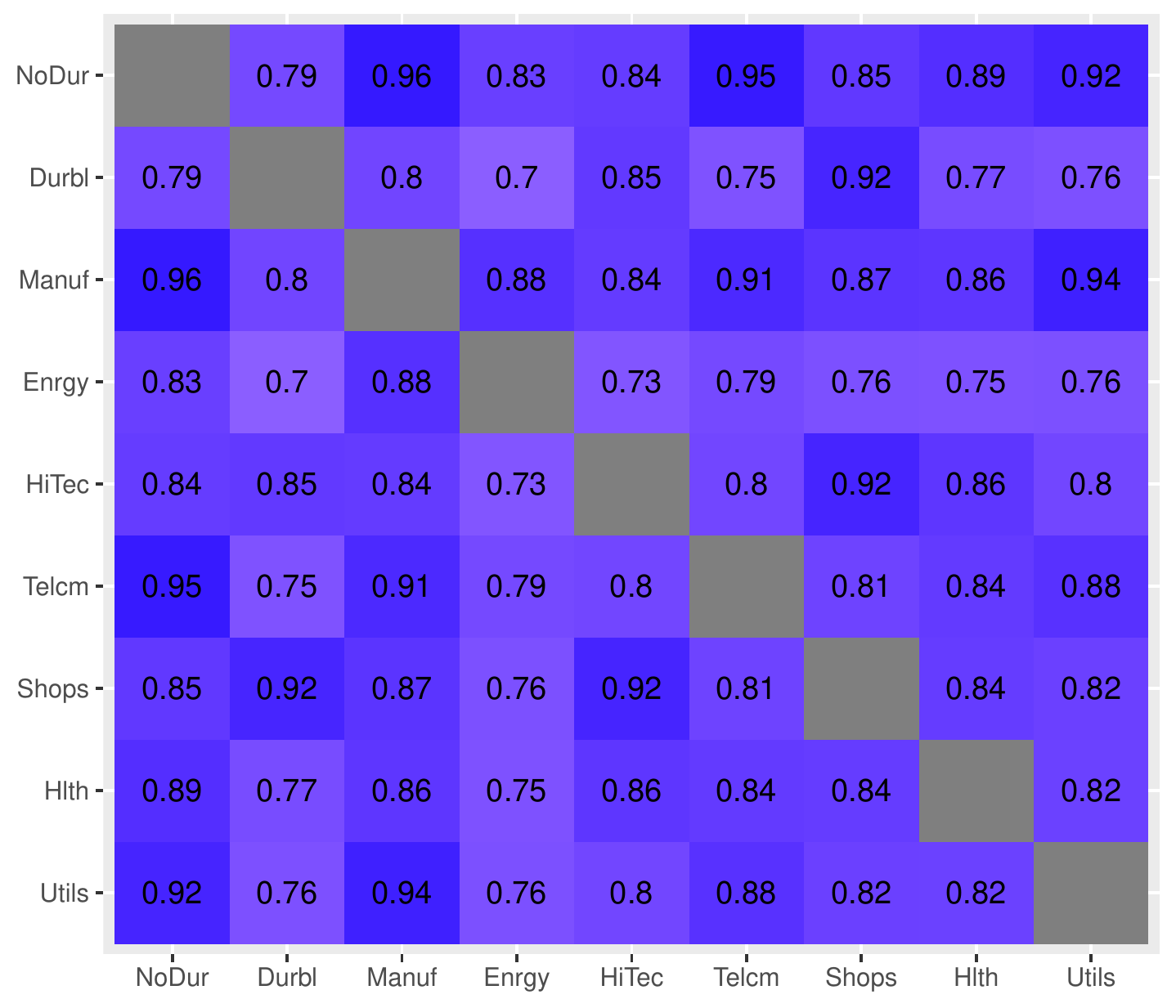}
\end{subfigure}
\begin{subfigure}[b]{0.5\textwidth}
\includegraphics[width=\textwidth]{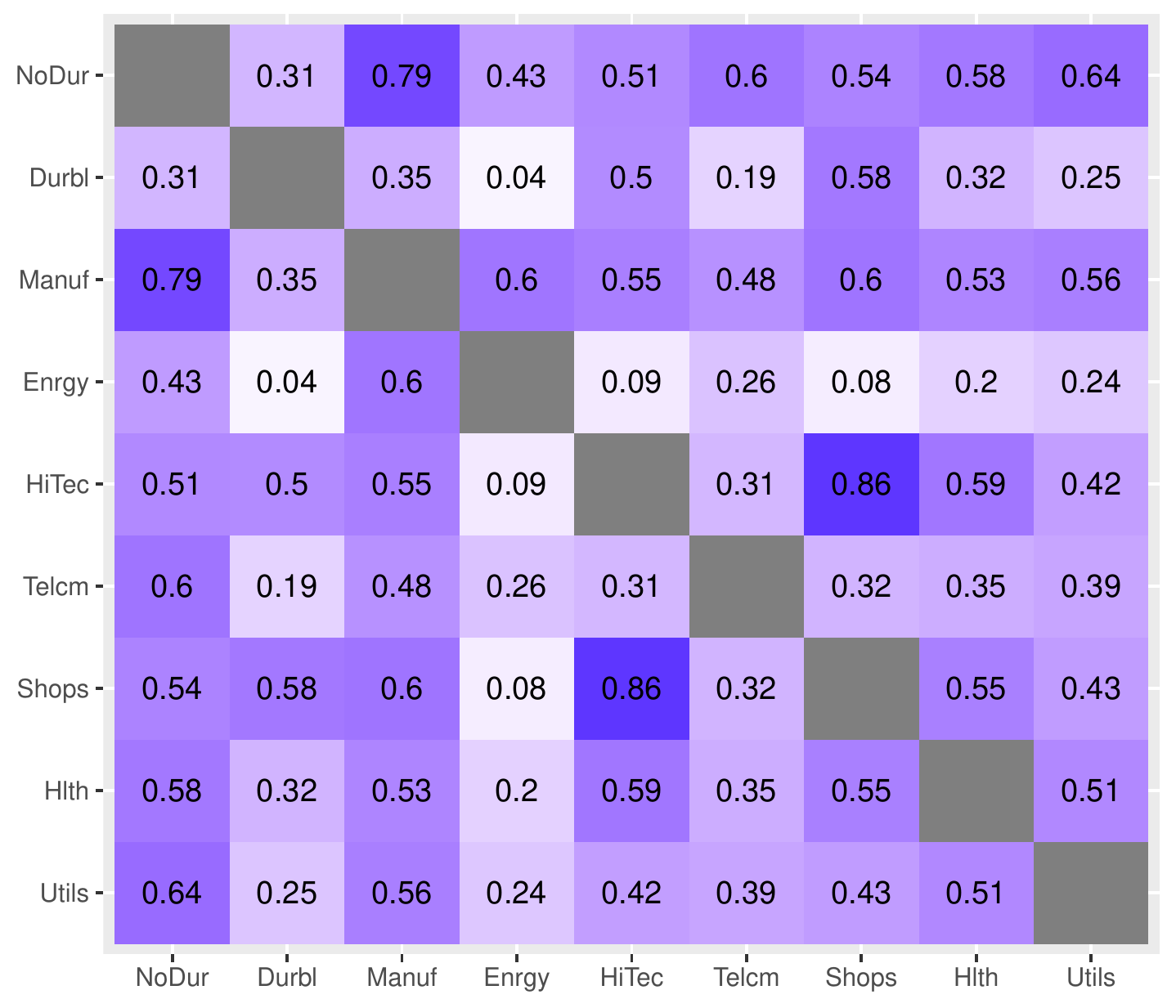}
\end{subfigure}
\caption{ Posterior estimates of the correlation matrices for weekly logarithmic returns. Posterior estimates are obtained from the estimates in Figure~\ref{real_sigma}.}.
\end{figure}

\renewcommand{\thefigure}{C.2.3}
\begin{figure}[H]
\centering
\begin{subfigure}[b]{0.5\textwidth}
\includegraphics[width=\textwidth]{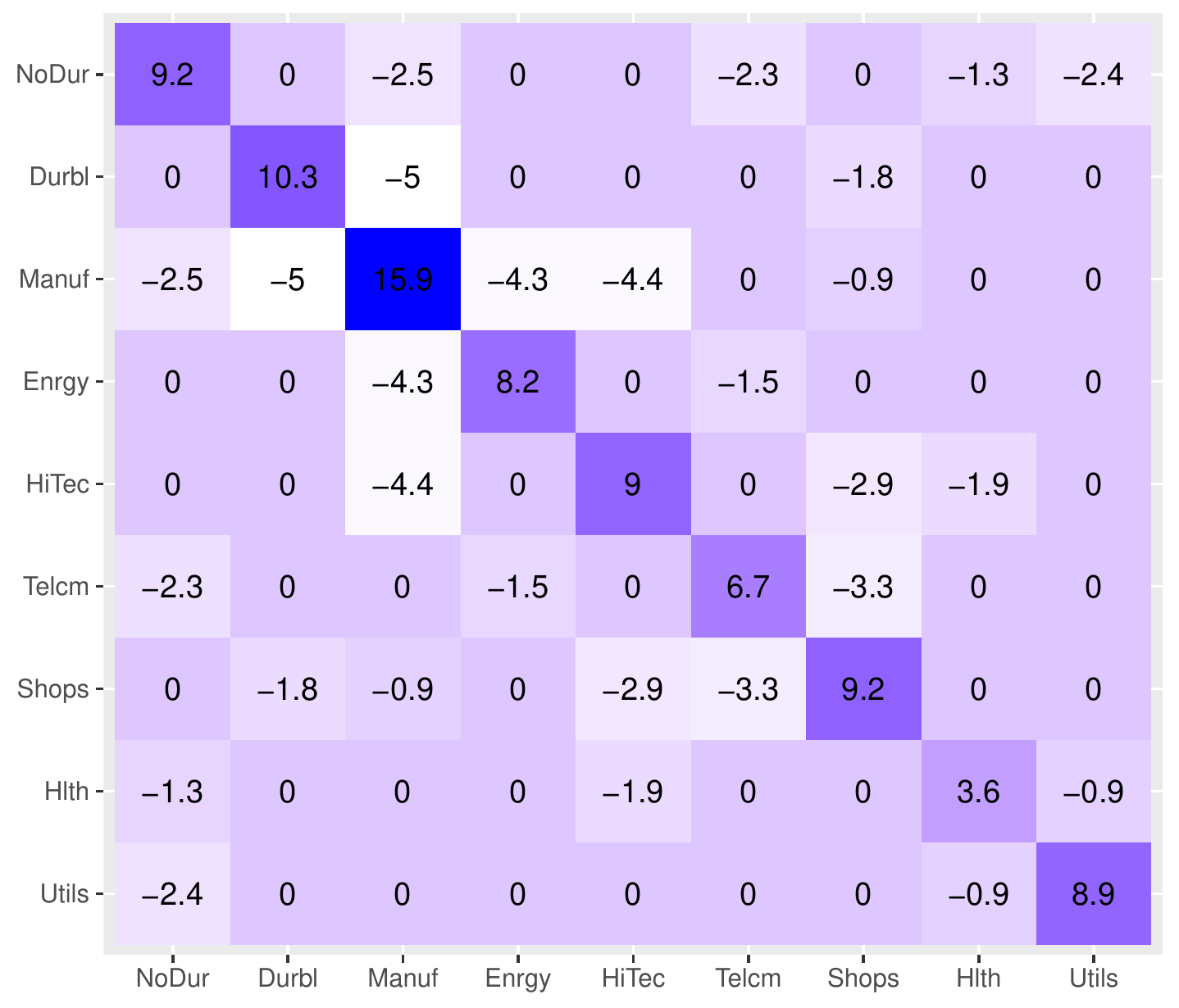}
\end{subfigure}
\begin{subfigure}[b]{0.5\textwidth}
\includegraphics[width=\textwidth]{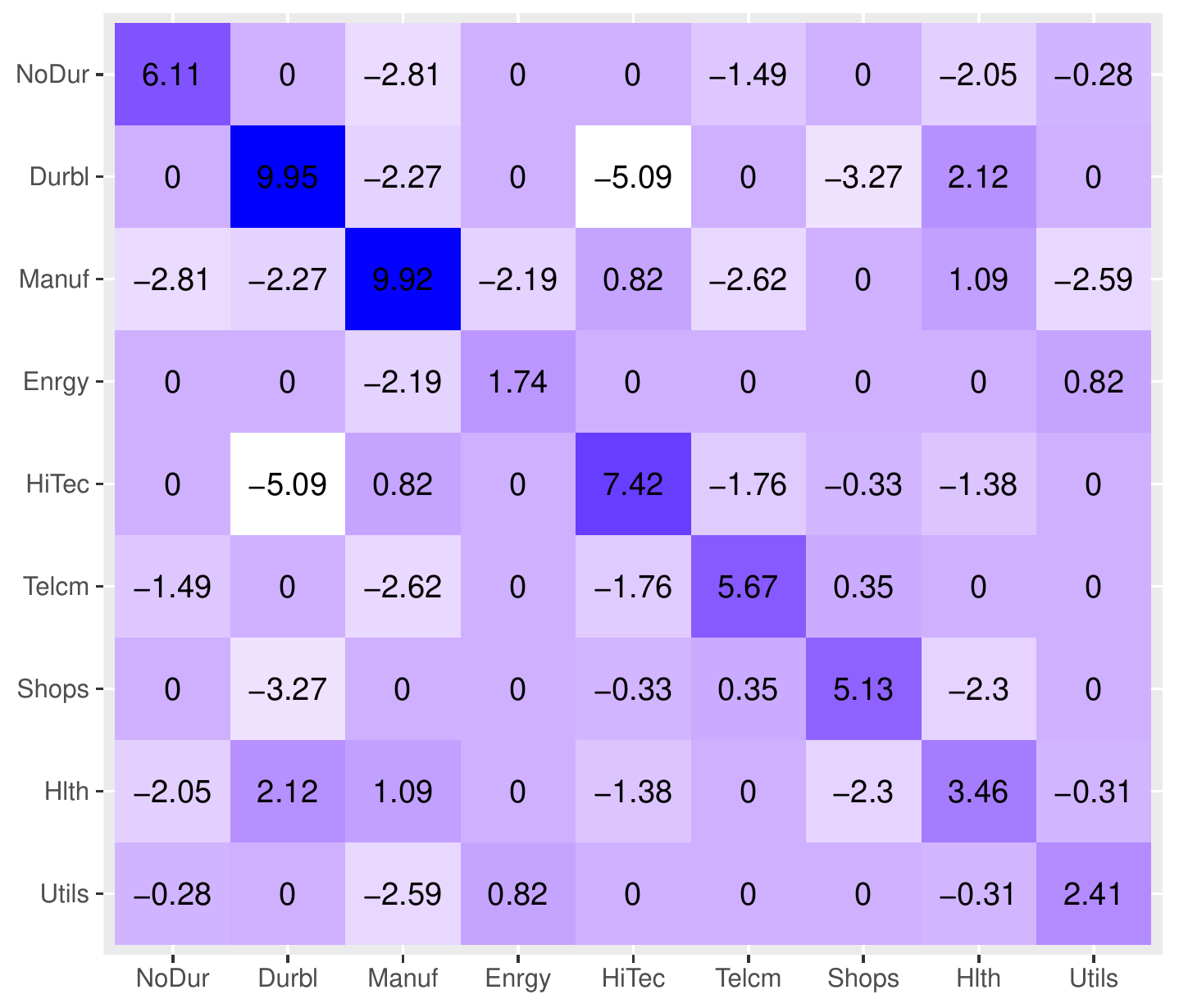}
\end{subfigure}
\begin{subfigure}[b]{0.5\textwidth}
\includegraphics[width=\textwidth]{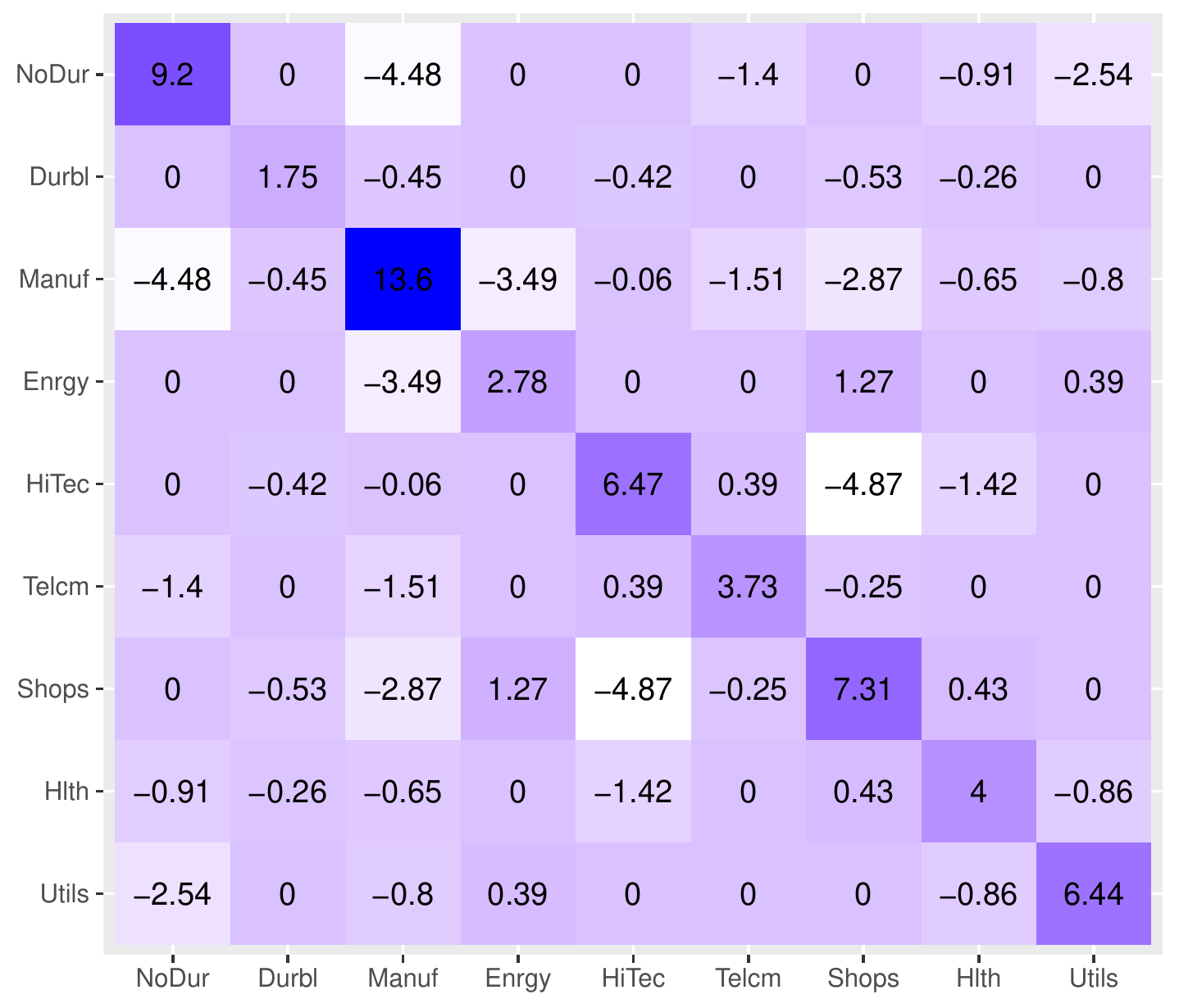}
\end{subfigure}
\caption{ Posterior estimates of the precision matrices for weekly standardized logarithmic returns. Posterior estimates are the entry-wise posterior expected value conditional on the estimated graph structure.}
\end{figure}

\clearpage

\subsection*{C.3 \enskip Results obtained with GFGL and \textsf{loggle}}

\renewcommand{\thefigure}{C.3.1}
\begin{figure}[H]
\centering
\begin{subfigure}[b]{0.3\textwidth}
\includegraphics[width=\textwidth]{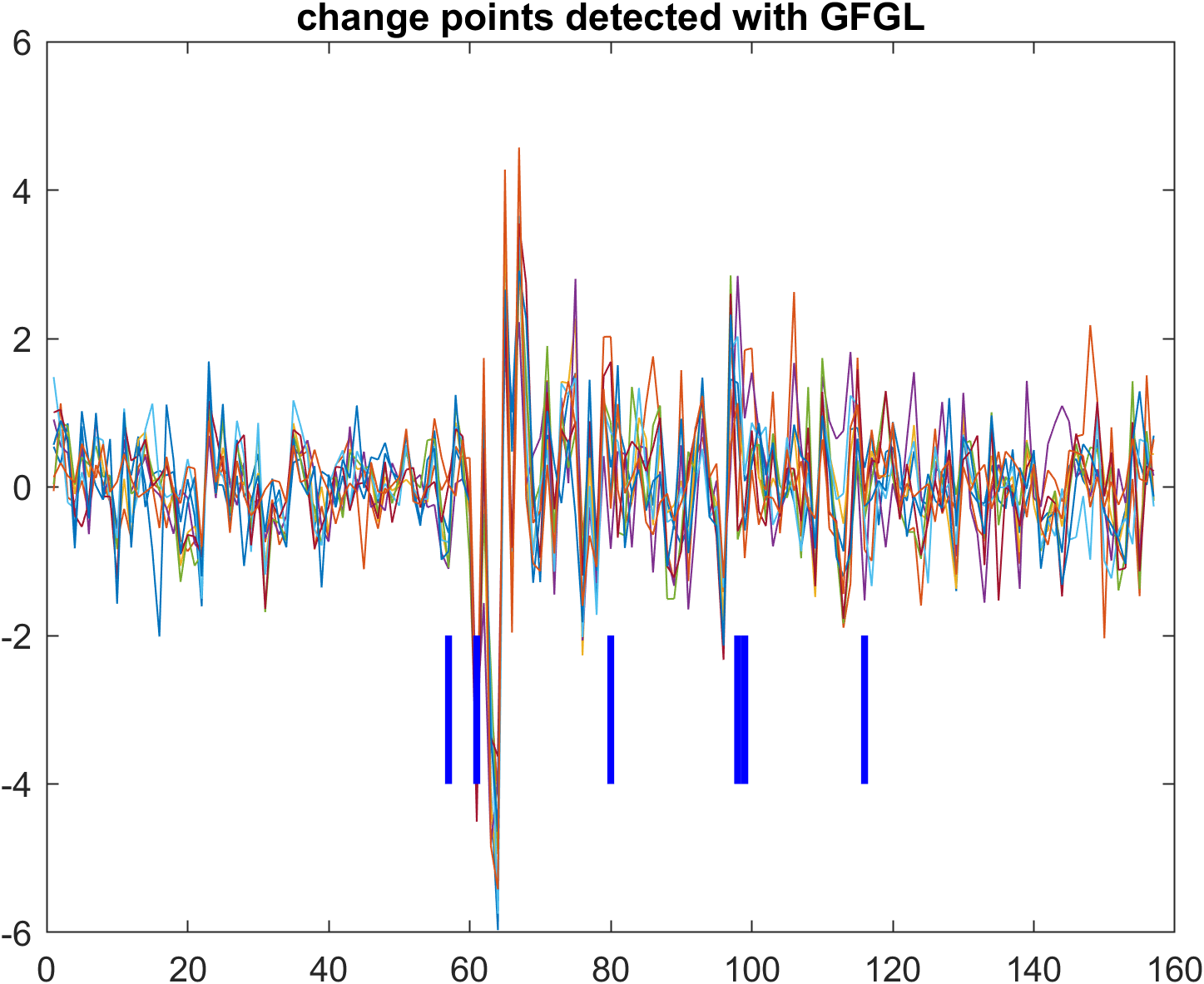}
\caption{\footnotesize The change points identified with $\lambda_1=0.25$ and $\lambda_2=10$ are $\{57,61,80,98,99,116\}$.}
\end{subfigure}
\hfill
\begin{subfigure}[b]{0.3\textwidth}
\includegraphics[width=\textwidth]{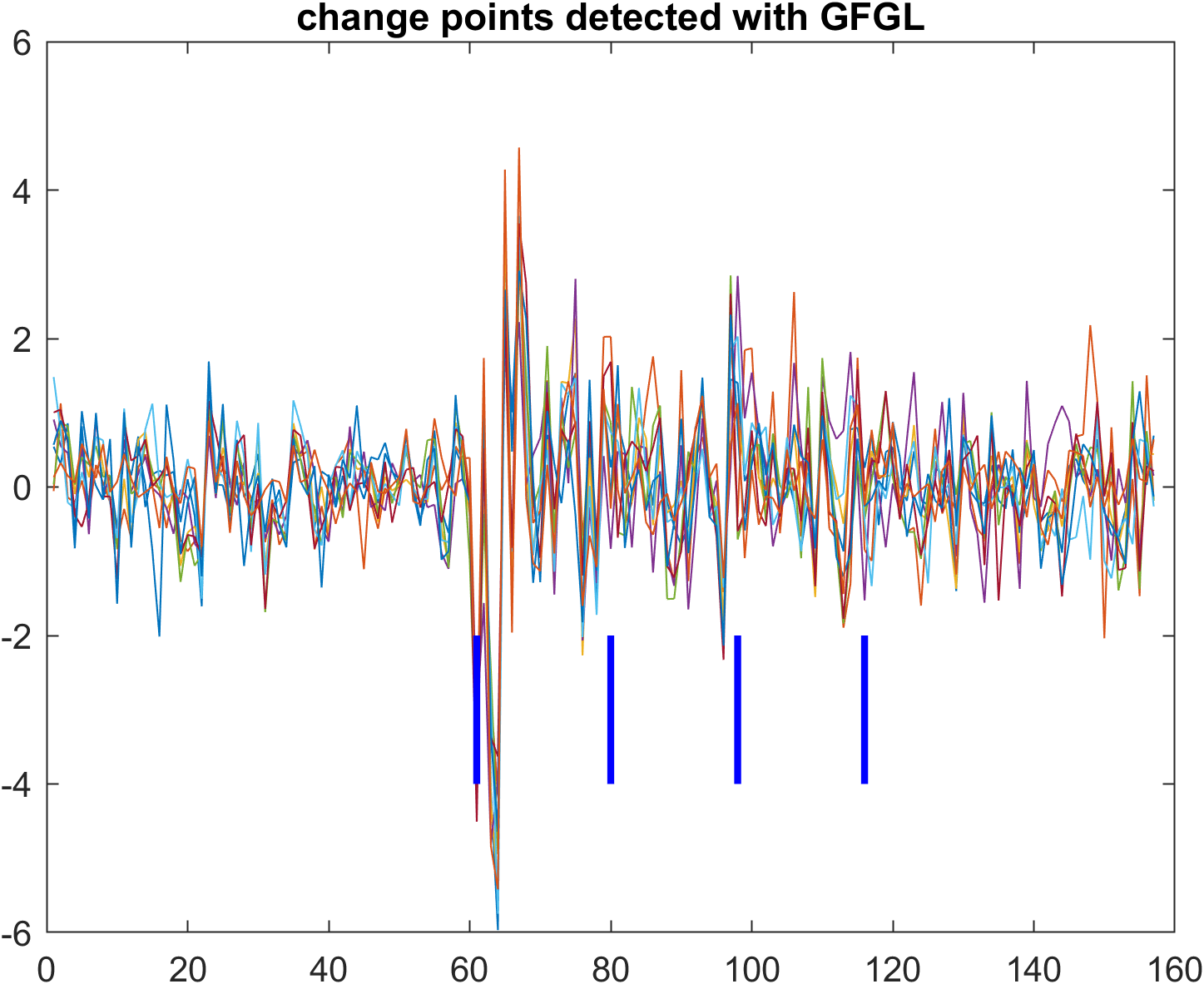}
\caption{ \footnotesize The change points identified with $\lambda_1=0.35$ and $\lambda_2=10$ are $\{61,80,98,116\}$.}
\end{subfigure}
\hfill
\begin{subfigure}[b]{0.3\textwidth}
\includegraphics[width=\textwidth]{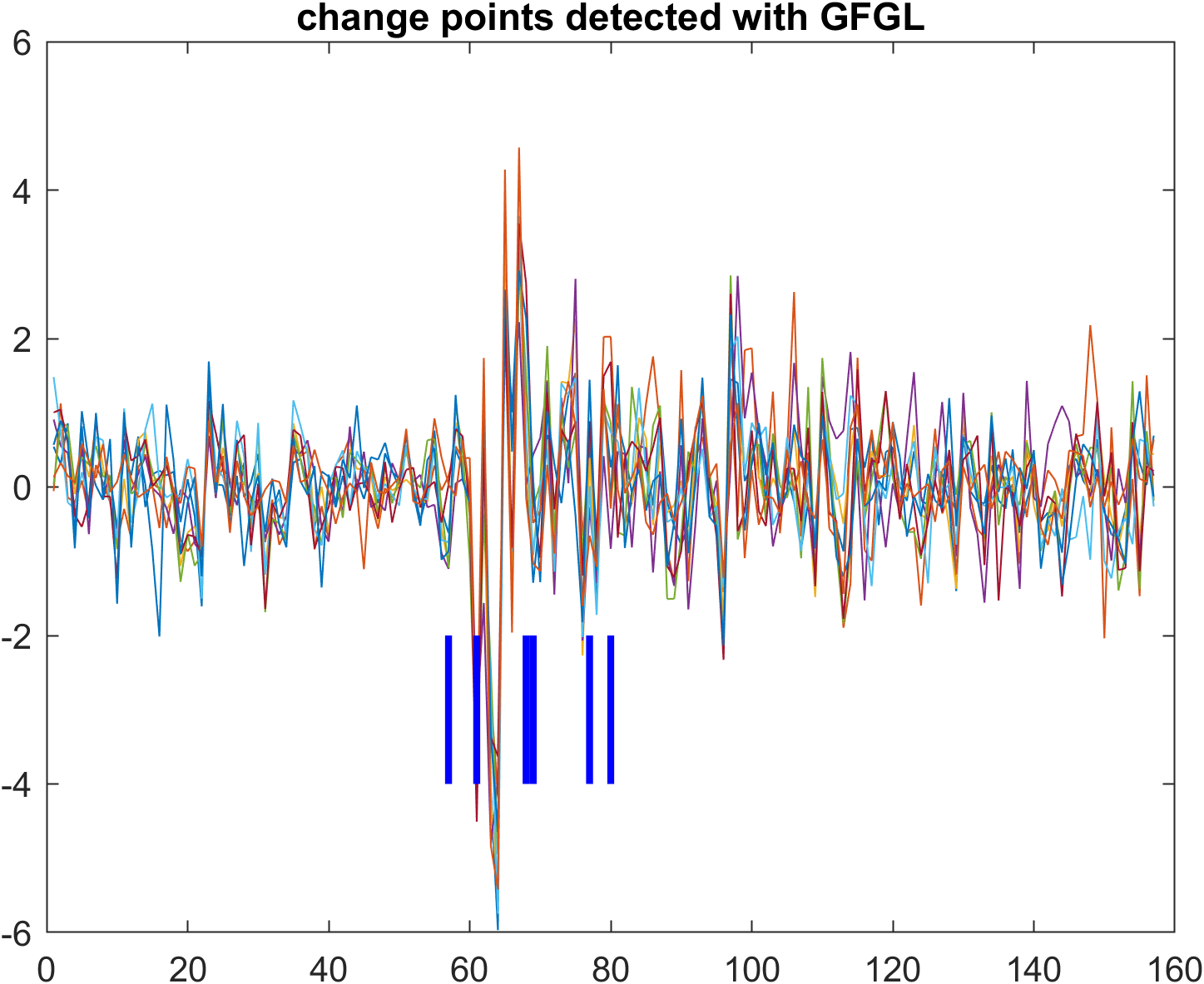}
\caption{ \footnotesize The change points identified with $\lambda_1=0.55$ and $\lambda_2=10$ are $\{57,61,68,69,77,80\}$.}
\end{subfigure}
\hfill
\begin{subfigure}[b]{0.3\textwidth}
\vspace{0.1cm}
\includegraphics[width=\textwidth]{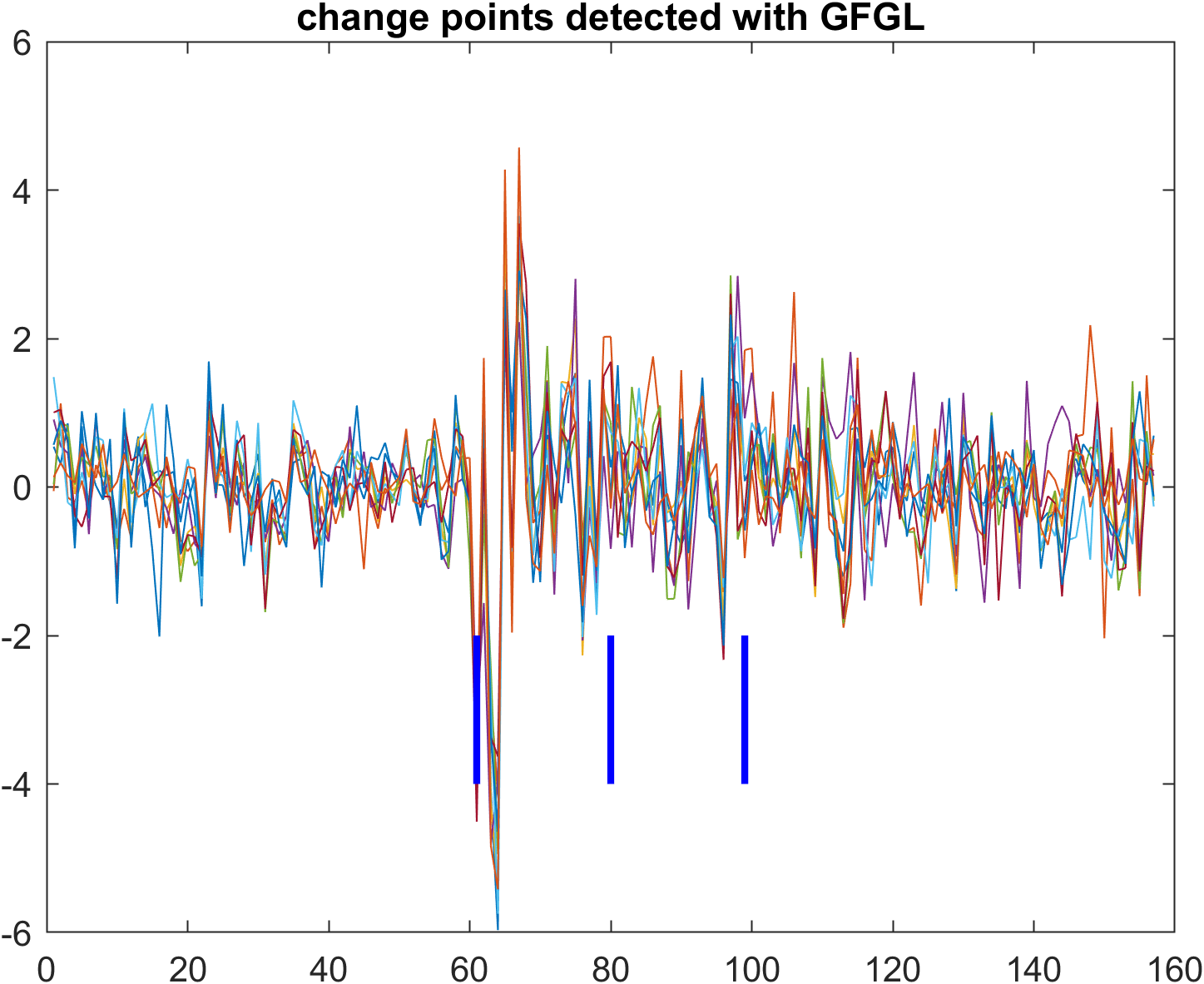}
\caption{\footnotesize The change points identified with $\lambda_1=0.25$ and $\lambda_2=20$ are $\{61,80,98\}$.}
\end{subfigure}
\hfill
\begin{subfigure}[b]{0.3\textwidth}
\includegraphics[width=\textwidth]{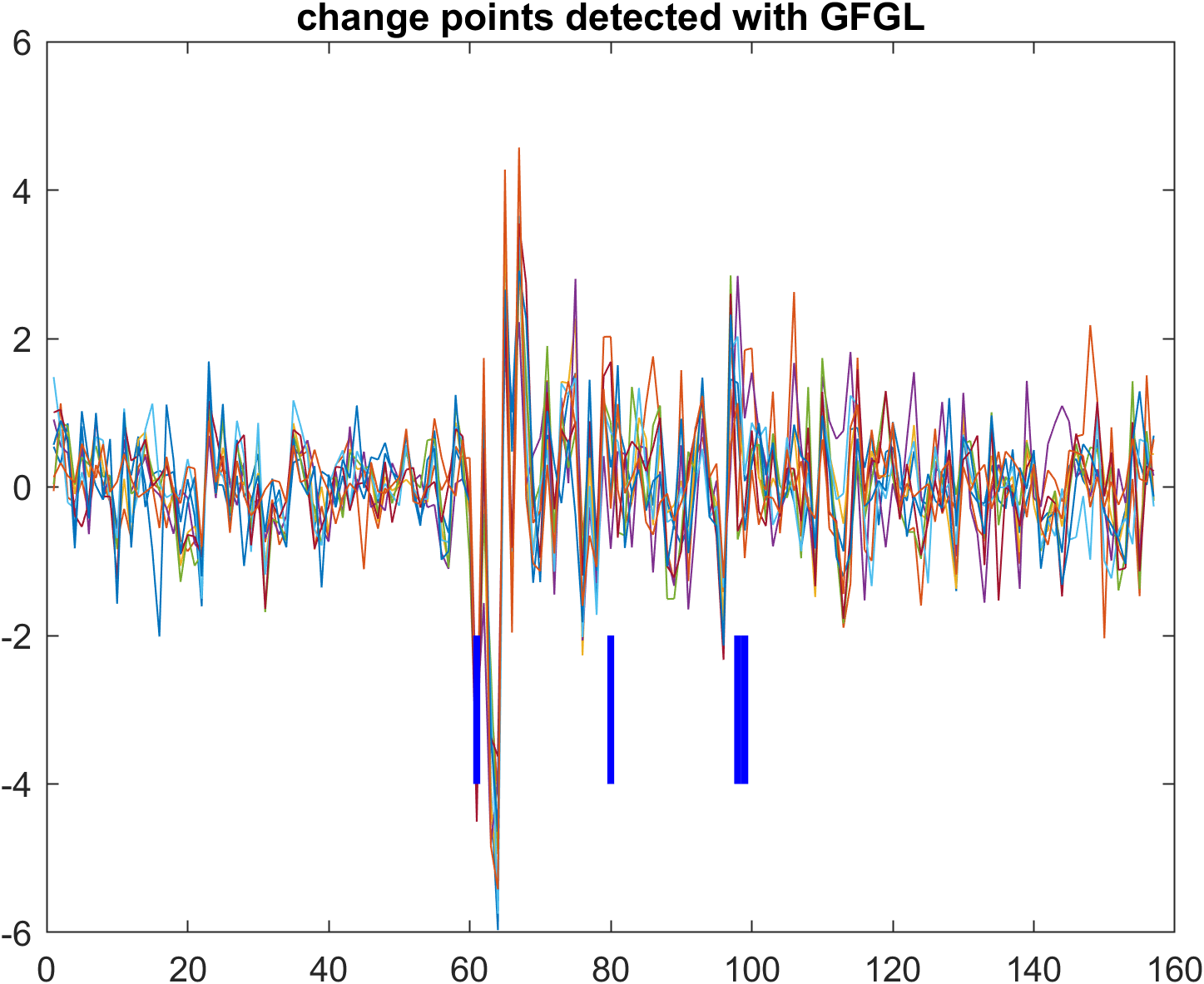}
\caption{ \footnotesize The change points identified with $\lambda_1=0.35$ and $\lambda_2=20$ are $\{61,80,98,99\}$.}
\end{subfigure}
\hfill
\begin{subfigure}[b]{0.3\textwidth}
\includegraphics[width=\textwidth]{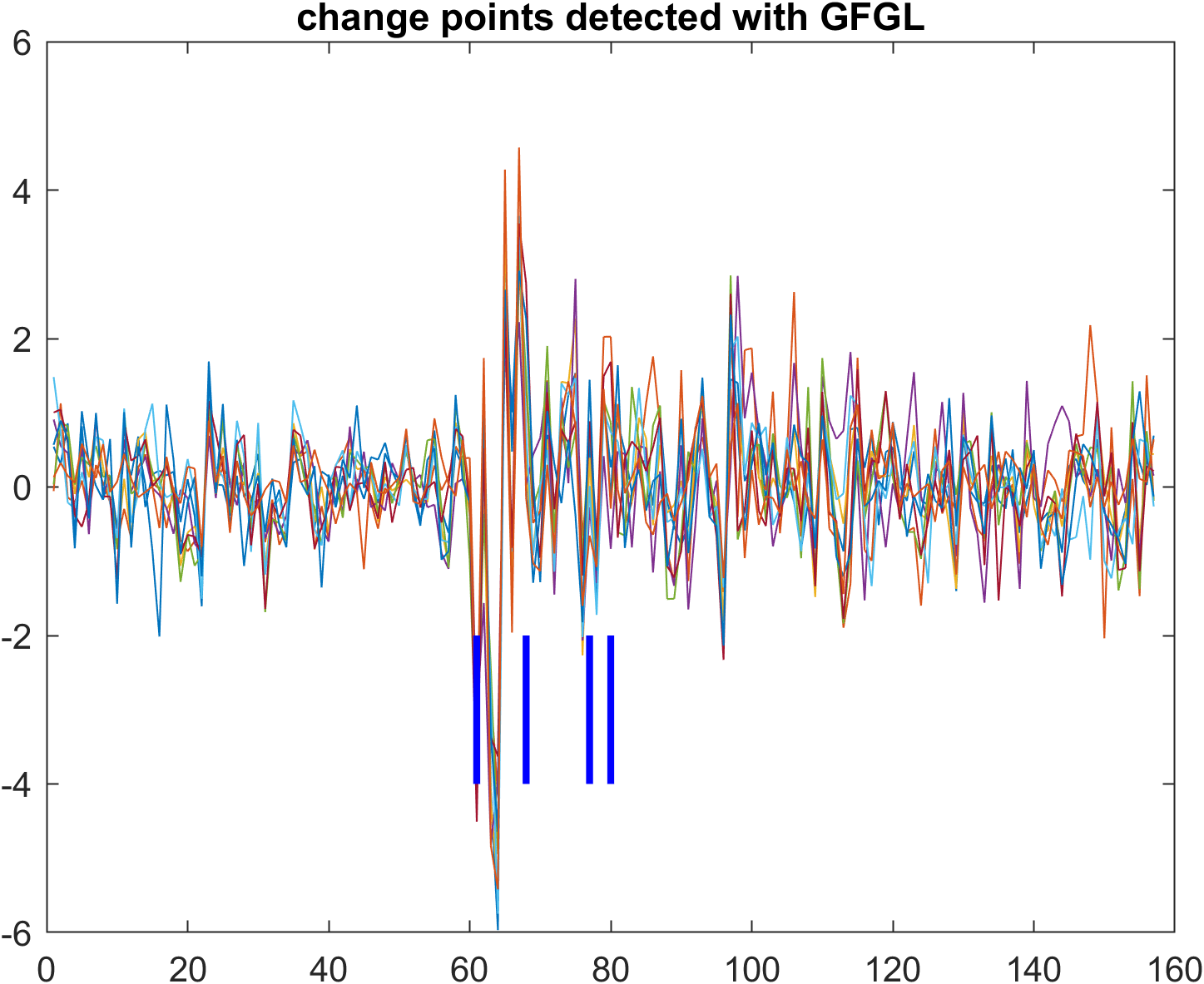}
\caption{ \footnotesize The change points identified with $\lambda_1=0.5$ and $\lambda_2=20$ are $\{61,68,77,80\}$.}
\end{subfigure}
\hfill
\begin{subfigure}[b]{0.3\textwidth}
\vspace{0.1cm}
\includegraphics[width=\textwidth]{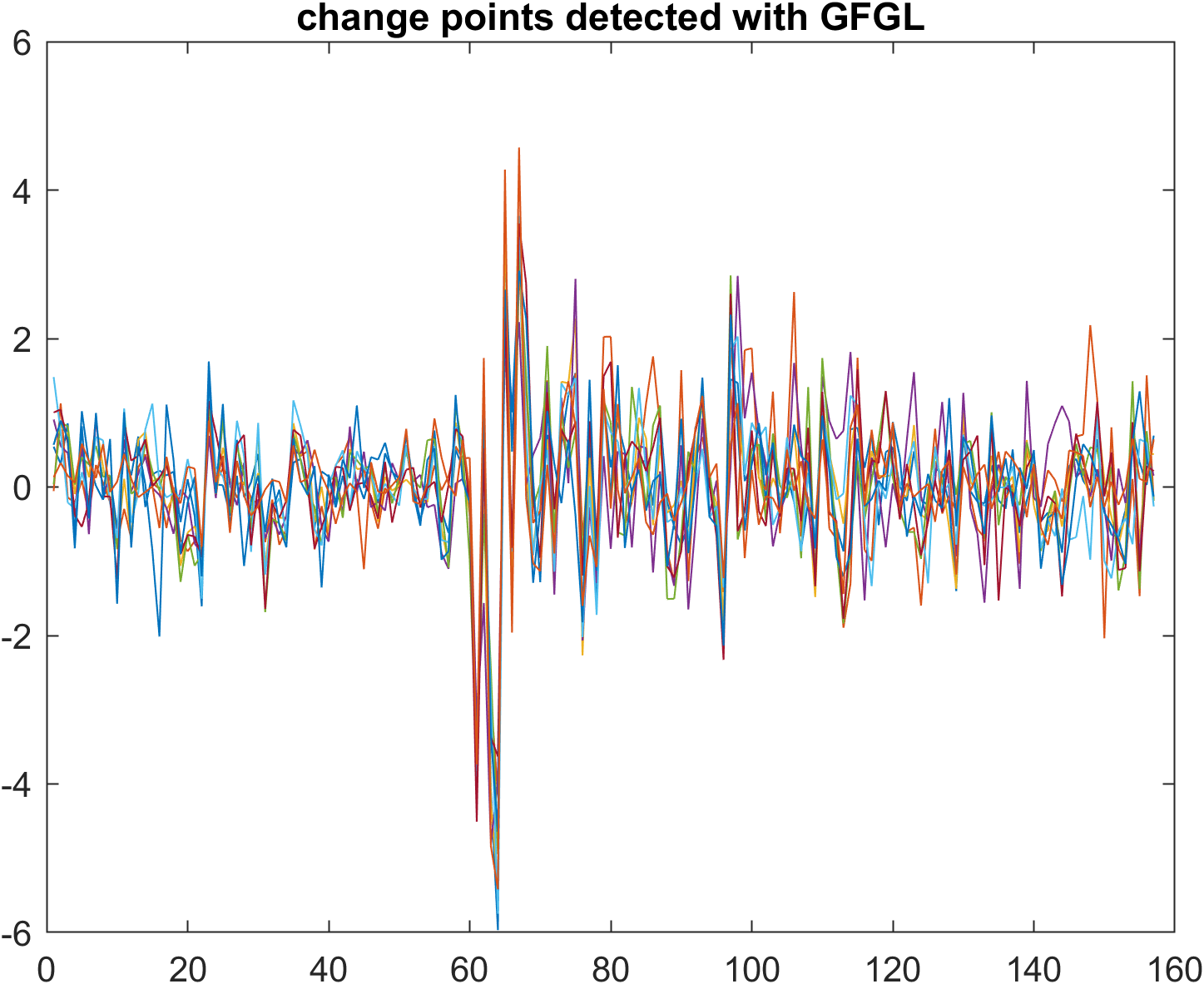}
\caption{ \footnotesize No change point is identified with $\lambda_1=0.25$ and $\lambda_2=60$. \textcolor{white}{xxxxxxxxxxxx}}
\end{subfigure}
\hfill
\begin{subfigure}[b]{0.3\textwidth}
\includegraphics[width=\textwidth]{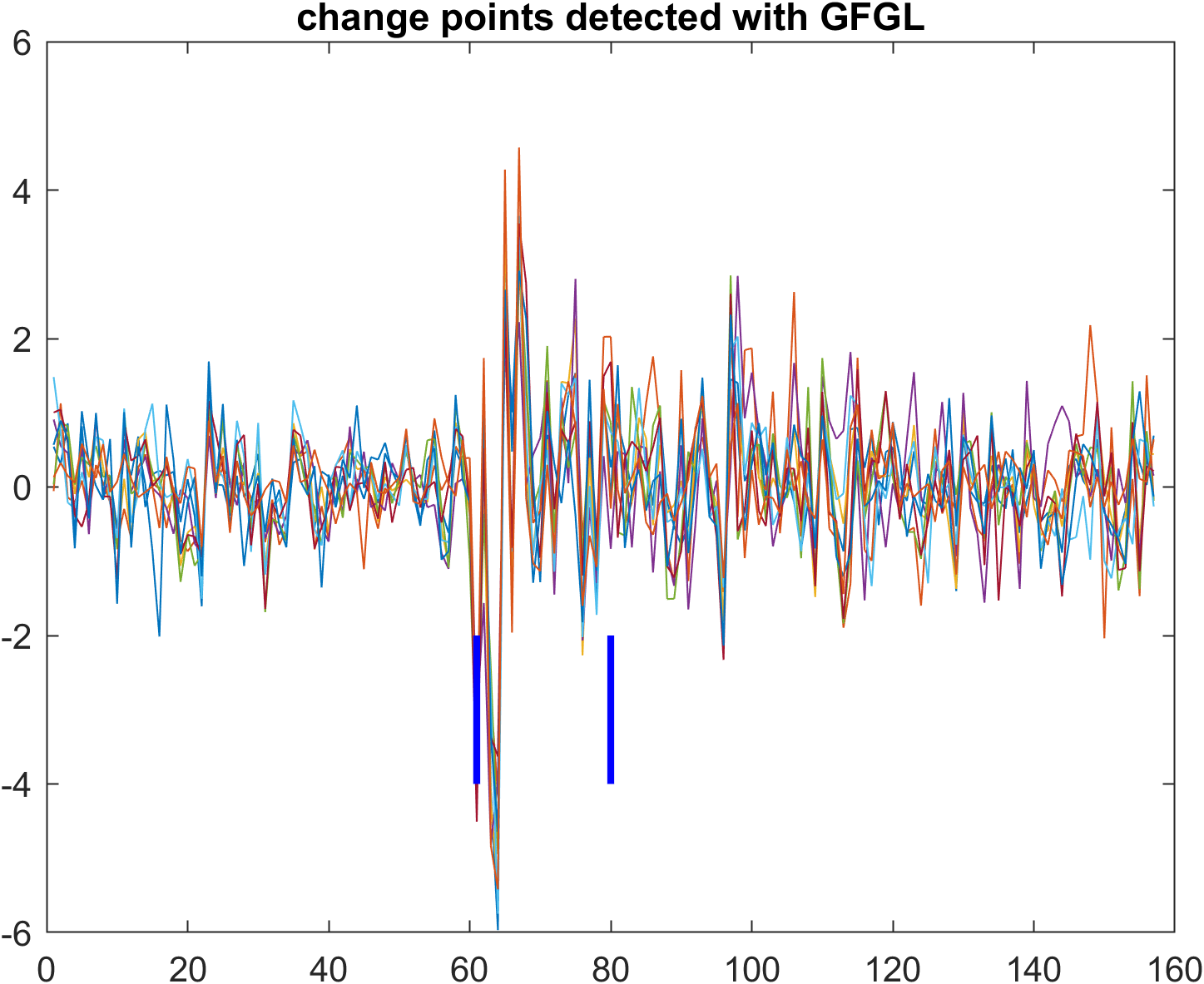}
\caption{ \footnotesize The change points identified with $\lambda_1=0.35$ and $\lambda_2=60$ are $\{61,80\}$.}
\end{subfigure}
\hfill
\begin{subfigure}[b]{0.3\textwidth}
\includegraphics[width=\textwidth]{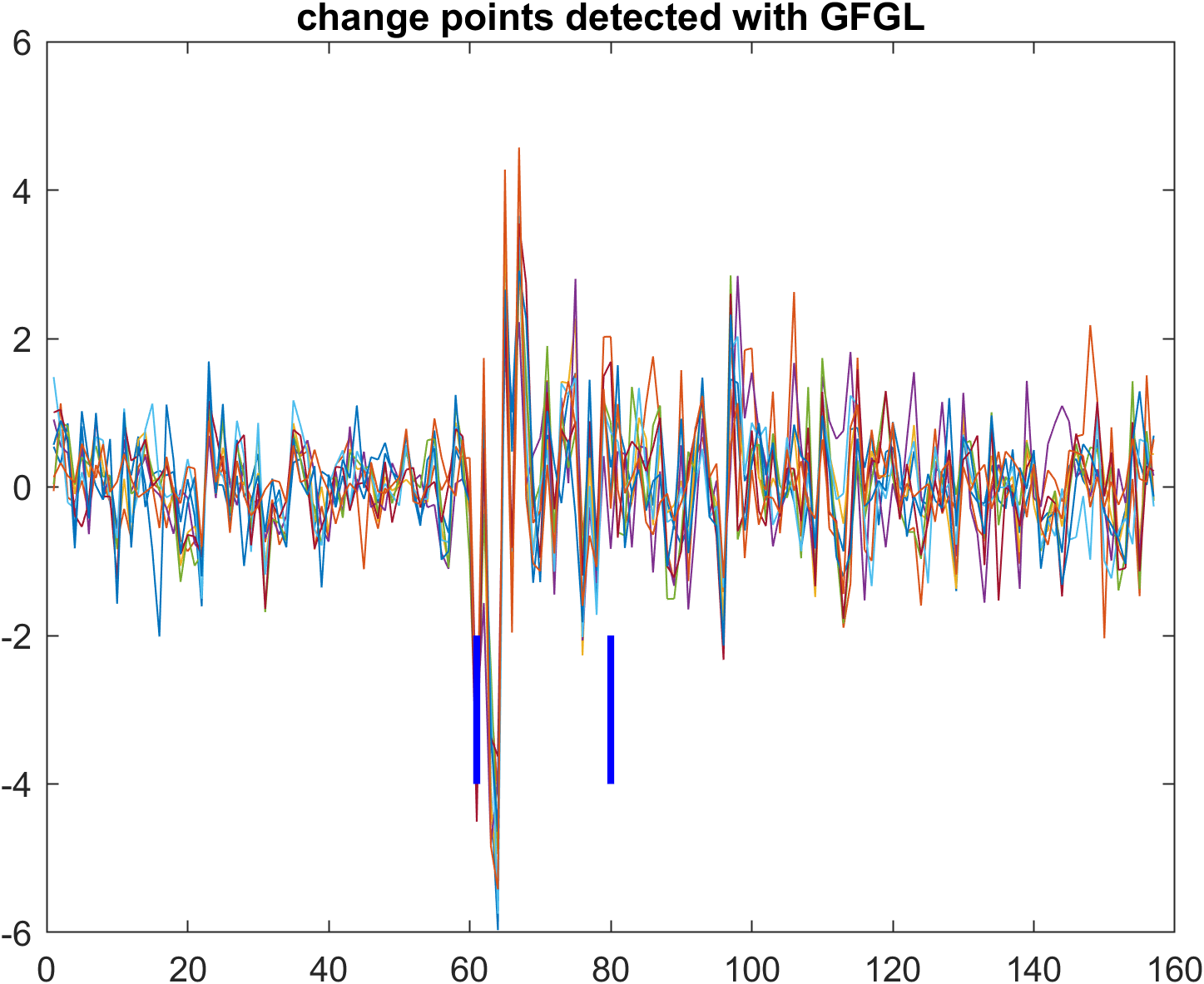}
\caption{ \footnotesize The change points identified with $\lambda_1=0.5$ and $\lambda_2=60$ are $\{61,80\}$.}
\end{subfigure}
\hfill
\begin{subfigure}[b]{0.3\textwidth}
\vspace{0.1cm}
\includegraphics[width=\textwidth]{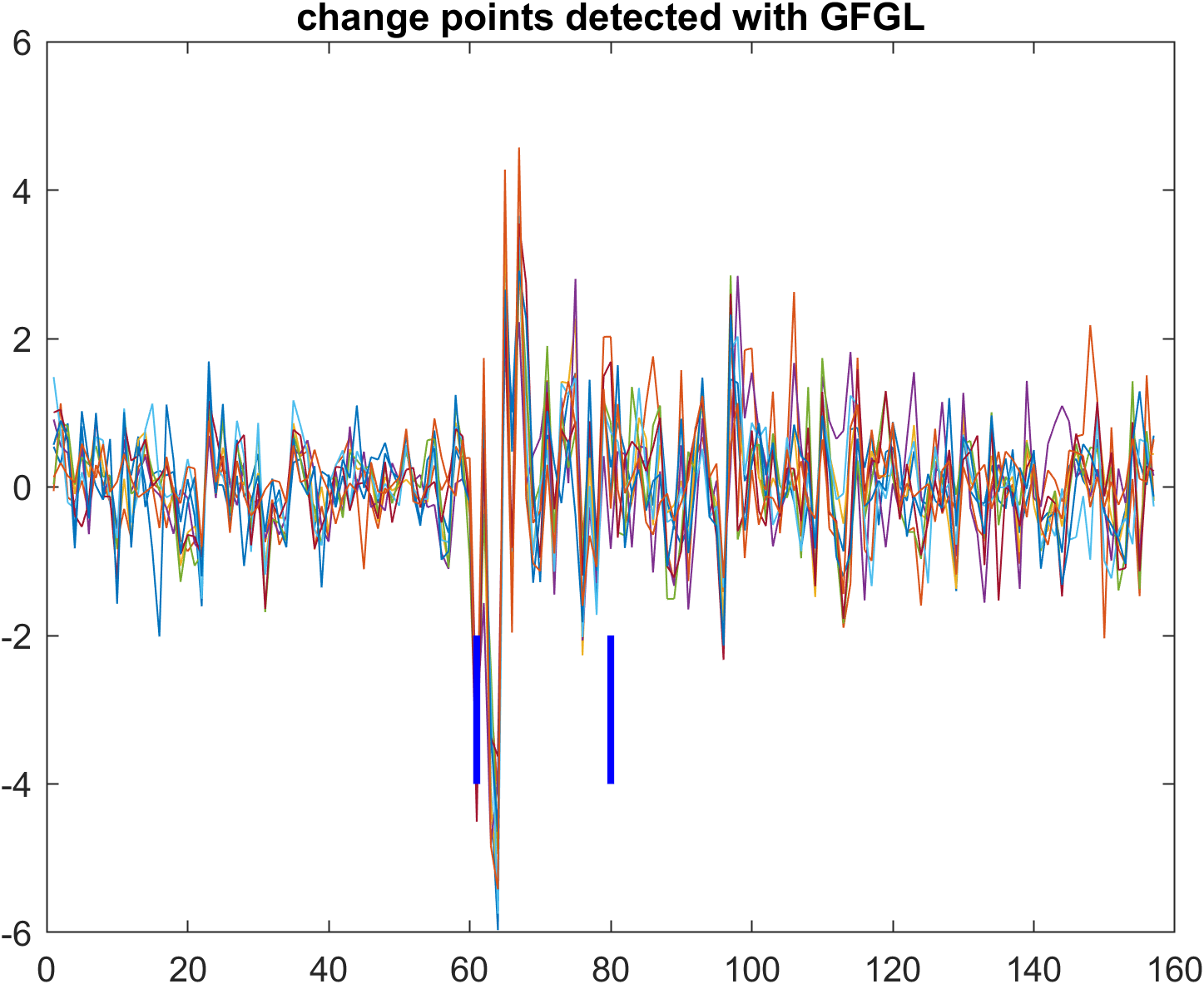}
\caption{ \footnotesize The change points identified with $\lambda_1=0.25$ and $\lambda_2=55$ are $\{61,80\}$.}
\end{subfigure}
\hfill
\caption{ Change points (as blue vertical lines) detected by the GFGL model. }
\end{figure}

\renewcommand{\thefigure}{C.3.2}
\begin{figure}[H]
\centering
\begin{subfigure}[b]{0.85\textwidth}
\includegraphics[width=\textwidth]{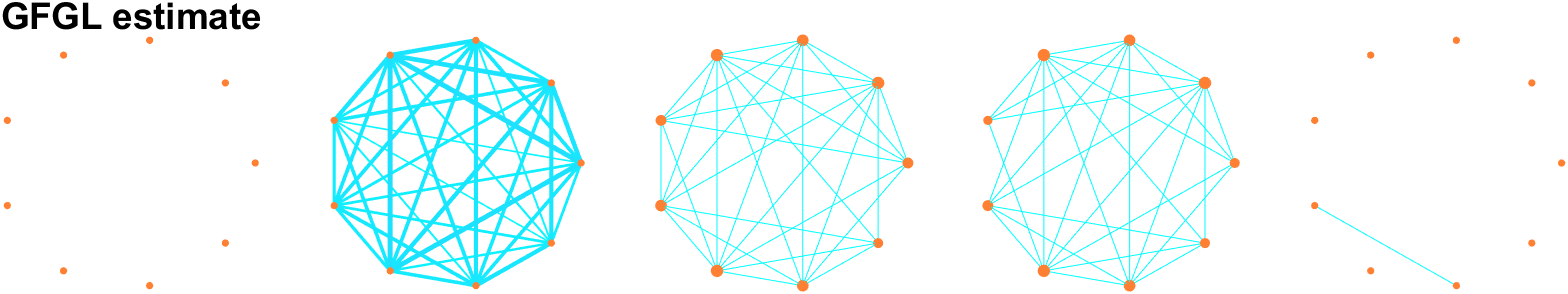}
\caption{The change points identified with $\lambda_1=0.35$ and $\lambda_2=10$ are $\{61,80,98,116\}$.}
\end{subfigure}
\begin{subfigure}[b]{0.85\textwidth}
\vspace{0.3cm}
\includegraphics[width=\textwidth]{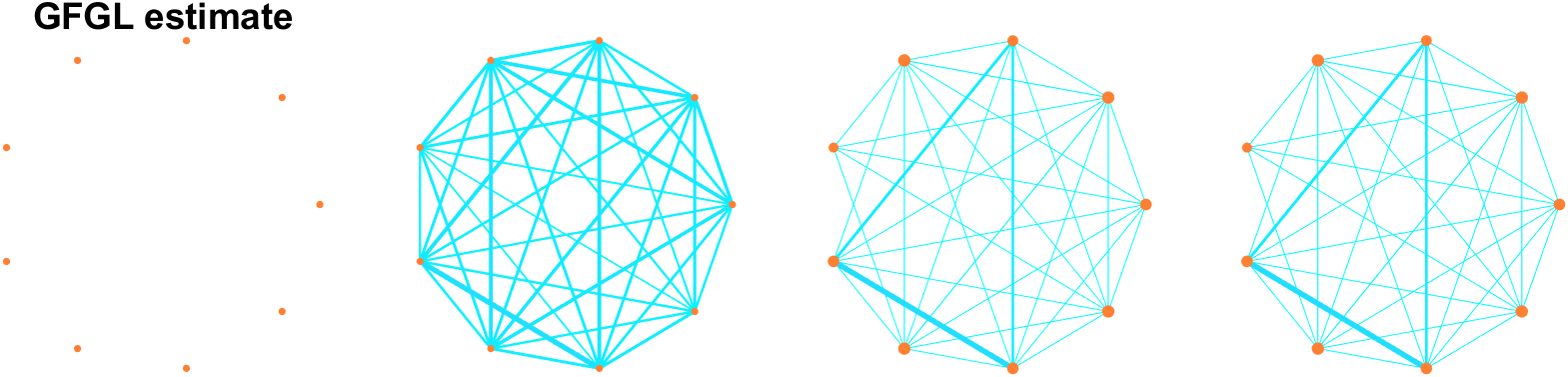}
\caption{The change points identified with $\lambda_1=0.25$ and $\lambda_2=20$ are $\{61,80,98\}$.}
\end{subfigure}
\hfill
\begin{subfigure}[b]{0.85\textwidth}
\vspace{0.3cm}
\includegraphics[width=\textwidth]{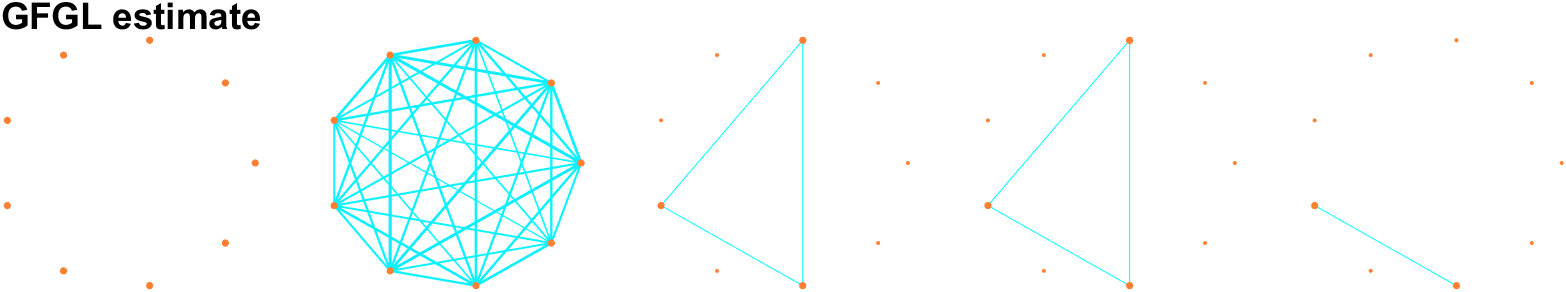}
\caption{The change points identified with $\lambda_1=0.35$ and $\lambda_2=20$ are $\{61,80,98,99\}$.}
\end{subfigure}
\hfill
\begin{subfigure}[b]{0.85\textwidth}
\vspace{0.3cm}
\includegraphics[width=\textwidth]{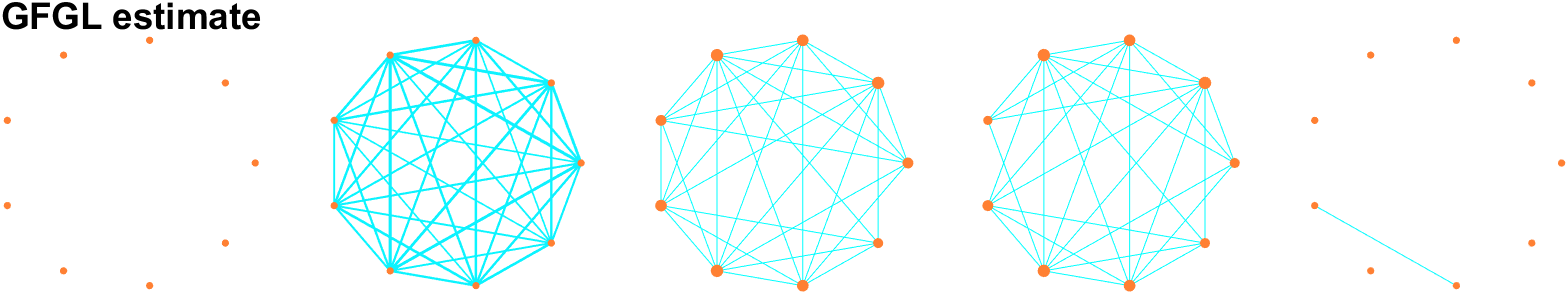}
\caption{The change points identified with $\lambda_1=0.5$ and $\lambda_2=20$ are $\{61,68,77,80\}$.}
\end{subfigure}
\hfill
\begin{subfigure}[b]{0.70\textwidth}
\vspace{0.3cm}
\includegraphics[width=\textwidth]{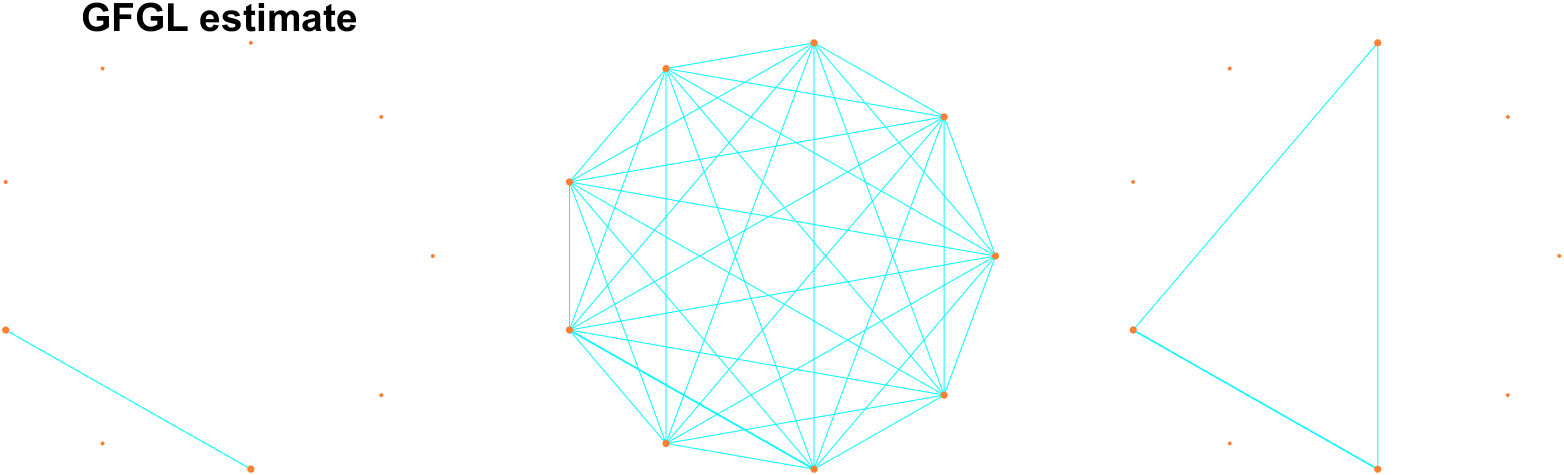}
\caption{The change points identified with $\lambda_1=0.35$ and $\lambda_2=60$ are $\{61,80\}$.}
\end{subfigure}
\hfill
\begin{subfigure}[b]{0.70\textwidth}
\vspace{0.3cm}
\includegraphics[width=\textwidth]{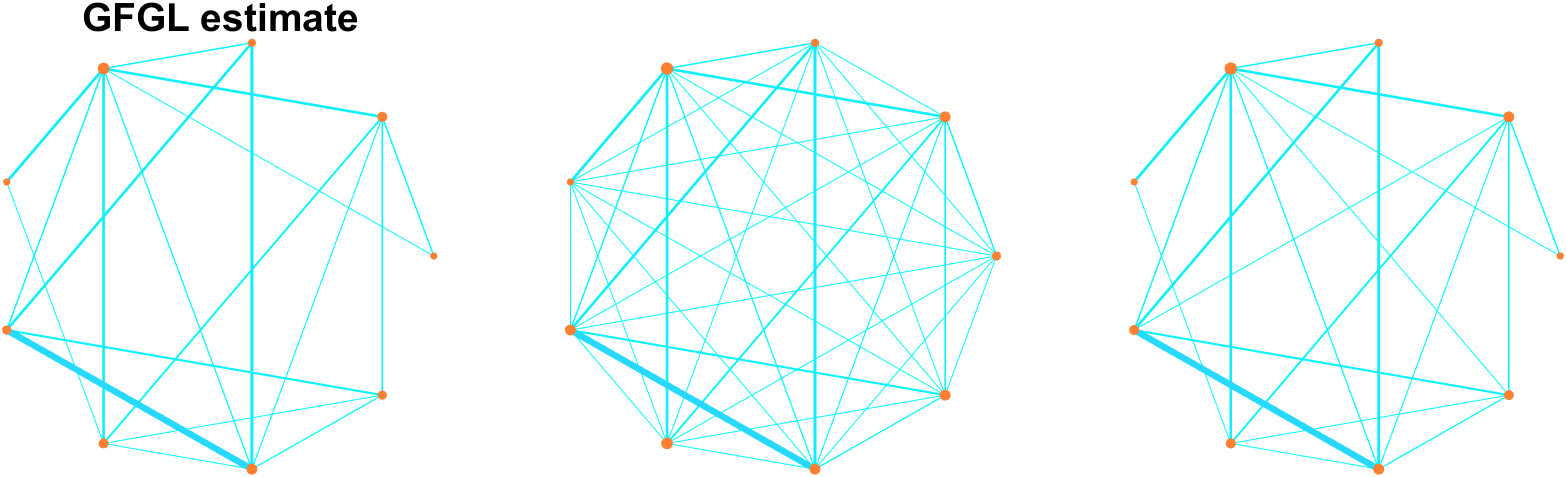}
\caption{The change points identified with $\lambda_1=0.25$ and $\lambda_2=55$ are $\{61,80\}$.}
\end{subfigure}
\hfill
\caption{ Graphs estimated by the group-fused graphical lasso model on real data for different values of the hyperparameters.}
\end{figure}

\renewcommand{\thefigure}{C.3.3}
\begin{figure}[H]
\centering
\includegraphics[width=\textwidth]{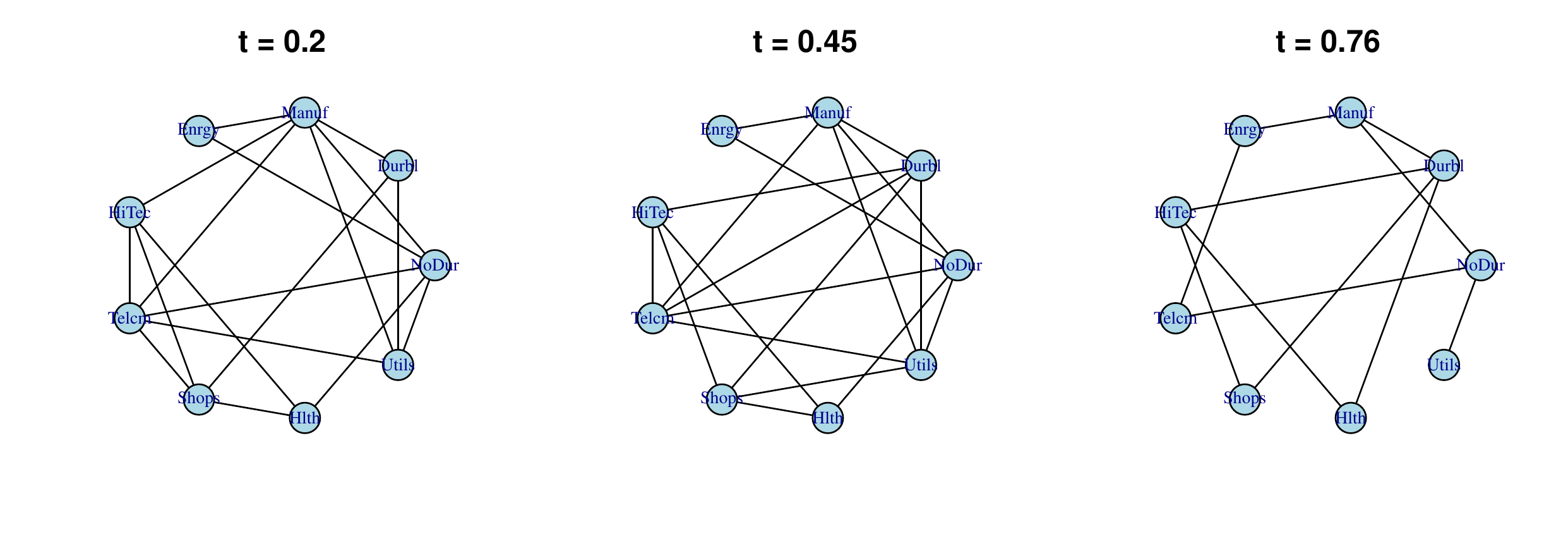}
\caption{ Estimated graph using the ``oracle version" of \textsf{loggle}.}
\end{figure}

\subsection*{C.4 \enskip Pooled estimate}
\renewcommand{\thefigure}{C.4.1}
\begin{figure}[H]
\centering
\includegraphics[width=0.5\textwidth]{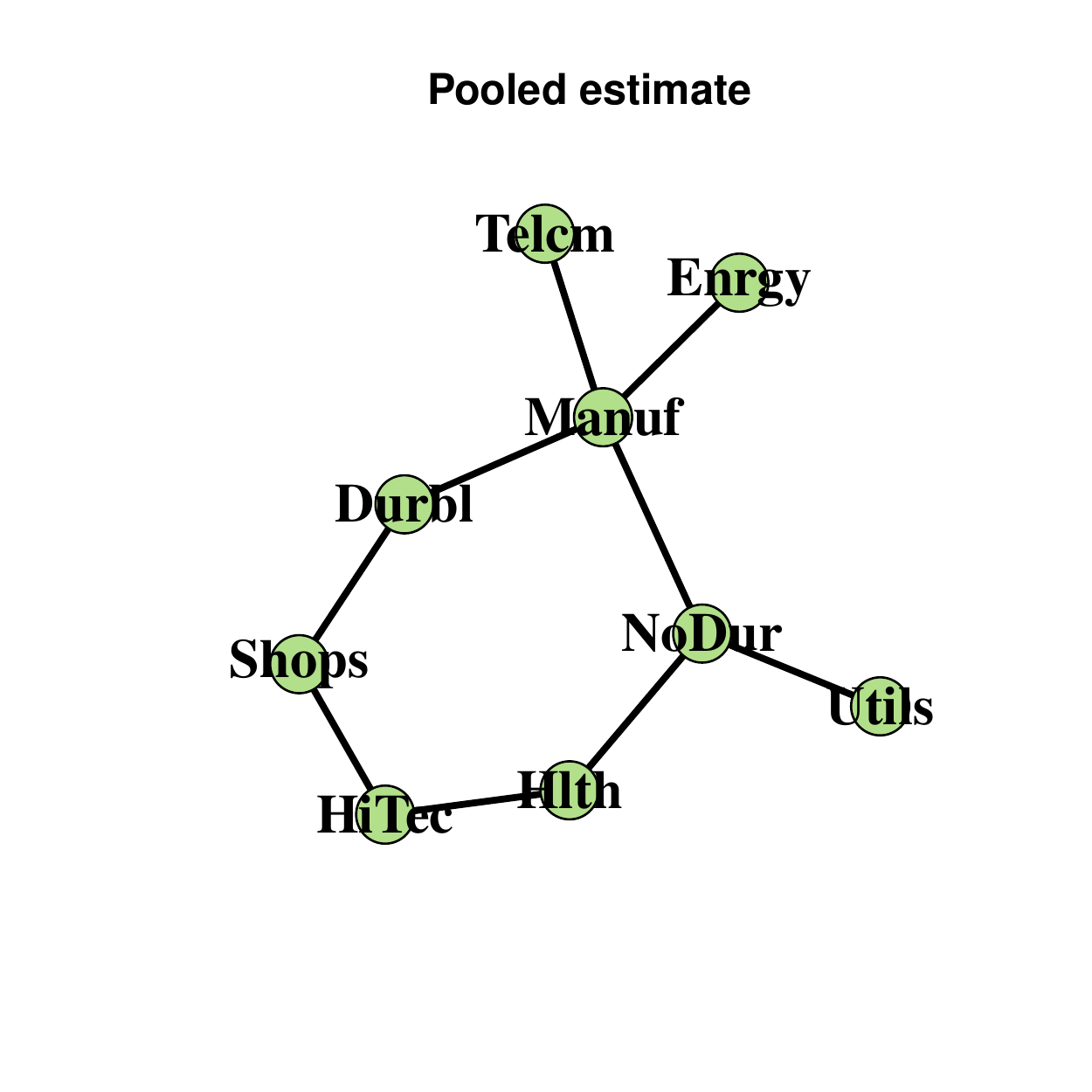}
\caption{ Estimated graph using the adaptive lasso approach as implemented in the \textsf{R} package \textsf{GGMselect} \citep{GGMselect} assuming no change points.}
\end{figure}

\subsection*{C.5 \enskip Sensitivity to the hyperparameter $p_0$ controlling the prior on the number of change points and the hyperparameter $z$ controlling edges' (de)activation}

\renewcommand\thetable{C.5.1}
\begin{table}[H]
    \centering
    \begin{tabular}{l|c|c|c}
        hyperparameter&prob. $\kappa = 2$&MAP configuration&MAP prob.\\
        \hline
        \hline
        $p_0 = 0.20$, $z = 0.1$&0.998&(61 79)&0.3735\\
        $p_0 = 0.50$, $z = 0.1$&0.997&(61 79)&0.3922\\
        $p_0 = 0.80$, $z = 0.1$&0.997&(61 79)&0.3810\\
        $p_0 = 0.20$, $z = 0.2$&0.997&(61 79)&0.3742\\
        $p_0 = 0.20$, $z = 0.4$&0.975&(61 79)&0.3614\\
    \end{tabular}
    \caption{Prior sensitivity: posteriors estimates of change point configuration.}
    \vspace{-\baselineskip}
\end{table}

\renewcommand{\thefigure}{C.5.1}
\begin{figure}[H]
\centering
\includegraphics[width=\textwidth]{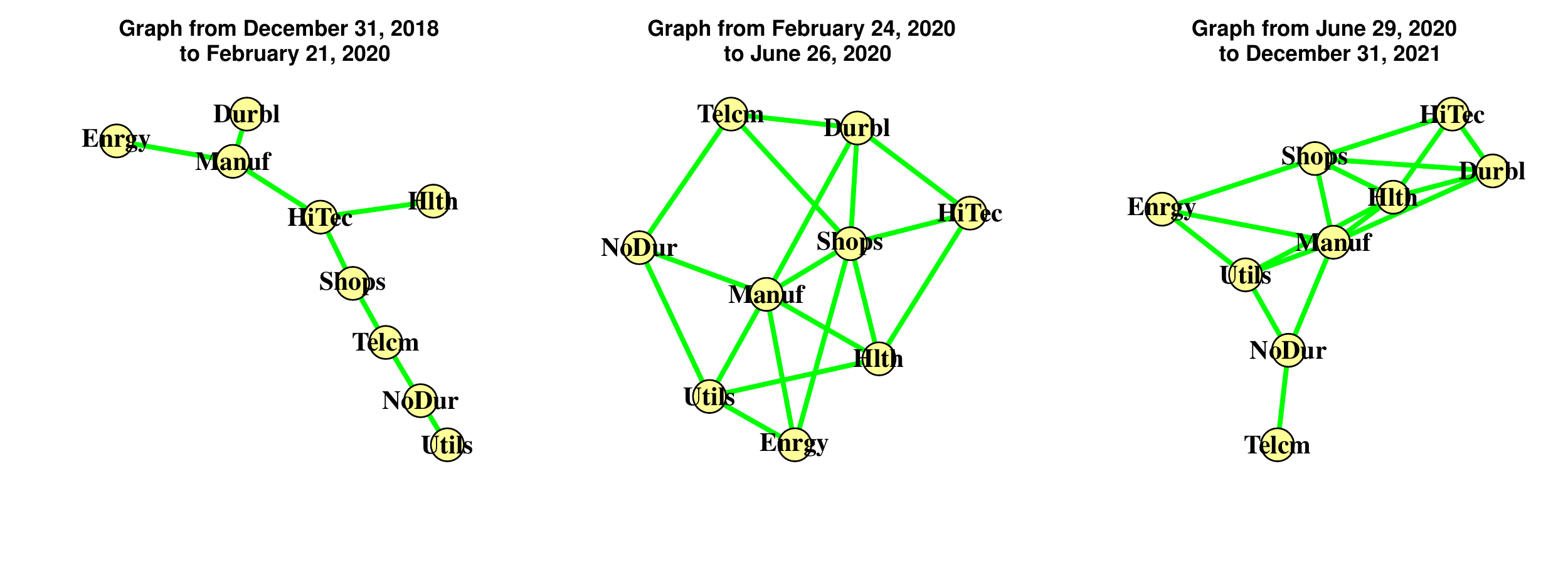}
\caption{ Posterior estimates of the graphs obtained with  $z = 0.2$. 
Threshold of inclusion is set to achieve an expected posterior specificity of at least 95\%.}
\end{figure}

\renewcommand{\thefigure}{C.5.2}
\begin{figure}[H]
\centering
\includegraphics[width=\textwidth]{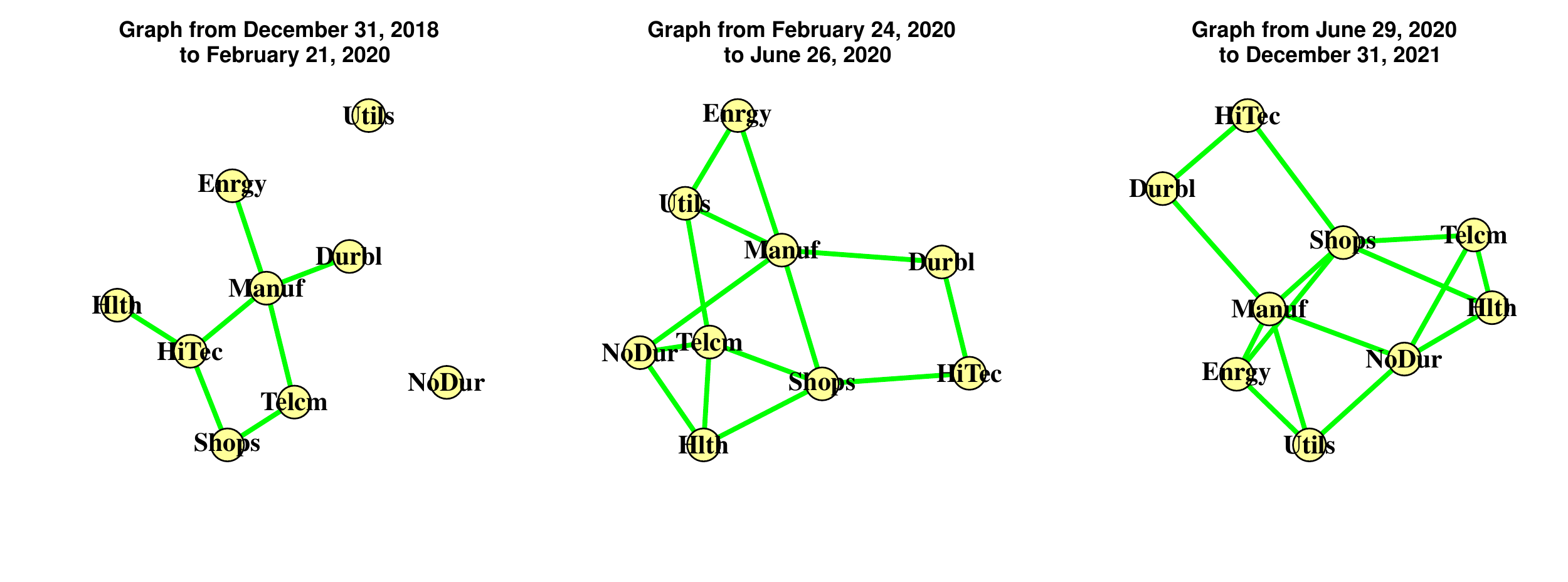}
\caption{ Posterior estimates of the graphs  obtained with  $z = 0.4$. 
Threshold of inclusion is set to achieve an expected posterior specificity of at least 95\%.}
\end{figure}

\begin{table}[!h]
    \centering
    \begin{tabular}{c|c|cc|cc|cc|}
        &&\multicolumn{6}{c}{$z=0.1$}\\
        \hline
        &&\multicolumn{2}{c}{$G_{c_0}$}&\multicolumn{2}{c}{$G_{c_1}$}&\multicolumn{2}{c}{$G_{c_2}$}\\
        \hline 
        &&Active&Inactive&Active&Inactive&Active&Inactive\\
        \hline
        $z=0.2$&Active&8&0&10&8&12&5\\
        &Inactive&0&73&1&62&1&63\\
        \hline
        $z=0.4$&Active&6&8&10&18&10&20\\
        &Inactive&2&65&1&52&3&48\\
    \end{tabular}
    \caption{Edge detection comparison for $z=0.1$, $z=0.2$, and $z=0.4$.}
    \vspace{-\baselineskip}
\end{table}

\subsection*{C.6 \enskip Goodness-of-fit: Posterior predictive checking}
The following figures showcase the posterior predictive checking for the real data. For brevity, plots refer to the first portfolio, however analogous results are observed for all  portfolios under investigation. Posterior predictive checking consists in simulating data from the posterior predictive distribution of a new data point, in our case an entire realization of the time series, given the observed time series.  
\renewcommand{\thefigure}{C.6.1}
\begin{figure}[H]
\centering
\includegraphics[width=0.85\textwidth]{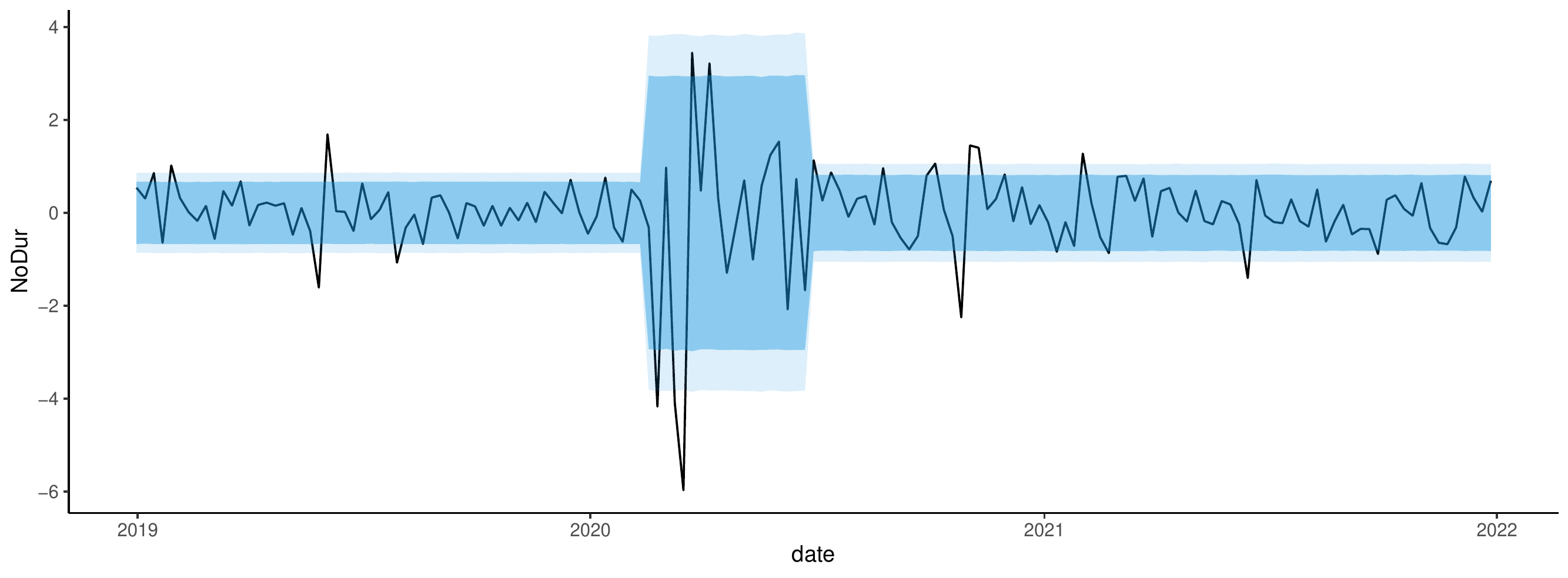}
\caption{ Posterior predictive checking for NoDur portfolio, conditionally on the change points configuration. Dark and light shaded areas correspond respectively to 90\% and 95\% credible intervals. The posterior predictive is obtained simulating 100 000 replicates of the data keeping the change points fixed. }
\end{figure}

\renewcommand{\thefigure}{C.6.2}
\begin{figure}[H]
\centering
\includegraphics[width=0.85\textwidth]{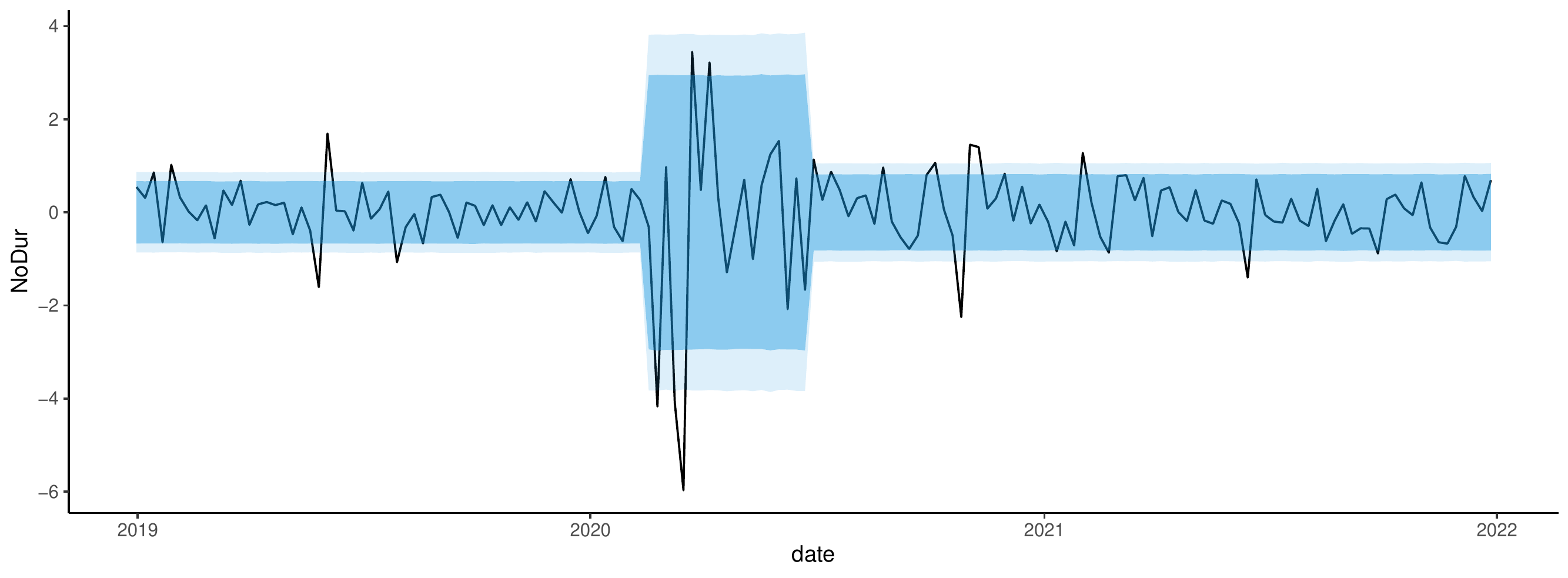}
\caption{ Posterior predictive checking for NoDur portfolio, conditionally on the graph topology. Dark and light shaded areas correspond respectively to 90\% and 95\% credible intervals. The black continuous line is the real observed standardised time series of returns. The posterior predictive is obtained simulating 100 000 replicates of the data keeping the graphs fixed. }
\end{figure}

\renewcommand{\thefigure}{C.6.3}
\begin{figure}[H]
\centering
\includegraphics[width=0.85\textwidth]{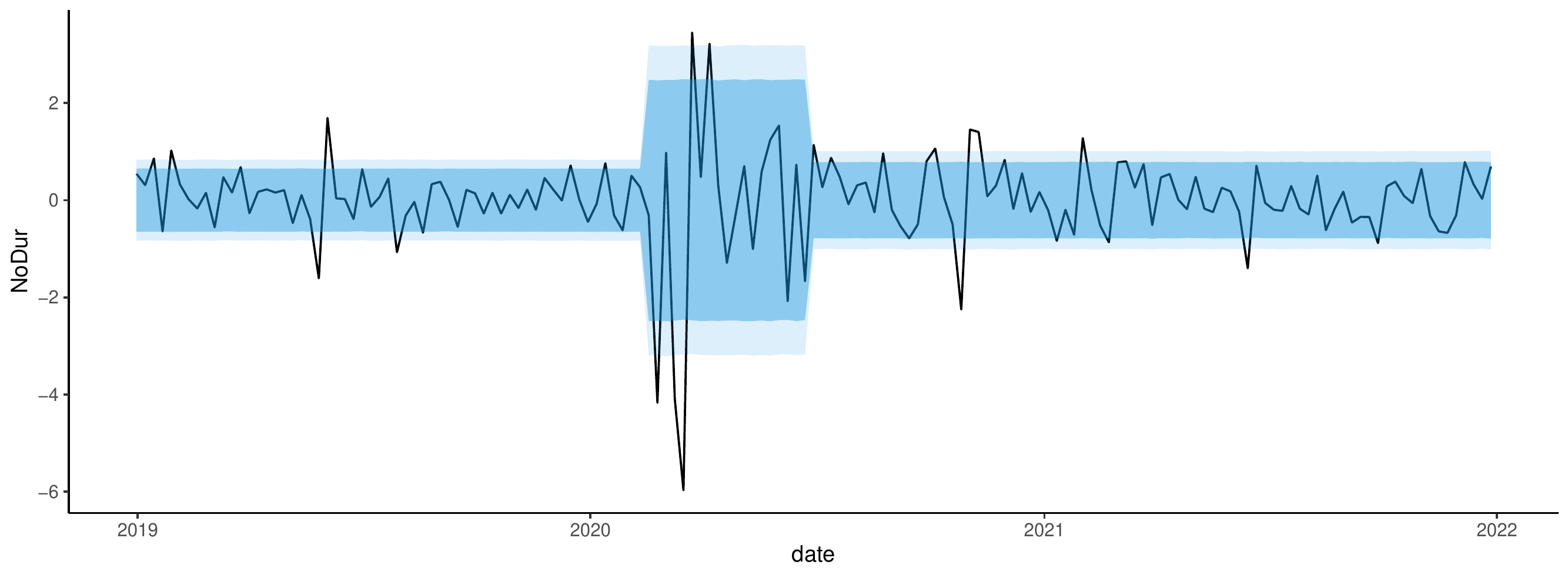}
\caption{ Posterior predictive checking for NoDur portfolio, conditionally on the covariance matrices. Dark and light shaded areas correspond respectively to 90\% and 95\% credible intervals. The black continuous line is the real observed standardised time series of returns. The posterior predictive is obtained simulating 100 000 replicates of the data keeping the covariance matrices fixed. }
\end{figure}

\clearpage
\section*{D \enskip Algorithmic mixing and computational time}
\subsection*{D.1 \enskip Mixing performance}
\renewcommand{\thefigure}{D.1.1}
\begin{figure}[H]
\centering
\includegraphics[trim={0cm 3.5cm 0cm 0cm},clip,width=\textwidth]{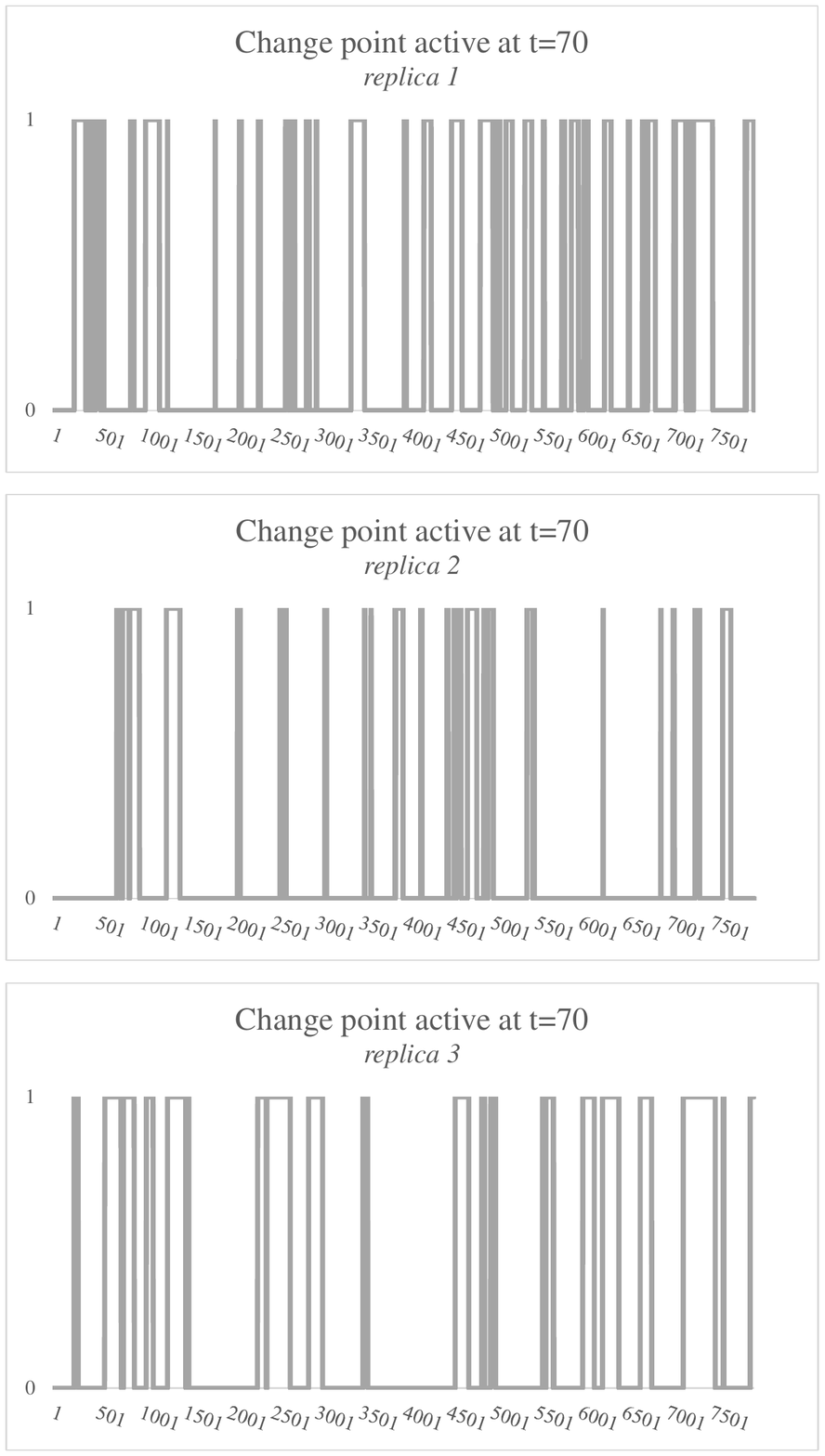}
\caption{Outer algorithm - simulated data - Scenario 3: Trace plots of the indicator variable of time point 70 being a change point. 8000 iterations after burn-in.}
\end{figure}

\renewcommand{\thefigure}{D.1.2}
\begin{figure}[H]
\centering
\includegraphics[trim={4cm 2cm 4cm 2cm},clip,width=\textwidth]{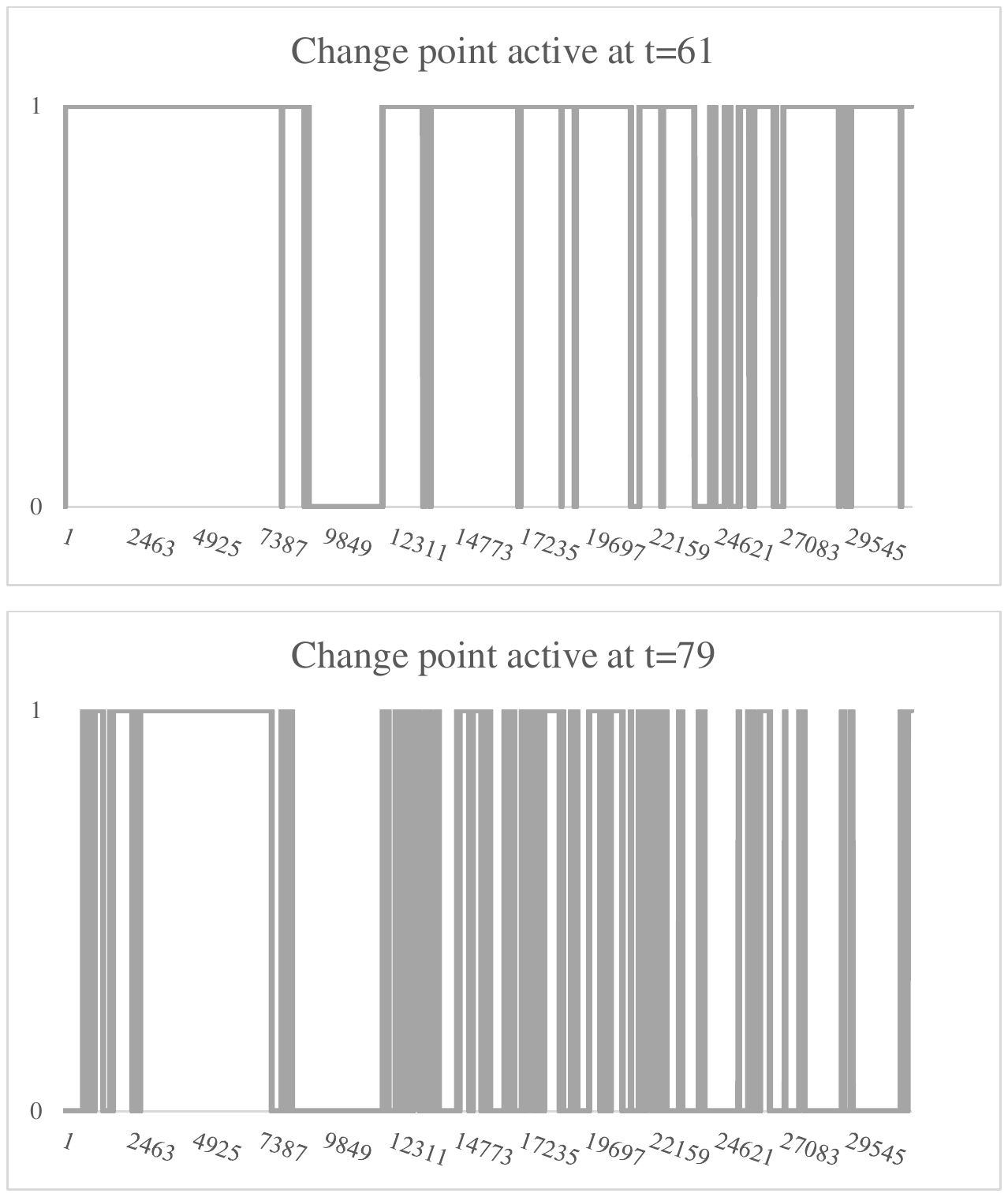}
\caption{Outer algorithm - real data: Trace plots of the indicator variables of time points 61 and 79 being change points in the real data analysis. 32000 iterations, including burn-in.}
\end{figure}

\subsubsection*{D.2 \enskip Computational time}
The effective computational time needed to estimate the model obviously depends on the dataset dimension, i.e., number of nodes $p$ and number of time points $T$, and the SMC parameters, i.e., number of particles $N$ and number of mutation steps $M$. However, it depends also on the true data generating process, i.e., true number of change points and true graph topology, as well as on the posterior distribution of the number of change points. Dependence from the latter is due to the fact that exploring configurations with a high number of change points is computationally more expensive. Moreover, we empirically observe that the cost for computing the marginal likelihood with the inner algorithm increases while exploring configurations distant from the posterior mode and the true change point configuration.  
In Table D.2.1 we report the recorded average computational time for the inner algorithm in the simulation studies and the real data analysis. Table D.2.2 contains the computational time per one iteration of the inner algorithm fixing a specific number of equally spaced change points proposed by the outer algorithm. Computational times tend to be higher in  Table D.2.2 than in Table D.2.1, because the configurations evaluated in the former are far from the posterior mode and the true configurations, which present one change point. In this sense computational times in Table D.2.2 may be intended as a ``worst case" scenario. The algorithm is coded in \textsf{R}, does not exploit parallelization, and is performed with an Intel Xeon W-1250 processor.

\renewcommand{\thetable}{D.2.1}
\begin{table}[H]
    \centering
    \begin{tabular}{c|c|c|c|}
       Data&Time in seconds\\
       \hline
       Scenario 1&0.25\\
       Scenario 2&0.43\\
       Scenario 3&0.80\\
       Scenario 4&3.90\\
       Scenario 5&0.82\\
       Financial data&1.07\\
    \end{tabular}
    \caption{Average computational time in seconds for one iteration of the inner algorithm for the simulation studies and the application. Algorithm is coded in \textsf{R}, does not employ parallelization, and is performed with an Intel Xeon W-1250 processor.}
    \vspace{-\baselineskip}
\end{table}

\renewcommand{\thetable}{D.2.2}
\begin{table}[H]
    \centering
    \begin{tabular}{c|c|c|c|}
    Nodes&Particles&proposed $\kappa$&Time\\
    \hline
       &&$\kappa=2$&0.34 sec\\\cline{3-4}
       &N = 50&$\kappa=3$&0.35 sec\\\cline{3-4}
       &&$\kappa=4$&0.49 sec\\\cline{2-4}
       &&$\kappa=2$&0.59 sec\\\cline{3-4}
       $p=10$&N = 100&$\kappa=3$&0.79 sec\\\cline{3-4}
       &&$\kappa=4$&0.81 sec\\\cline{2-4}
       &&$\kappa=2$&1.15 sec\\\cline{3-4}
       &N = 200&$\kappa=3$&1.68 sec\\\cline{3-4}
       &&$\kappa=4$&2.16 sec\\\cline{1-4}
       \hline
       &&$\kappa=2$&2.10 sec\\\cline{3-4}
       &N = 50&$\kappa=3$&2.67 sec\\\cline{3-4}
       &&$\kappa=4$&3.36 sec\\\cline{2-4}
       &&$\kappa=2$&4.10 sec\\\cline{3-4}
       $p=20$&N = 100& $\kappa=3$&5.52 sec\\\cline{3-4}
       &&$\kappa=4$&6.81 sec\\\cline{2-4}
       &&$\kappa=2$& 8.21 sec\\\cline{3-4}
       &N = 200&$\kappa=3$& 10.95 sec\\\cline{3-4}
       &&$\kappa=4$& 13.78 sec\\\cline{1-4}
       \hline
      &&$\kappa=2$& 28.19 sec\\\cline{3-4}
       &N = 50&$\kappa=3$&1.06 min \\\cline{3-4}
       &&$\kappa=4$&1.66 min\\\cline{2-4}
       &&$\kappa=2$&24.65 sec\\\cline{3-4}
       $p=50$&N = 100&$\kappa=3$&2.67 min\\\cline{3-4}
       &&$\kappa=4$&3.32 min\\\cline{2-4}
       &&$\kappa=2$&1.94 min\\\cline{3-4}
       &N = 200&$\kappa=3$&5.27 min\\\cline{3-4}
       &&$\kappa=4$&6.62 min\\\cline{1-4}
    \end{tabular}
    \caption{Computational time for one iteration of the inner algorithm for different values of the number of nodes $p$, the number of particles $N$, and the number of change points $\kappa$ in the configuration proposed by the outer algorithm. Here data are simulated for $T=200$ total times points. The truth presents one change point and both graphs (before and after the change point) have $p-1$ activated edges. Algorithm is coded in \textsf{R}, does not employ parallelization, and is performed with an Intel Xeon W-1250 processor.}
    \vspace{-\baselineskip}
\end{table}
\end{document}